\documentclass[aps,pre,reprint,groupedaddress,amsmath]{revtex4-1}
\pdfoutput=1
\usepackage{graphicx,color,dcolumn}

\newcommand{\mub}{\mu_{B}}
\newcommand{\kb}{k_{B}}

\newcommand{\va}{\vec a}

\newcommand{\vB}{\vec B}
\newcommand{\vb}{\vec b}

\newcommand{\ve}{\vec e}

\newcommand{\vj}{\vec j}

\newcommand{\vn}{\vec n}
\newcommand{\vp}{\vec p}
\newcommand{\vq}{\vec q}
\renewcommand{\vr}{\vec r}

\newcommand{\vS}{\vec S}
\newcommand{\vs}{\vec s}
\newcommand{\vu}{\vec u}
\newcommand{\vv}{\vec v}
\newcommand{\vnabla}{\vec \nabla}

\newcommand{\vpsi}{\mbox{\boldmath $\psi$}}

\newcommand{\cE}{{\cal E}}
\newcommand{\cDD}{\boldsymbol{\cal D}}

\newcommand{\cK}{{\cal K}}
\newcommand{\cG}{{\cal G}}
\newcommand{\cR}{{\cal R}}
\newcommand{\cS}{{\cal S}}
\newcommand{\cU}{{\cal U}}
\newcommand{\limp}{\lambda_{\rm imp}}

\renewcommand{\vec}[1]{\mathbf{#1}}

\begin{document}

\title{Theory of the low-temperature longitudinal spin Seebeck effect}

\author{Rico Schmidt and Piet W. Brouwer}
\affiliation{Dahlem Center for Complex Quantum Systems and Fachbereich Physik, Freie Universit\"at Berlin, 14195 Berlin, Germany}

\date{\today}

\begin{abstract}
Using a simplified microscopic model of coupled spin and lattice excitations in a ferromagnetic insulator we evaluate the magnetic-field dependence of the spin Seebeck effect at low temperatures. The model includes Heisenberg exchange coupling, a harmonic lattice potential, and a pseudo-dipolar exchange interaction. Our approach goes beyond previous work [Phys.\ Rev.\ B {\bf 98}, 134421 (2018)] in that it does not rely on the {\em a priori} assumption of a fast equilibration of the magnon and phonon distributions. Our theory shows that singular features in the magnetic-field dependence of the spin Seebeck effect at low temperatures observed by Kikkawa {\em et al.} [Phys.\ Rev.\ Lett.\ {\bf 117}, 207203 (2016)] are independent of the relative strength of magnon-impurity and phonon-impurity scattering.
\end{abstract}

\maketitle

\section{Introduction}

The spin Seebeck effect refers to the phenomenon that an applied temperature gradient causes the flow of a spin current \cite{Uchida-2008,Jaworski-2010,Uchida-2010,Bauer-2012}. This effect takes a central position in the field of ``spin caloritronics'', the study of the interplay of spin degrees of freedom and heat. A particularly pure form of the spin Seebeck effect exists in magnetic insulators, because in this case spin transport takes place exclusively via spin waves or ``magnons'', whereas the electronic degrees of freedom are frozen out. Since phonons are the dominant carriers of heat in an insulator, spin caloritronic effects in magnetic insulators depend strongly on the magnon-phonon interaction. Because of its low magnetic damping and high acoustic quality, most experimental studies of the magnon-driven spin Seebeck effect focus on the synthetic ferromagnetic insulator Yttrium Iron Garnet Y$_3$Fe$_5$O$_{12}$ (YIG) \cite{Geller-1957,Geller-1958}. 

The important role of magnon-phonon coupling for the spin Seebeck effect was already pointed out in the initial theoretical works \cite{Xiao-2010,Adachi-2010,Schreier-2013}, where it was suggested that a so-called ``phonon-drag'' is the cause of the significant enhancement of the spin Seebeck voltage at low temperatures, which follows the temperature dependence of the phonon thermal conductivity \cite{Adachi-2010}. The experimental demonstration of ``acoustic spin pumping'', the generation of a spin current by injection of acoustic waves, instead of the application of a temperature gradient, is another indicator of the importance of magnon-phonon coupling in the spin Seebeck effect \cite{Uchida-2011,Weiler-2012,Polzikova-2018}.
Specific evidence of the strong coupling between the two subsystems was the discovery of distinct peaks in the magnetic field dependence of the spin Seebeck voltage at two ``critical'' magnetic fields, at which the acoustic magnon and phonon dispersions have touching points \cite{Kikkawa-2016}. These features were associated with the formation of ``magnon polarons'' \cite{Kikkawa-2016,Flebus-2017,Cornelissen-2017}, coherent superpositions of magnon and phonon excitations formed near the ``resonant'' frequencies at which their dispersions cross \cite{Kittel-1958,Akhiezer-1959,Schloemann-1960}. (Strictly speaking, magnon polarons are formed at all frequencies, but at generic frequencies the magnon polaron modes are either ``magnon-like'' or ``phonon-like'', {\em i.e.}, their weight exists mainly in either the spin or the lattice sector, with a small admixture of the other subsystem.) Magnon polarons were also observed outside the context of the spin Seebeck effect. References \cite{Bozhko-2017} and \cite{Hayashi-2018} report an accumulation of magnon polarons and anomalies in the spin pumping efficiency in the spectral region near the magnon-phonon resonance following parametric magnon excitation, respectively. Reference \cite{Holanda-2018} reported the direct observation of wave-like excitations in the lattice subsystem after excitation of the spin subsystem.

In a recent article together with Wilken and Nunner \cite{Schmidt-2018}, we have shown that the observed peak structure of the spin Seebeck voltage as a function of the magnetic field can also be explained using an incoherent picture, arising from a critical enhancement of the magnon-phonon scattering rate when their dispersions have a touching point. Both the incoherent theory of Ref.\ \cite{Schmidt-2018} and the magnon-polaron theory of Refs.\ \cite{Kikkawa-2016,Flebus-2017,Cornelissen-2017} make the assumption that magnon-magnon and phonon-phonon relaxation processes are strong enough that the distribution functions of magnons and phonons or magnon polarons are given by Planck- or Bose-Einstein-type local equilibrium distributions at all times. A Planck-type local equilibrium distribution for magnons is justified if the relaxation processes are dominated by number-non-conserving three-magnon confluence or splitting processes \cite{Rezende-2014,Schmidt-2018}; A Bose-Einstein-type distribution function is applicable if number-conserving four-magnon processes dominate \cite{Cornelissen-2016}. At low temperatures and for magnetic fields in the vicinity of the critical values, inelastic magnon-magnon and phonon-phonon scattering are suppressed and relaxation is dominated by the interaction with impurities \cite{Douglass-1963,Walton-1973,Heremans-2014,Rueckriegel-2014}. Impurity scattering is elastic and cannot relax distribution functions to the Planck- or Bose-Einstein form. Instead, at low temperatures, one expects that the distribution function of magnon polarons has a singular frequency dependence near the magnon-phonon resonance frequencies and that it cannot be approximated by a Planck-type or Bose-Einstein-type distribution function.

In this article, we present a theory of the longitudinal spin Seebeck effect in a ferromagnetic insulator and for close-to-critical magnetic fields that is tailored to temperatures low enough and/or system lengths small enough that no {\em a priori} assumption of strong relaxation to a Planck-type or Bose-Einstein-type magnon-polaron distribution function can be made. This includes the range of temperatures and system sizes that were considered in the experiment of Ref. \cite{Kikkawa-2016}. We consider elastic scattering from impurities as well as inelastic processes and describe the full crossover between the extreme low-temperature regime, in which elastic scattering dominates the spin Seebeck effect, and the intermediate-temperature regime, in which relaxation by inelastic processes imposes a local-equilibrium form of the distribution functions, so that the distribution can be characterized by a ``magnon temperature'' or a ``magnon chemical potential'' \cite{Kikkawa-2016,Flebus-2017,Cornelissen-2017,An-2016,Schmidt-2018,Agrawal-2013,Xi-2020}. Our theory is based on the solution of the Boltzmann equation for the distribution function of magnon-polaron modes. Whereas the dominance of impurity scattering at low and intermediate temperatures allows us to use a simplified {\em ansatz} for the angle dependence of the distribution functions, the full frequency dependence of the distribution functions is kept at all stages of the calculation.

A central role in the Boltzmann theory is played by transition rates for elastic scattering from impurities as well as for inelastic interactions of magnons and phonons. Although they are often treated as phenomenological parameters, to capture parameter dependences, such as the dependence on an external magnetic field or on temperature, it is necessary to obtain transition rates from a microscopic picture. For YIG, which is a synthetic ferrimagnetic insulator of complex structure, elaborate effective spin Hamiltonians have been developed \cite{Gurevich-1996,Sparks-1964,Cherepanow-1993,Boothroyd-2017,Barker-2017,Kreisel-2009}, which have been found to predict the experimentally observed magnon spectrum well. At low temperatures, however, only a single magnon band is relevant, and an effective model of spins on a cubic lattice already provides an accurate description of the magnon spectrum, consistent with experiments \cite{Plant-1983,Cherepanow-1993}. 
Building on the success of this simplified description of YIG, we here link the magnonic transition rates in the Boltzmann theory as well as the magnon-polaron dispersion to a simple model of spins on a cubic lattice and with nearest-neighbor interactions only. The phonon system is included by a simple harmonic potential between nearest and next-nearest-neighboring atoms. The magnon-phonon coupling is included by accounting for the dependence of these interactions on the displacement of lattice sites \cite{Kaganov-1959,Kaganov-1961-2}. Although the spin-spin interactions are predominantly of the isotropic Heisenberg exchange type, an additional weak anisotropic interaction, such as a pseudodipolar anisotropic exchange interaction or the relativistic Dzyaloshinskii-Moryia interaction \cite{Dzyaloshinsky-1958,Moriya-1960}, must be included to generate the magnon number-non-conserving processes required to obtain magnon-polaron modes and to reproduce the observed low-temperature phenomenology of the magnetic field-dependent spin Seebeck effect. The same phenomenology can also be derived upon replacing the microscopic model by a phenomenological ``magneto-elastic'' Hamiltonian \cite{Kittel-1949,Kaganov-1959,Kaganov-1961-2,Rueckriegel-2014,Streib-2019} and we compare the two approaches in the appendix.

Adjusting the parameters in the microscopic model to reproduce low-temperature magnetic and acoustic (transport) properties of YIG, we find that for system sizes up to $L = 10\,\mu{\rm m}$ the magnon-polaron distribution is well approximated by completely neglecting inelastic processes for all temperatures at which our model description is valid, $T \lesssim 30\,{\rm K}$. Even for larger system sizes $L \lesssim 100\,\mu{\rm m}$ --- which is far beyond the range of system sizes investigated experimentally ---, we find that a theory based on elastic impurity scattering only remains an excellent approximation for $T \lesssim 10\,{\rm K}$. For these system sizes and temperatures, a theory with elastic scattering only gives a strongly frequency dependent distribution function, in which the population of magnon-like magnon-polaron modes has a sharp singularity in the immediate vicinity of the magnon-phonon resonance. Such distribution functions are not at all well approximated by a Planck-like or Bose-Einstein-like form. Indeed, our theory leads to a number of predictions that differ qualitatively from previous theories of the magnon-polaron-mediated spin Seebeck effect. Most notably, we find that the spin current always shows a peak at the critical magnetic fields at which magnon and phonon dispersions touch. In contrast, Refs.\ \cite{Kikkawa-2016,Flebus-2017} predict a peak only if the sample is of a better acoustic quality than magnetic quality (mean free path $l_{\rm pi}$ for phonon-impurity scattering larger than mean free path $l_{\rm mi}$ for magnon-impurity scattering). This implies that the experimental observation of peaks in the magnetic-field dependence of the spin Seebeck coefficient in Ref.\ \cite{Kikkawa-2016} can not be used to determine the relative magnitude of $l_{\rm mi}$ and $l_{\rm pi}$.

The ferromagnetic insulator--normal metal (FN) interface is a crucial ingredient to the magnonic spin Seebeck effect. In a theory based on magnon polarons, the key processes at the FN interface are the conversion of magnon polarons in the ferromagnetic insulator (F) into phonons in the normal metal (N) \cite{Kamra-2015,Latcham-2019} and ``spin pumping'' \cite{Tserkovnyak-2002}, the excitation of spin current in the normal metal by a precessing magnetization. It is the spin pumping process that facilitates the conversion of a non-equilibrium accumulation of magnon polarons at the FN interface into a spin current in N. However, spin pumping also has an important inverse effect on the magnon-polaron distribution: It equilibrates the population of magnon-like magnon polarons in F to the equilibrium distribution of the conduction electrons in N. This inverse effect is absent in a perturbative treatment of the FN interface, in which the distribution of magnon-like magnon polarons is calculated with reflecting boundary conditions at the FN interface \cite{Xiao-2010,Cornelissen-2016,Schmidt-2018}.

The remainder of this paper is organized as follows. In Sec.\ \ref{sec:model} we present a microscopic model of a ferromagnetic insulator based on a simple cubic lattice and show that the model has magnon-polaron modes as its elementary excitations. In Sec.\ \ref{sec:boltzmann} we then review the Boltzmann transport theory of magnon-polaron modes, discuss the relevant relaxation processes, and address the boundary conditions at the interface between the ferromagnetic insulator and a non-magnetic insulator (which serves as the heat source that causes the thermal gradient) and a non-magnetic metal. In Sec.\ \ref{sec:results} we apply our theory to a spin Seebeck heterostructure, using typical material parameters for the ferrimagnetic insulator YIG attached to a thin Platinum (Pt) film. We conclude in Sec.\ \ref{sec:conclusions}.

\section{Model}
\label{sec:model}

\begin{figure}
\includegraphics[width=.9\linewidth]{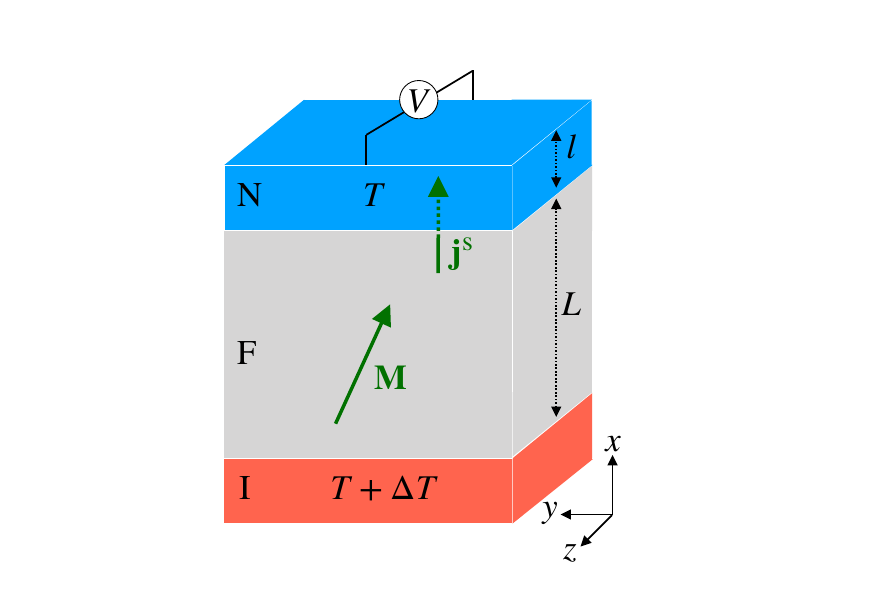}
\caption{\label{fig:setup} Geometry for the longitudinal spin Seebeck effect: A ferromagnetic insulator F of length $L$ (center, gray) is placed between an insulator I (bottom, red) and a normal metal N of thickness $l$ (top, blue), which also act as heat reservoirs held at a temperature difference $\Delta T$. Via magnon-phonon coupling, the applied temperature gradient leads to a nonequilibrium magnon distribution, which causes the flow of a spin current $\vj^{\rm s}$ into the normal metal. The spin current can be measured in the normal metal by means of the inverse spin Hall effect.}
\end{figure}

We consider the conventional setup for the longitudinal spin Seebeck effect, which consists of a ferromagnetic insulator--normal metal heterostructure as illustrated in Fig.\ \ref{fig:setup}. The system is coupled to heat baths to the top and bottom, which are held at a temperature difference $\Delta T$. We assume that the system is isotropic and choose coordinate axes such that the temperature gradient and the resulting spin currents are in the $x$-direction, see Fig.\ \ref{fig:setup}. We focus on the low-temperature regime in which Umklapp scattering and excitation of optical magnons and phonons  is strongly suppressed. Spin and heat transport in the ferromagnetic insulator is governed by the interaction of acoustic magnons and phonons as well as by scattering from impurities. 

\subsection{Lattice model and continuum limit}
\label{sec:mp1}

We first present a minimal lattice model of classical spins, which serves as a microscopic starting point for the derivation of the continuum theory of coupled magnon and phonon modes. The subsequent continuum theory may also be derived from phenomenological considerations, such as the magneto-elastic theory of Refs.\ \cite{Kittel-1949,Kaganov-1959,Kaganov-1961-2,Rueckriegel-2014}.

{\em Lattice model.}---
The guiding principle for the construction of the minimal model is the accepted wisdom that at low temperatures YIG may be well described as a ferromagnetic insulator with effective spins $\vS_j$ of magnitude $S$, located at the sites $\vr_j$ of a simple cubic lattice with lattice constant $a$ \cite{Cherepanow-1993}. The lattice ions have mass $m$, displacement $\vu_j$, and momentum $\vp_j$. We consider the classical Hamiltonian 
\begin{equation}
  H = H^{\rm pho} + H^{\rm mag} + H^{\rm mag-pho}, \label{eq:Hb}
\end{equation}
in which the three terms $H^{\rm mag}$, $H^{\rm pho}$, and $H^{\rm mag-pho}$ describe classical ``magnons'' and ``phonons'', collective small excursions from the equilibrium state of the spins and the lattice, respectively, and the interaction between these.

For term $H^{\rm pho}$, which describes lattice vibrations, we impose a harmonic lattice potential \cite{Bruus-2004} with coupling constants $K_1$ and $K_2$, involving couplings between nearest-neighbor lattice atoms $\langle i,j \rangle$ and between next-nearest neighbors $\langle\!\langle i,j\rangle\!\rangle$,
\begin{align}
  H^{\rm pho} =&\, \sum_{j} \frac{|\vp_j|^2}{2 m} + \frac{K_1}{2} \sum_{\langle i,j \rangle} |\vu_{ij} \cdot \ve_{ij}|^2 
   \nonumber \\ &\, \mbox{} + \frac{K_2}{2} \sum_{\langle i,j \rangle}  | \vu_{ij}|^2 + 
  \frac{K_1}{2}\sum_{\langle\!\langle i,j \rangle\!\rangle} |\vu_{ij} \cdot \ve_{ij}|^2 , \label{eq:hp}
\end{align}
where
$\ve_{ij} = (\vr_i - \vr_j) / |\vr_i -\vr_j|$ is the unit vector pointing from $\vr_i$ to $\vr_j$, and we abbreviated $\vu_{ij} = \vu_i - \vu_j$. In the summations over nearest neighbors and next-nearest neighbors every pair is summed over only once. The use of two coupling constants $K_1$ and $K_2$ is necessary to obtain different velocities for longitudinal and transverse phonon modes; the next-nearest-neighbor coupling term is required to reproduce an isotropic phonon dispersion in the long-wavelength limit \cite{Kittel-1963}.

The term $H^{\rm mag}$, which describes the collective excitations of the spin system, derives from the Zeeman coupling to an external magnetic field $\vB$ and the Heisenberg exchange interaction,
\begin{equation}
  H^{\rm mag} = - J \sum_{\langle i,j \rangle} \vS_{i} \cdot \vS_{j} - \mu \vB \cdot \sum_j \vS_j, \label{eq:hex}
\end{equation}
where $\mu = g \mub$ is the magnetic moment of the spins, with $\mub$ and $g$ the Bohr magneton and Land\'e factor, respectively.
To derive the magnon Hamiltonian we take $\ve$ to be the unit vector pointing in the direction of the external magnetic field $\vB = B \ve$, and parameterize (similar to the Holstein-Primakoff transformation for quantum spins \cite{Holstein-1940})
\begin{equation}
  \vS_j =  \ve\, \sqrt{S^2 - S n_j^2} + \vn_j\, \sqrt{S}, \label{eq:expansion}
\end{equation}
where
$\vn_j \perp \ve$ is the (suitably renormalized) transverse magnetization amplitude and $n_j = |\vn_j|$.
Expanding $H^{\rm ex}$ to quadratic order in the amplitudes $\vn_j$ gives the magnon Hamiltonian
\begin{equation}
  H^{\rm mag} = \frac{J S}{2} \sum_{\langle i,j \rangle} |\vn_i - \vn_j|^2 + \frac{\mu B}{2} \sum_j \vn_j^2 . \label{eq:hm}
\end{equation}

Taking the exchange constant $J$ in the Heisenberg Hamiltonian (\ref{eq:hm}) to depend on the displacements $\vu_j$ of the lattice ions one obtains a magnon-phonon interaction. This interaction, however, conserves the magnon number, so that it alone does not allow for a steady-state spin Seebeck effect. Moreover, since its leading contribution is of (combined) cubic order in the small amplitudes $\vn_j$ and $\vu_j$, the effect of the exchange-based magnon-phonon coupling is strongly suppressed at low temperatures. Instead, at low temperatures the magnon-phonon coupling is dominated by relativistic corrections to the Hamiltonian, which give corrections to the Hamiltonian of (combined) quadratic order in $\vn_i$ and $\vu_i$. As an example of such a relativistic correction we here consider the Van Vleck pseudo-dipolar exchange interaction \cite{Vanvleck-1937}. It results from the combination of Heisenberg exchange and spin-orbit coupling \cite{Moriya-1960,Keffer-1962} and can be written as
\begin{equation}
H^{\rm pd} = \sum_{\langle i,j \rangle} D_{ij} (\vS_{i} \cdot \ve_{ij}) (\vS_{j} \cdot \ve_{ij}) . \label{eq:hpd}
\end{equation}
Again, we consider nearest-neighbor interactions only. To leading order in $S$ the contribution of $H^{\rm pd}$ to the magnon Hamiltonian in Eq.\ (\ref{eq:hm}) causes a weakly anisotropic shift of the magnon frequency, which we neglect because typically $D_{ij} \ll J$ for neighboring spins. To obtain the magnon-phonon coupling Hamiltonian $H^{\rm mag-pho}$, we take the dipolar exchange constant $D_{ij}$ to depend on the relative displacement $\vu_{ij} = \vu_i - \vu_j$ of nearest-neighbor atoms and expand to linear order $\vu_{ij}$,
\begin{equation}
  D_{ij} \to D + D' \vu_{ij} \cdot \ve_{ij}.
\end{equation}
We also expand the unit vectors $\ve_{ij}$ to linear order in $\vu_{ij}$,
\begin{equation}
\ve_{ij} \to \ve_{ij} + \frac{\vu_{ij} - (\vu_{ij} \cdot \ve_{ij})\, \ve_{ij} }{a}.
\end{equation}
Expanding $H^{\rm pd}$ to linear order in both $\vu_{i}$ and $\vn_{i}$ then gives the magnon-phonon Hamiltonian
\begin{equation}
H^{\rm mag-pho} = \sum_{\langle i,j \rangle} (\vn_{i} + \vn_{j}) \cdot {\cal D}_{ij} (\vu_{i}-\vu_{j}), \label{eq:hpd_mp}
\end{equation}
where the $3 \times 3$ matrix ${\cal D}_{ij}$ reads (with dyadic products $\ve_{ij} \ve^{\rm T}$ and $\ve_{ij} \ve_{ij}^{\rm T}$)
\begin{align}
{\cal D}_{ij} =&\, S^{3/2} (\ve\cdot \ve_{ij})  \left[D'  \ve_{ij} \ve_{ij}^{\rm T} + \frac{D}{a} (\openone - 2 \ve_{ij} \ve_{ij}^{\rm T}) \right] \nonumber \\
&\, + S^{3/2} \frac{D}{a} \ve_{ij} \ve^{\rm T} .
\end{align}

The equations of motion for the displacements $\vu_j$ and the magnetization amplitudes $\vn_j$ take the familiar Hamilton form
\begin{equation}
  \dot \vu_j = \frac{\partial H}{\partial \vp_j},\ \ \dot \vp_j = - \frac{\partial H}{\partial \vu_j}, \ \
  \dot \vn_{j} = -\ve \times \frac{\partial H}{\partial \vn_{j}}.
  \label{eq:equationsOfMotionReal}
\end{equation}
The property that $\vn_j \perp \ve$ is conserved under these equations of motion.

{\em Fourier transform and continuum limit.---}
To obtain a formulation in terms of classical phonons and magnons we introduce the Fourier transforms
\begin{align}
  \vu_{j} =&\, \frac{1}{\sqrt{N}} \sum_{\vq} \vu_{\vq} e^{i \vq \cdot \vr_{j}}, \nonumber \\
  \vp_{j} =&\, \frac{1}{\sqrt{N}} \sum_{\vq} \vp_{\vq} e^{i \vq \cdot \vr_{j}}, \nonumber \\
  \vn_{j} =&\, \frac{1}{\sqrt{N}} \sum_{\vq} \vn_{\vq} e^{i \vq \cdot \vr_{j}},
  \label{eq:Fourier}
\end{align}
where we assume a lattice with $N=V/a^3$ lattice sites and periodic boundary conditions. In terms of the Fourier-transformed amplitudes, the Hamiltonian reads
\begin{align}
  H =&\, \frac{1}{2} \sum_{\vq} 
  \Big\{ \frac{|\vp_{\vq}|^2}{m} + {\cal B}(\vq) |\vn_{\vq}|^2
  + \vu_{\vq}^* \cdot {\cal K}(\vq) \vu_{\vq}
  \nonumber \\ &\, \mbox{}
  + 2 \mbox{Re}\, \vn_{\vq}^* \cdot {\cal D}(\vq) \vu_{\vq}
  \Big\}.
  \label{eq:HFourier}
\end{align}
In the long-wavelength limit $q \to 0$, which is the relevant limit at low temperatures, one has
\begin{equation}
\cK(\vq) =  \frac{m}{a^3} \left[ c_{\rm t}^2 q^2 \openone + (c_{\rm l}^2 - c_{\rm t}^2) \vq \vq^{\rm T} \right], \label{eq:Kdef}
\end{equation}
where $c_{\rm l}$ and $c_{\rm t}$ are the velocities of the longitudinal and transverse phonon modes in the lattice model (\ref{eq:Hb}),
\begin{equation}
  c_{\rm l}^2 = \frac{a^2}{m} (K_1 + K_2),\ \
  c_{\rm t}^2 = \frac{a^2}{m} (3 K_1 + K_2).
\end{equation}
In the limit $q \to 0$, the energy of the magnon modes is
\begin{equation}
  {\cal B}(\vq) = \mu B + J S a^2 q^2 \label{eq:Bdef}
\end{equation}
and the magnon-phonon coupling is described by the $3 \times 3$ matrix
\begin{align}
  {\cal D}(\vq) =&\, 2i \left(\frac{S}{a}\right)^{3/2}
  \sum_{\alpha} q_{\alpha} \left[ D \ve_{\alpha} \ve^{\rm T}
  \right. \nonumber\\ & \left.
   + (\ve \cdot \ve_{\alpha}) (a D' \ve_{\alpha} \ve_{\alpha}^{\rm T}  + D (\openone-2 \ve_{\alpha} \ve_{\alpha}^{\rm T})) \right],
  \label{eq:Ddef}
\end{align}
where $\ve_{\alpha}$ denotes the unit vector in the spatial directions $\alpha=x,y,z$, and $\ve_{\alpha} \ve_{\alpha}^{\rm T}$ is the dyadic product. A real-space formulation in the long-wavelength limit $q \to 0$ can be obtained by inverse Fourier transform of Eq.\ (\ref{eq:HFourier}). This amounts to the replacement of the lattice amplitudes $\vu_j$, $\vp_j$, and $\vn_j$ by smooth functions $\vp(\vr)$, $\vu(\vr)$, and $\vn(\vr)$ of the position $\vr$ and the substitution $\vq \to -i \vnabla$ in the Hamiltonian (\ref{eq:HFourier}). Expressions for ${\cal K}(\vq)$, ${\cal B}(\vq)$, and ${\cal D}(\vq)$ for the full lattice model (\ref{eq:Hb}), without the approximation $q \to 0$, can be found in Appendix \ref{app:a}.

{\em Boundary conditions.---}
In the lattice model, the magnetic medium F exists for $0 < x < L$ with $L = N_x a$, $N_x$ being the number of lattice sites in the $x$-direction, see Fig.\ \ref{fig:setup}. At $x=L$ there is a boundary to a non-magnetic metal N; at $x=0$ there is a boundary to a non-magnetic insulator I. In both the non-magnetic insulator I and the normal metal N we consider phonon degrees of freedom only, described by the Hamiltonian $H^{\rm pho}$ of Eq.\ (\ref{eq:hp}). At the boundaries at $x=0$ and $x=L$ the magnon Hamiltonian $H^{\rm mag}$ and the magnon-phonon coupling $H^{\rm mag-pho}$ are truncated by omitting any on-site terms or nearest-neighbor contributions involving lattice sites in the non-magnetic metal N or the non-magnetic insulator I. In the long-wavelength limit $q \to 0$, one can show that this amounts to the boundary conditions \cite{Kamra-2014,Kamra-2015}
\begin{equation}
  \label{eq:boundaryInsulator1}
  \vu(0^{-}) = \vu(0^{+}) , \ \ \ 0 = \frac{\partial \vn(0^{+})}{\partial x}, 
\end{equation}
and
\begin{equation} \label{eq:boundaryInsulator2}
  \frac{\partial \cK(\vq)}{\partial q_x} \vu(0^{-}) = \frac{\partial \cK(\vq)}{\partial q_x} \vu(0^{+}) + 2 \frac{\partial \cDD(\vq)^{\dagger}}{\partial q_x} \vn(0^{+}),
\end{equation}
for the IF interface at $x=0$, with $q_x$ replaced by $-i \partial/\partial x$. For the FN boundary at $x=L$, the boundary conditions for the displacement field $\vu(\vr)$,
\begin{align}
\vu(L^{-}) =&\, \vu(L^{+}), \\
  \frac{\partial \cK(\vq)}{\partial q_x} \vu(L^{-}) =&\,
  \frac{\partial \cK(\vq)}{\partial q_x} \vu(L^{+}) + 2 \frac{\partial \cDD(\vq)^{\dagger}}{\partial q_x} \vn(L^{+}),
  \label{eq:boundaryF1}
\end{align}
are the same as at the interface between the ferromagnetic insulator and the normal metal. The boundary condition for the spin wave amplitude at $x=L$ is different for the FN interface, because magnons can excite conduction electrons in the normal metal \cite{Tserkovnyak-2002,Hoffman-2013},
\begin{align}
  \label{eq:boundaryF}
  - a^2 J S\, \ve \times \frac{\vn(L^-)}{\partial x} =&\, \frac{\mu}{4\pi M_{\rm s}}
  \\ \nonumber &\, \mbox{} \times
  \left[ \sigma_{\uparrow\downarrow}' \ve \times \dot \vn(L^-) - \sigma_{\uparrow\downarrow}'' \dot \vn(L^-) \right] .
\end{align}
Here $M_{\rm s} = \mu S/a^3$ is the magnetic moment per unit volume and $\sigma_{\uparrow\downarrow} = \sigma_{\uparrow\downarrow}' + i \sigma_{\uparrow\downarrow}''$ is the spin-mixing conductance per unit area.

\subsection{Phonons, magnons, and magnon polarons}
\label{sec:mp2}

{\em Classical phonons and magnons.---}
A formulation in terms of classical phonons and magnons is obtained upon switching to complex phasor variables $b_{\vq,\lambda}$ with $\lambda=1,2,3,4$,
\begin{align}
  \vu_{\vq} =&\, \sum_{\lambda=1}^{3} \sqrt{\frac{\hbar}{2 m \omega^0_{\vq,\lambda}}} (b_{\vq,\lambda} + b_{-\vq,\lambda}^*) \ve_{\vq,\lambda}, \nonumber \\
  \vp_{\vq} =&\, -i \sum_{\lambda=1}^{3} \sqrt{\frac{\hbar  m \omega^0_{\vq,\lambda}}{2}} (b_{\vq,\lambda} - b_{-\vq,\lambda}^*) \ve_{\vq,\lambda}, \nonumber \\
  \vn_{\vq} =&\, \sqrt{\hbar} \left( b_{\vq,4} \ve_- +
  b_{-\vq,4}^* \ve_+ \right), \label{eq:normal}
\end{align}
where we introduced Planck's constant $\hbar$ to obtain a formal analogy with a quantum-mechanical treatment of the same problem. The phasor variables $b_{\vq,\lambda}$ with $\lambda=1$ and $\lambda=2,3$ describe the longitudinal and transverse phonon modes, respectively. In the long-wavelength limit, the phonon frequencies are
\begin{equation}
\omega^0_{\vq,1} = c_{\rm l} q, \ \
\omega^0_{\vq,2} = \omega^0_{\vq,3} = c_{\rm t} q. \label{eq:wpho}
\end{equation}
The polarization vectors $\ve_{\vq,\lambda} = \ve_{-\vq,\lambda}^*$ are the corresponding eigenvectors of ${\cal K}(\vq)$. In the limit $q \to 0$, the unit vector $\ve_{\vq,1}$ is collinear with $\vq$; $\ve_{\vq,2}$ and $\ve_{\vq,3}$ are orthogonal to $\vq$. Since the transverse phonon modes $\omega^0_{\vq,2}$ and $\omega^0_{\vq,3}$ are degenerate, the polarization vectors $\ve_{\vq,2}$ and $\ve_{\vq,3}$ are not uniquely determined at this stage. The magnon frequency is
\begin{equation}
  \omega^0_{\vq,4} = {\cal B}(\vq). \label{eq:wmag}
\end{equation}
The magnon polarization vectors $\ve_+ = \ve_-^*$ are complex unit vectors satisfying the property $\ve_{\pm} \times \ve = \pm i \ve_{\pm}$.

After this variable transformation, the Hamiltonian (\ref{eq:Hb}) and the equations of motion (\ref{eq:equationsOfMotionReal}) can be written in the compact form \cite{Kikkawa-2016,Flebus-2017}
\begin{equation}
  H = \frac{1}{2} \sum_{\vq} \vb_{\vq}^{\dagger} \cdot H_{\vq} \vb_{\vq}, \label{eq:HF}
  \ \ 	\dot b_{\vq,\lambda} = - \frac{i}{\hbar} \frac{\partial H}{\partial b_{\vq,\lambda}^*},
\end{equation}
where $\vb_{\vq}$ is the eight-component column vector
\begin{equation}
  \vb_{\vq} = \begin{pmatrix}
  b_{\vq,\lambda} \\ b_{-\vq,\lambda}^*
  \end{pmatrix}_{\lambda=1,2,3,4}
  \label{eq:bqdef}
\end{equation}
and $H_{\vq}$ the $8 \times 8$ hermitian matrix
\begin{equation}
  H_{\vq} = 
  \begin{pmatrix}
  \hbar \omega^0_{\vq,\lambda} \delta_{\lambda\lambda'} & \Delta^*_{\vq,\lambda} & 0 & \Delta_{-\vq,\lambda} \\
  \Delta_{\vq,\lambda'} & \hbar \omega^0_{\vq,4} & \Delta_{\vq,\lambda'} & 0 \\
  0 & \Delta^*_{\vq,\lambda} & \hbar \omega^0_{\vq,\lambda} \delta_{\lambda\lambda'} & \Delta_{-\vq,\lambda} \\
  \Delta^*_{-\vq,\lambda'} & 0 & \Delta^*_{-\vq,\lambda'} & \hbar \omega^0_{\vq,4}
  \end{pmatrix}_{\lambda,\lambda'=1,2,3}.
  \label{eq:Hq}
\end{equation}
The diagonal elements of $H_{\vq}$ contain the frequencies of the phonon and magnon modes; the off-diagonal elements $\Delta_{\vq,\lambda}$, $\lambda=1,2,3$, describe the magnon-phonon coupling,
\begin{equation}
\Delta_{\vq,\lambda} =  \sqrt{\frac{\hbar a^3}{2 m \omega^0_{\vq,\lambda}}} \ve_{+} \cdot {\cal D}(\vq) \ve_{\vq,\lambda},
\end{equation}
where the $3 \times 3$ matrix ${\cal D}(\vq)$ was defined in Eq.\ (\ref{eq:Ddef}).

{\em Magnon-polaron modes.---}
The magnon-polaron modes are the eigenmodes of the full magnon-phonon Hamiltonian (\ref{eq:HF}). To find their dispersion we perform a canonical transformation that diagonalizes the matrix $H_{\vq}$ of Eq.\ (\ref{eq:HF}),
\begin{equation}
  H_{\vq} = V_{\vq}\, \hbar \Omega_{\vq}\, V_{\vq}^{\dagger},
  \label{eq:Vq}
\end{equation}
where the entries of the diagonal matrix
\begin{equation}
	\Omega_{\vq} = \begin{pmatrix} \omega_{\vq,\nu} & 0 \\ 0 & \omega_{-\vq,\nu} \end{pmatrix}_{\nu=1,2,3,4}
\end{equation}
are the frequencies of the magnon-polaron modes and the symplectic transformation matrix $V_{\vq}$ satisfies the condition $\Sigma_3 V_{\vq}^{\dagger} \Sigma_3 = V_{\vq}^{-1}$ with $\Sigma_3 = {\rm diag}(\openone_4,-\openone_4)$. The phasor variables $b_{\vq,\lambda}$ of the phonon and magnon modes are related to the phasor variables $a_{\vq,\nu}$ of the magnon-polaron modes via
\begin{align}
	\va_{\vq} = V^{\dagger}_{\vq} \vb_{\vq} , \ \ \ \
	\va_{\vq} = \begin{pmatrix}
	a_{\vq,\nu} \\ a_{-\vq,\nu}^*
	\end{pmatrix}_{\nu=1,2,3,4} .
        \label{eq:ab}
\end{align}
This transformation brings $H$ to diagonal form
\begin{equation}
	H = \sum_{\vq, \nu} \hbar \omega_{\vq,\nu} a_{\vq,\nu}^* a_{\vq,\nu} . \label{eq:Ha}
\end{equation}
The equations of motion for the phasor variables $a_{\vq,\nu}$ read
\begin{equation}
  \dot a_{\vq,\nu} = - \frac{i}{\hbar}
  \frac{\partial H}{\partial a_{\vq,\nu}^*},
  \ \ \ \ \nu=1,2,3,4. \label{eq:mot}
\end{equation}
To construct a quantum theory, one simply replaces the complex amplitudes $a_{\vq,\nu}$ and $a_{\vq,\nu}^*$ by operators $\hat a_{\vq,\nu}$ and $\hat a_{\vq,\nu}^\dagger$ with commutation relations $[\hat a_{\vq,\nu}, \hat a_{\vq,\nu}^\dagger] = 1$.

The magnon-polaron modes are linear superpositions of wave-like excitations of the spins and of the lattice, {\em i.e.}\, of magnons and phonons. The precise form of the superposition is described by the matrix $V_{\vq}$ that transforms between the formulation (\ref{eq:HF}) in terms of phonon and magnon modes and the formulation (\ref{eq:Ha}) in terms of magnon-polaron modes, see Eq.\ (\ref{eq:ab}). At generic frequencies, the mixing of spin and lattice degrees of freedom is small. One of the magnon-polaron modes is magnon-like, with a small admixture of longitudinal and transverse phonon modes, whereas three of the magnon-polaron modes are phonon-like. Two of the phonon-like modes have a small magnon admixture; the third mode is a pure transverse phonon mode. (This follows because the perturbation proportional to $\Delta$ in Eq.\ (\ref{eq:Hq}) does not have maximal rank.) At the resonant frequencies at which magnon and phonon dispersions cross, two of the magnon-polaron modes have significant spin and lattice components, one mode is phonon-like with a small magnon component, and one mode is a pure transverse phonon mode.

The magnon-polaron frequencies $\omega_{\vq,\nu}$ and the matrices $V_{\vq}$ that diagonalize the magnon-polaron Hamiltonian satisfy the symmetry constraints
\begin{equation}
  \omega_{\vq,\nu} = \omega_{-\vq,\nu} \label{eq:omegasymmetry}
\end{equation}
and
\begin{equation}
  V_{-\vq} =  I V_{\vq},\ \
  I = \mbox{diag}\,(\openone_3, -1,\openone_3,-1). \label{eq:Vsymmetry}
\end{equation}
For a more elaborate discussion of the symmetry properties of the $8 \times 8$ matrices appearing in this discussion we refer to App.\ \ref{app:b}.

{\em Numerical values.---} 
To obtain numerical values for YIG, we take the material parameters from Table \ref{tab:values}. Since the pseudo-dipolar exchange interaction (\ref{eq:hpd}) describes an anisotropic spin-spin interaction, we must specify the polarization direction $\ve$ of the ferromagnetic ground state. Following Ref.\ \cite{Barker-2017} we choose the polarization direction to be the $(111)$ direction. To determine the strength of the pseudo-dipolar anisotropic exchange coupling $D$ and its derivative $D' \sim D/a$ we compare the amplitudes of the magnon-phonon processes with the results of the phenomenological magneto-elastic energy, see appendix \ref{app:pheno}, which gives $D / J \approx 8.3\times10^{-2}$. This confirms that the pseudo-dipolar contribution to the magnon dispersion is indeed small compared to the Heisenberg exchange coupling for these parameters.
\begin{table}
\center{YIG continuum theory parameters}
\begin{ruledtabular}
\begin{tabular}{llcr}
 & \textrm{Quantity} & \textrm{Value} & \textrm{Ref.} \\
\colrule
lattice constant & $a$ & 1.24 nm & \cite{Gurevich-1996} \\
exchange stiffness & $J S a^2$ & 8.5 $\times$ 10$^{-40}$ J m$^2$ & \cite{Cherepanow-1993}  \\
exchange coupling & $J'$ & $J/a$ & \\
mass density & $m/a^3$ & 5170 kg/m$^3$  & \cite{Cherepanow-1993} \\
pseudo-dipolar exchange & $D/J$ & $8.3 \times 10^{-3}$ & \cite{Strauss-1968}\\
pseudo-dipolar exchange & $D'$ & $D/a$ & \\
saturation magnetization & $M_{\rm s}$ &1.4 $\times$ 10$^{5}$ A/m & \cite{Gurevich-1996}\\
long.\ sound velocity & $c_{\rm l}$ & 7209 m/s & \cite{Rueckriegel-2014} \\
trans.\ sound velocity & $c_{\rm t}$ & 3843 m/s & \cite{Rueckriegel-2014} \\
anharmonicity & $K'$ & 2 $\times$ 10$^{10}$ J/m$^3$ & \cite{Ziman-1960} \\ 
\end{tabular}

\center{Pt and YIG/Pt interface parameters}
\begin{tabular}{llcr}
 & \textrm{Quantity} & \textrm{Value} & \textrm{Ref.} \\
\colrule
spin mixing conductivity & $\sigma_{\uparrow\!\downarrow}$ & $1.3\times 10^{18}$ 1/m$^2$ & \cite{Qiu-2013} \\
spin Hall angle & $\theta_{\rm sh}$ & 0.0037 & \cite{Maekawa-2007} \\
spin diffusion length & $\lambda_{\rm sf}$ & 7.3 nm & \cite{Du-2015} \\
electrical resistivity & $\varrho$ & 0.91$\times$10$^{-6}$ $\Omega$/m & \cite{Uchida-2010-2} \\
sample dimensions & $l \times w$ & 5 nm $\times$ 2 mm & \\
\end{tabular}
\caption{\label{tab:values} Parameter values used for the numerical evaluation for the spin Seebeck effect in a YIG|Pt bilayer, together with the relevant references where these values were obtained.}
\end{ruledtabular}
\end{table}

\begin{figure}[t]
\includegraphics[width=\linewidth]{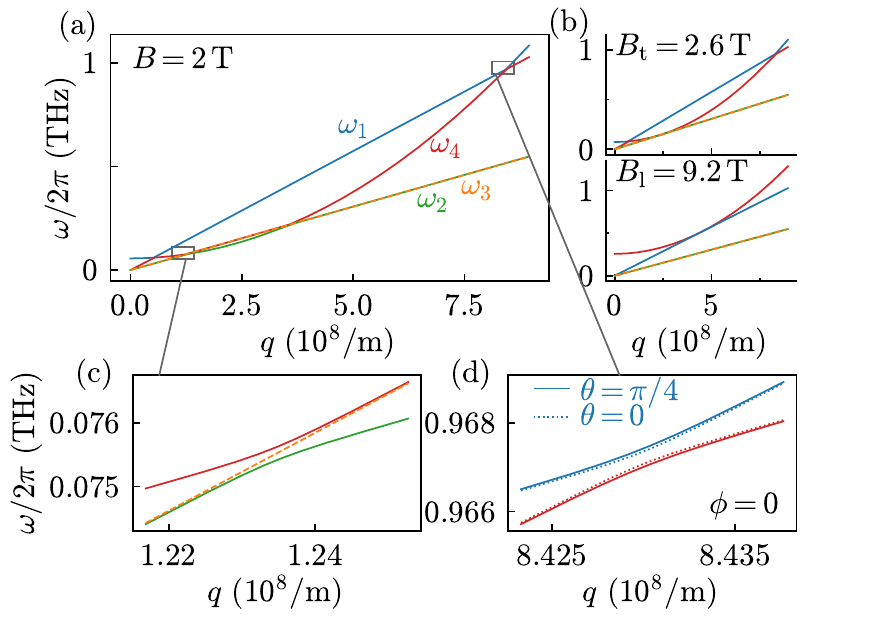}
\caption{\label{fig:dispersion} (a) Magnon-polaron frequency dispersions: $\omega_1$ (blue), $\omega_2$ (green), transverse phonon dispersion $\omega_3$ (orange, dashed), and $\omega_4$ (red). In panel (b) the  polaron dispersions for the ``critical'' applied magnetic fields $B_{\rm t} = 2.6\,{\text T}$ and $B_{\rm l} = 9.2\,\text{T}$ are shown. Magnifications of the dispersions: (c) around the crossing of the magnon-polaron mode with the transverse phonon mode and (d) longitudinal phonon mode for different angles $\theta$ defined by $q_x = q \cos \theta$.}
\end{figure}

Figure \ref{fig:dispersion} shows the magnon-polaron dispersions $\omega_{\vq,\nu}$ as a function of the wavevector $q$. The magnon and longitudinal phonon dispersions cross at wavevectors
\begin{equation}
  q^{\pm}_{\rm l,t} = \frac{1}{2} \left(q^0_{\rm l,t} \pm \sqrt{(q^0_{\rm l,t})^2 - 4 \mu B/J S a^2} \right)
  \label{eq:crossing}
\end{equation}
where $q^0_{\rm l,t} = \hbar c_{\rm l,t} / J S a^2$ is the crossing of the magnon and longitudinal/transverse phonon dispersion without an applied magnetic field. The hybridization of magnons and phonons is strongest at these intersection points. Without magnon-phonon interaction the transverse phonon branches $\omega^0_{\vq,2}$ and $\omega^0_{\vq,3}$ are degenerate. This degeneracy is lifted by the magnon-phonon interaction. Note, that only one of the two transverse phonon modes interacts with the magnons to form a magnon-polaron mode. The range of wavevectors $\vq$ with strong magnon-phonon interaction is significantly enhanced when reaching ``critical'' magnetic fields $B_{\rm l,t} = J S a^2(q^0_{\rm l,t})^2 /4 \mu$ \cite{Kikkawa-2016,Flebus-2017}.

\section{Boltzmann theory}
\label{sec:boltzmann}

We describe the four propagating magnon-polaron modes in the magnetic insulator in terms of a distribution function $n_{\vq,\nu}(x)$, which counts the occupation of the (quantized) magnon-polaron mode $(\vq,\nu)$, $\nu = 1,2,3,4$. The distribution function is related to the energy current carried by the magnon-polaron modes as
\begin{equation}
  j_x = \sum_{\vq,\nu} \hbar \omega_{\vq,\nu} v_{\vq,\nu x} n_{\vq,\nu}(x),
\end{equation}
where $v_{\vq,\nu x} = \partial \omega_{\vq,\nu} / \partial q_x$ is the group velocity of the magnon-polaron mode. (The $x$ direction is the direction along the applied temperature gradient.) In equilibrium, {\em i.e.}, without applying a temperature gradient across the magnetic insulator, the distribution function $n_{\vq,\nu} = n^0(\omega_{\vq,\nu})$ is given by the Planck distribution 
\begin{equation}
  n^0(\omega) = \frac{1}{e^{\hbar \omega/k_{\rm B} T} - 1}. 
\end{equation}
Because of the symmetry condition (\ref{eq:omegasymmetry}), the frequency $\omega_{\vq,\nu}$ and the velocity $v_{\vq,\nu x}$ are even and odd functions of $\vq$, respectively, ensuring that $j_x = 0$ in equilibrium.

The out-of-equilibrium distribution function $n_{\vq,\nu}(x)$ can be solved from the steady-state Boltzmann equation, which has the general form
\begin{equation}
v_{\vq,\nu x} \frac{\partial n_{\vq,\nu}(x)}{\partial x} = I_{\vq,\nu}, \label{eq:bo}
\end{equation}
where $I_{\vq,\nu}$ is the collision integral. The role of the collision integral $I_{\vq,\nu}$ is to regulate the relaxation of the distribution function towards a local equilibrium. We discuss a simplified ansatz of the distribution function $n_{\vq,\nu}$ in the linear-response regime in Sec.\ \ref{sec:linear} and the specific form of the collision integral in Sec.\ \ref{sec:scattering}. The boundary conditions at interfaces of the ferromagnetic insulator F with the non-magnetic insulator I and the normal metal N are considered in Sec.\ \ref{sec:reflection}, together with the spin current that is emitted into N.

\subsection{Qualitative considerations}

\begin{figure*}
\includegraphics[width=\linewidth]{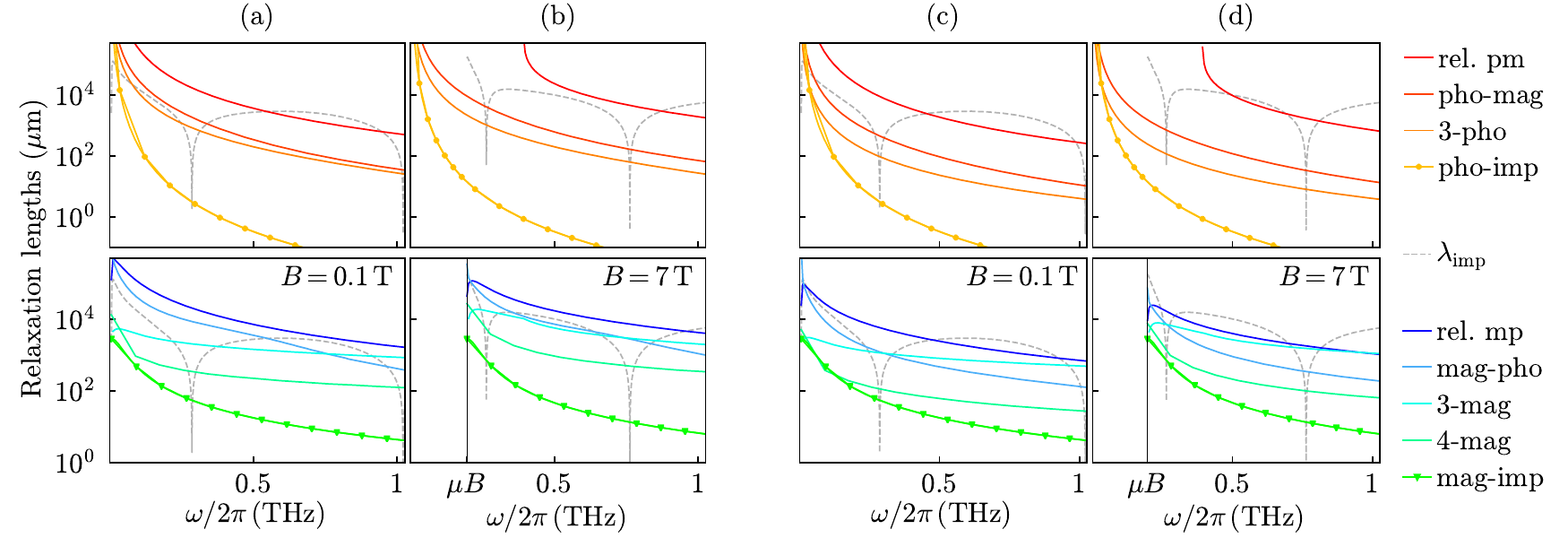}
\caption{\label{fig:relaxationlengths} Relaxation lengths for various scattering mechanisms in a magnetic insulator for temperatures $T = 10\,{\rm K}$ (a and b) and $T = 30\,{\rm K}$ (c and d). Panels (a) and (c) are for a magnetic field $B = 0.1\,{\rm T}$; panels (b) and (d) are for $B = 7\,{\rm T}$. Top and bottom panels show relaxation lengths for phonons and magnons, respectively. Material parameters for YIG are taken from Table \ref{tab:values}. The microscopic model for impurity scattering and for the inelastic scattering processes is discussed in Sec.\ \ref{sec:scattering}; values for the impurity potential are taken from Table \ref{tab:impurity}, center column. The relaxation lengths for phonon modes are averaged over polarization. The scattering processes shown in the figure are: magnon-impurity and phonon-impurity scattering, three-magnon, four-magnon, three-phonon, exchange-based magnon-phonon scattering, and relativistic or dipole-dipole-based inelastic magnon-phonon scattering. These processes are shown schematically in Table \ref{tab:diagrams}. The dashed curve shows the length scale $\limp$ for impurity-mediated inter-mode scattering of magnon polarons.}
\end{figure*}

Before we enter into a quantitative description of the formalism, we discuss the relevant relaxation processes and length scales qualitatively.
In Fig.\ \ref{fig:relaxationlengths} we show relaxation lengths for magnons and phonons --- {\em i.e.}, without taking into account magnon-polaron formation --- at two different temperatures and magnetic fields. The relaxation lengths shown in the figure are based on the material parameters of Table \ref{tab:values} and the collision integrals that will be discussed in Sec.\ \ref{sec:scattering}. The relevant elastic and inelastic scattering processes are shown schematically in Table \ref{tab:diagrams}.

The key observation underlying our theoretical analysis is that at low temperatures, elastic impurity scattering dominates over the inelastic processes. Elastic scattering not only causes a quick relaxation of the propagation direction, but it also causes scattering between different magnon-polaron modes. The length scale $\limp$ for such impurity-mediated inter-mode scattering of magnon polarons is shown by the dashed curve in Fig.\ \ref{fig:relaxationlengths}. The impurity-mediated inter-mode scattering is strongest close to the ``resonance frequencies'', because there magnon-polaron modes have significant magnon and phonon content. The relaxation length $\limp$ remains shorter than the inelastic scattering lengths for a small but finite window around the resonance frequencies. Moreover, it remains shorter than the length scale for relativistic or dipole-dipole-based inelastic phonon-to-two-magnon conversion at all frequencies for temperatures $T \lesssim 30\,{\rm K}$. Although it is significantly weaker than intra-mode impurity scattering at generic frequencies, impurity-mediated inter-mode scattering will be found to be the dominant source of the spin Seebeck effect at low temperatures.

\begin{table}
\begin{ruledtabular}
\begin{tabular}{lcc}
\underline{magnon-impurity} & \includegraphics[scale=.3]{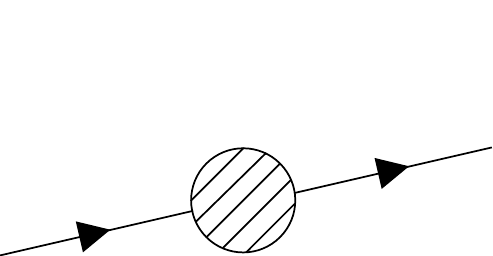} \\
\underline{phonon-impurity} & \includegraphics[scale=.3]{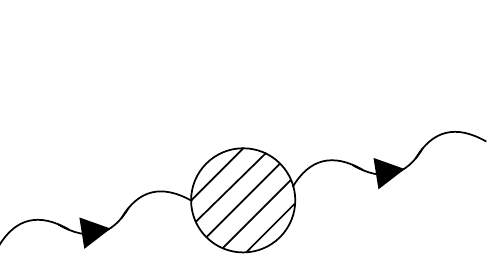} \\ ~\\
\hline\\
\underline{phonon-phonon} & \\
three-phonon & \includegraphics[scale=.3]{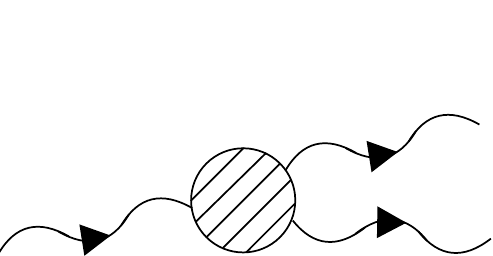} \ \includegraphics[scale=.3]{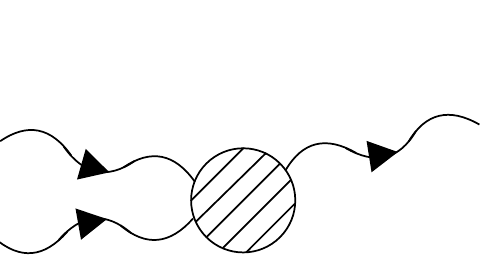} \\ ~\\ \hline\\
\underline{magnon-magnon} & \\
four-magnon & \includegraphics[scale=.3]{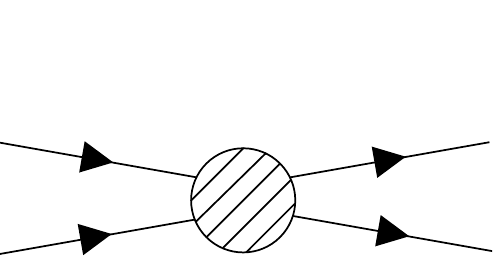} \\
three-magnon & \includegraphics[scale=.3]{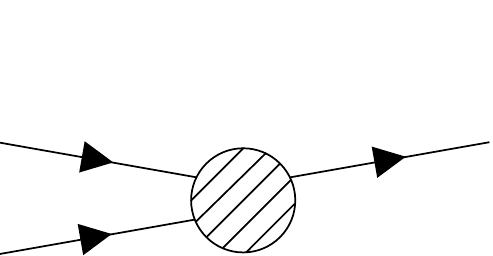} \ \includegraphics[scale=.3]{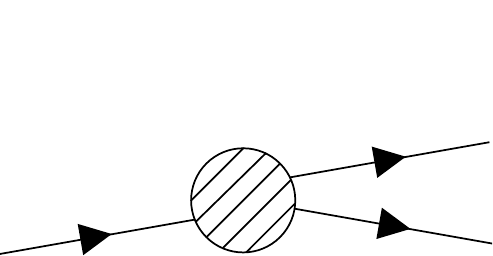} \\ ~\\
\hline\\
\underline{magnon-phonon} & \\
phonon emission/absorption & \includegraphics[scale=.3]{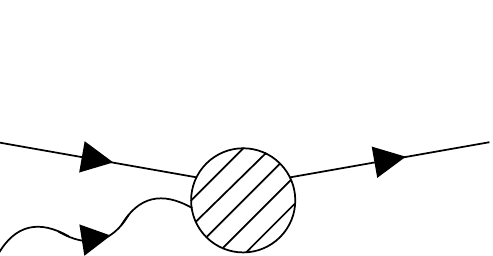} \ \includegraphics[scale=.3]{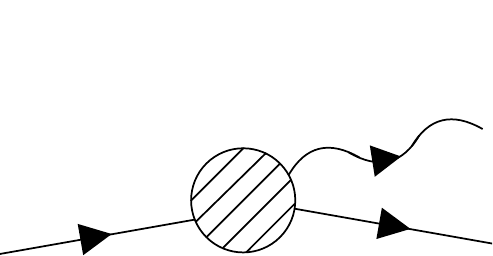} \\
(mainly exchange-based) \\
2-magnon conversion & \includegraphics[scale=.3]{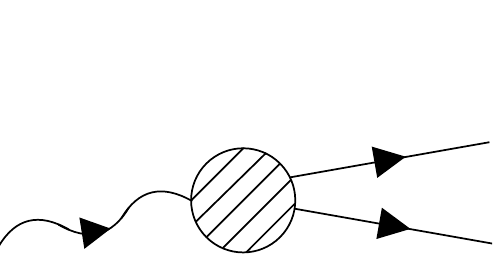} \ \includegraphics[scale=.3]{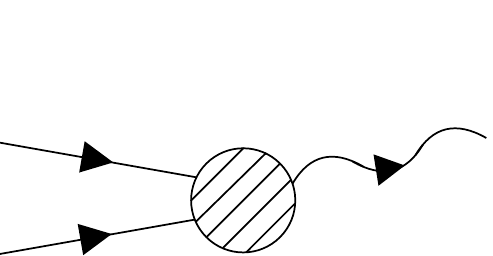} \\
(relativistic or dipole-dipole) \\
\end{tabular}
\caption{\label{tab:diagrams} Schematic representation of the relevant microscopic scattering processes for magnon polarons: magnon-impurity, phonon-impurity, three-magnon, three-phonon, four-magnon, and magnon-phonon scattering. Solid arrows represent magnons; wavy arrows represent phonons.}
\end{ruledtabular}
\end{table}

The strong frequency dependence of the degree of mixing of magnon and phonon modes implies a strong frequency dependence of the distribution function $n_{\vq,\nu}$, especially at temperatures low enough that the system size $L$ is not much larger than the inelastic relaxation lengths. As a consequence, a frequency-averaged description in terms of a (mode-dependent) ``temperature'' or ``chemical potential'' \cite{Kikkawa-2016,Flebus-2017,Cornelissen-2017,An-2016,Schmidt-2018,Agrawal-2013,Xi-2020} is unlikely to be an accurate characterization of the magnon-polaron distribution at low temperatures. Instead, the full frequency dependence of the distribution must be retained in a theoretical description. At the same time, the dominance of intra-mode impurity scattering justifies a simplified description of the distribution function in which the full angle dependence is replaced by one isotropic and one anisotropic moment only. These considerations are the basis for the approach we outline in Secs.\ \ref{sec:linear} and \ref{sec:scattering}.

The insignificance of inelastic relaxation processes at low temperatures means that coherent superpositions of magnon-polaron modes could in principle be long lived. Such coherent positions occur naturally, {\em e.g.} when a magnon polaron scatters from a magnetic impurity, which couples to its spin content only. In that case, the excitation that exists immediately after the scattering event is a {\em coherent} superposition of magnon polarons at the same frequency, with amplitudes that are such that excitation is (initially, in this case) of pure magnon type. For such a coherent superposition of magnon polarons, the phonon and magnon content of the excitation undergo Rabi-like oscillations upon propagation. The length scale for these oscillations is $l_{\rm coh}(\omega) \sim 1/\min_{\lambda}(\Delta q_{\lambda}(\omega))$, where $\Delta q_{\lambda}(\omega)$ is the difference of the wavenumbers of magnon-like and phonon-like magnon-polaron modes at the same frequency $\omega$. Only after a propagation length much larger than $l_{\rm coh}$ the excitation can be described as a ``classical'' mixture of different magnon-polaron modes. Like any theory that describes excitations in terms of their distribution function, the Boltzmann theory of this Section does not include coherence effects. This means that the Boltzmann approach is valid only on length scales larger than $l_{\rm coh}(\omega)$. For generic frequencies $\omega$, $l_{\rm coh}(\omega)$ is of the order of the wavelength, so that this condition is not a serious limitation on the applicability of the Boltzmann approach. However, close to the resonance frequencies, $l_{\rm coh}(\omega)$ may become appreciable and the Boltzmann theory could possibly overestimate the rate of impurity-mediated scattering between different magnon-polaron modes. Indeed, taking numerical values from Table \ref{tab:values}, we estimate that close to the resonance frequencies, $l_{\rm coh}(\omega)$ may be several $\mu$m, which is only slightly below typical system sizes or impurity scattering lengths.

\subsection{Linear response}
\label{sec:linear}

{\em Distribution function.---} To simplify the analysis of the coupled Boltzmann equations for the magnon-polaron distribution functions we consider small deviations from equilibrium only and linearize the distribution functions $n_{\vq,\nu}(x)$ around their equilibrium distributions,
\begin{equation}
n_{\vq,\nu} = n^0(\omega_{\vq,\nu}) + \omega_{\vq,\nu} \left(-\frac{\partial n^0}{\partial \omega}\right) \psi_{\vq,\nu} , \label{eq:linearresponse}
\end{equation}
where $n^0(\omega_{\vq,\nu}) = 1/(e^{\hbar \omega_{\vq,\nu}/\kb T} - 1)$ is the Planck distribution. In local thermal equilibrium at temperature $T + \Delta T(x)$, one has $\psi_{\vq,\nu}(x) = \Delta T(x) / T$. Similarly, the distribution functions $n_{\rm I}(\omega)$ and $n_{\rm N}(\omega)$ in the non-magnetic regions to the left and right of the magnetic insulator are written
\begin{equation}
  n_{\rm I,N}(\omega) = n^0(\omega_{\vq,\nu}) +\omega_{\vq,\nu} \left(-\frac{\partial n^0}{\partial \omega}\right) \, \psi_{\rm I,N} ,
  \label{eq:nIN}
\end{equation}
where $\psi_{\rm I,N} = \Delta T_{\rm I,N}/T$. The Boltzmann equation (\ref{eq:bo}) for the linearized distribution function reads
\begin{equation}
  v_{\vq,\nu x} \frac{\partial \psi_{\vq,\nu}}{\partial x} =
  J_{\vq,\nu},
\label{eq:bo-lin}
\end{equation}
where $J_{\vq,\nu}$ is a linearized version of the collision integral, see Sec.\ \ref{sec:scattering}.

{\em Simplified angular dependence.---}
Anticipating that impurity scattering suppresses most variations of $\psi_{\vq,\nu}$ with the propagation direction $\vv_{\vq,\nu}$ of the magnon-polaron mode, for $\psi_{\vq,\nu}$ we assume a simplified dependence on the wavevector $\vq$ such that at each frequency $\omega$ there is one isotropic moment (even in $\vq$) and one anisotropic moment (odd in $\vq$),
\begin{equation}
  \psi_{\vq,\nu} = \psi_{0,\nu}(\omega_{\vq,\nu}) + v_{\vq,\nu x} \psi_{1,\nu}(\omega_{\vq,\nu}).
  \label{eq:psiansatz}
\end{equation}
The isotropic moment and the anisotropic moment are obtained from the full distribution function $\psi_{\vq,\nu}$ as
\begin{align}
  \psi_{0,\nu}(\omega) =&\, \frac{1}{V \cE_{2,\nu}(\omega)} \sum_{\vq} \psi_{\vq,\nu} v^2_{\vq,\nu x} \delta(\omega_{\vq,\nu} -\omega) , \nonumber \\
  \psi_{1,\nu}(\omega) =&\, \frac{1}{V \cE_{2,\nu}(\omega)} \sum_{\vq} \psi_{\vq,\nu} v_{\vq,\nu x} \delta(\omega_{\vq,\nu} -\omega) ,
  \label{eq:psimoments}
\end{align}
where the normalization factor $\cE_{n,\nu}(\omega)$ is defined as
\begin{equation}
\cE_{n,\nu}(\omega) = \frac{1}{V} \sum_{\vq} |v_{\vq,\nu x}|^n \delta(\omega_{\vq,\nu} - \omega),\ \ n=0,1,2,\ldots.
\label{eq:norm}
\end{equation}
For an isotropic dispersion one has
\begin{equation}
\cE_{n,\nu}(\omega) = \cE_{0,\nu}(\omega) \frac{v_{\nu}(\omega)^n}{n+1},
\end{equation}
where $v_{\nu}(\omega) = |\partial \omega_{\vq,\nu}/\partial \vq|$ is the group velocity of the magnon-polaron mode $\nu$ at frequency $\omega$. 

\subsection{Collision integral}
\label{sec:scattering}

The collision integral $J_{\vq,\nu}$ in the linearized Boltzmann equation (\ref{eq:bo-lin}) has the general form
\begin{equation}
  J_{\vq,\nu} = \frac{1}{V} \sum_{\vq',\nu'} \Gamma_{\vq,\nu;\vq',\nu'} (\psi_{\vq',\nu'} - \psi_{\vq,\nu}),
  \label{eq:Jgeneral}
\end{equation}
with an effective linearized collision rate $\Gamma_{\vq,\nu;\vq',\nu'}$ that describes both elastic and inelastic scattering processes,
\begin{equation}
  \Gamma_{\vq,\nu;\vq',\nu'} = \Gamma^{\rm el}_{\vq,\nu;\vq',\nu'} \delta(\omega_{\vq,\nu} - \omega_{\vq',\nu'})
  + \Gamma^{\rm inel}_{\vq,\nu;\vq',\nu'}.
\end{equation}
Magnon-impurity and phonon-impurity scattering contribute to the elastic term in the collision integral. The dominant microscopic inelastic scattering processes are ``three-magnon'' and ``three-phonon'' scattering --- splitting or confluence processes in which one phonon or magnon scatters into two or vice versa ---, ``four-magnon'' processes, exchange-based magnon-phonon interaction, and relativistic or dipole-dipole-based inelastic magnon-phonon scattering, which includes processes in which one phonon creates a pair of magnons and vice versa. These processes are illustrated schematically in Table \ref{tab:diagrams}. In this Section we describe representative model Hamiltonians for the dominant scattering mechanisms and give an overview of the structure of the corresponding linearized collision rates for the magnon-polaron system.

{\em Magnon-impurity scattering---} As a simple model for impurity scattering of magnetic modes, we consider the Heisenberg interaction Hamiltonian (\ref{eq:hex}) with a random fluctuating magnetic field $B_j = B + \delta B_j$ and a random value $S_j = S + \delta S_j$ of the magnitude of the spin at each lattice site, $\delta B_j$ and $\delta S_j$ being randomly distributed with zero mean and with variance $\langle \delta B^2 \rangle \equiv \langle \delta B_j^2 \rangle \ll B^2$ and $\langle \delta S^2 \rangle \equiv \langle \delta S_j^2 \rangle \ll S^2$. In terms of the magnon polarons, the magnon-impurity Hamiltonian reads
\begin{align}
	H^{\rm mi} = \frac{\hbar}{\sqrt{V}} \sum_{\vq,\vq'} \sum_{\nu,\nu'} U^{\rm mi}_{\vq \nu;\vq'\nu'} a_{\vq,\nu}^* a_{\vq',\nu'} , \label{eq:hmagimp-pol}
\end{align}
plus terms that create or annihilate two magnon polarons. Since such processes do not conserve energy they do not contribute to the collision integral. The matrix element for magnon-impurity scattering is
\begin{equation}
  U^{\rm mi}_{\vq \nu;\vq' \nu'} = U^{\rm mi,0}_{\vq;\vq'} \left( V_{\vq,4,\nu}^* V_{\vq',4,\nu'} + V_{\vq,8,\nu}^* V_{\vq',8,\nu'} \right) .
\end{equation}
with $U^{\rm mi,0}_{\vq;\vq'}$ the corresponding matrix element in the absence of magnon-phonon coupling. The matrix $V_{\vq}$ is the symplectic $8 \times 8$ matrix that diagonalizes the magnon-phonon Hamiltonian $H_{\vq}$ of Eq.\ (\ref{eq:Hq}), see Eq.\ (\ref{eq:Vq}). Statistically, the mean $\langle U^{\rm mi,0}_{\vq;\vq'} \rangle$ vanishes, whereas the fluctuations of $U^{\rm mi,0}_{\vq;\vq'}$ are given by
\begin{equation}
  \langle |U^{{\rm mi},0}_{\vq;\vq'}|^2 \rangle = \frac{a^7 J^2}{4} \langle \delta S^2 \rangle (q^2 + q'^2)^2
  + \mu^2 a^3 \langle \delta B^2 \rangle. \label{eq:magimp}
\end{equation}

\bigskip
{\em Phonon-impurity scattering---} As a simple model for impurity scattering of lattice vibrations, we consider the phonon Hamiltonian (\ref{eq:hp}) with a random value  $m_j = m + \delta m_j$ of the masses of the lattice ions. Again, we take $\delta m_j$ randomly distributed with zero mean and with variance $\langle \delta m^2\rangle \equiv \langle \delta m_j^2 \rangle \ll m^2$. In terms of the magnon-polaron modes we find the phonon-impurity Hamiltonian
\begin{equation}
	H^{\rm pi} = \frac{\hbar}{\sqrt{V}} \sum_{\vq,\vq'} \sum_{\nu,\nu'} U^{\rm pi}_{\vq \nu;\vq' \nu'} a^*_{\vq,\nu} a_{\vq',\nu'} \label{eq:hphoimp-pol}
\end{equation}
where
\begin{align}
  U^{\rm pi}_{\vq \nu;\vq' \nu'}
  =&\,
  \sum_{\lambda,\lambda'}
  (V_{\vq,\lambda,\nu}^* + V_{\vq,\lambda+4,\nu}^*)
  \\ \nonumber &\, \ \ \mbox{} \times
  (V_{\vq',\lambda',\nu'} + V_{\vq',\lambda'+4,\nu'})
  U^{\rm pi,0}_{\vq \lambda;\vq' \lambda'} .
\end{align}
We have again left out contributions that create or annihilate two magnon polarons, because these do not contribute to the collision integrals.
The statistical average of the phonon-impurity matrix element vanishes, $\langle U_{\vq\lambda;\vq' \lambda'}\rangle = 0$. The variance is \footnote{The phonon-impurity Hamiltonian of Ref. \cite{Flebus-2017} has a statistically independent mode-diagonal matrix elements $U^{\rm pi,0}_{\vq\lambda;\vq'\lambda'} \propto \delta_{\lambda\lambda'}$ for the three phonon modes that do not scatter between the two degenerate transverse phonon modes. This is unphysical, since the assignment of the polarization vectors for the degenerate transverse phonon modes is arbitrary. Moreover, in spite of its simplicity, the microscopic model (\ref{eq:phoimp}) clearly shows that impurity scattering approximately equally connects phonon modes of all polarizations, longitudinal as well as transverse.}
\begin{equation}
  \langle |U^{\rm pi,0}_{\vq,\lambda;\vq',\lambda'}|^2 \rangle = \frac{a^3 \langle \delta m^2 \rangle}{4 m^2} |\ve_{\vq,\lambda}^* \cdot \ve_{\vq',\lambda'}|^2 \omega^0_{\vq\lambda} \omega^0_{\vq'\lambda'}.
  \label{eq:phoimp}
\end{equation}

The linearized collision rate that is derived from the magnon-impurity interaction (\ref{eq:hmagimp-pol}) and phonon interaction (\ref{eq:hphoimp-pol}) reads
\begin{align}
\Gamma^{\rm el}_{\vq,\nu;\vq',\nu'} =&\, 2\pi 
 \left(
  \langle |U^{\rm mi}_{\vq \nu;\vq' \nu'}|^2
    \rangle +
  \langle |U^{\rm pi}_{\vq \nu;\vq' \nu'}|^2
    \rangle \right) . \label{eq:collisionsElastic}
\end{align}

{\em Numerical values.---} To obtain numerical values for the variances $\langle \delta B^2\rangle$ and $\langle \delta S^2\rangle$, we relate these to the corresponding magnon mean free path $l_{\rm mi}$, which determines low-temperature measurements of the respective magnon thermal conductivities,
\begin{align}
  l^{-1}_{\rm mi}(\omega) =&\, \frac{1}{4\pi J^2 S^2 a}  \left[\mu^2 \langle \delta B^2\rangle + \frac{\langle\delta S^2\rangle}{S^2} (\omega - \mu B)^2 \right].
  \label{eq:mfpm}
\end{align}
Similarly, we relate the variance $\langle \delta m^2 \rangle$ to the phonon mean free paths $l_{{\rm pi},\lambda}$, which is related to the phonon thermal conductivity,
\begin{align}
  l^{-1}_{{\rm pi},\lambda}(\omega) =&\, \tau_{\rm pi}(\omega)^{-1} c_{\lambda}^{-1},
  \label{eq:mfpp}
\end{align}
with
\begin{align}
  \tau_{\rm pi}(\omega)^{-1} =&\, \frac{a^3}{12 \pi \hbar^4} \frac{\langle \delta m^2 \rangle}{m^2} \sum_{\lambda'} \frac{1}{c^3_{\lambda'}} \omega^4,
  \label{eq:tau_pi}
\end{align}
where $c_1 = c_{\rm l}$ is the longitudinal phonon velocity and $c_{2} = c_{3} = c_{\rm t}$ the transverse phonon velocity.
Our microscopic model coincides with the shape of the best fit to the impurity rates in Ref.\ \cite{Walton-1973}. Comparison of Eqs.\ (\ref{eq:mfpm}) and (\ref{eq:mfpp}) with the mean free paths reported in Ref.\ \cite{Walton-1973} yields the variances $\langle \delta S^2 \rangle$, $\langle \delta B^2 \rangle$, and $\langle \delta m^2 \rangle$ given in the center column Table \ref{tab:impurity}. The corresponding mean free paths $l_{\rm mi}$ and $l_{\rm pi}$ are shown in Fig.\ \ref{fig:relaxationlengths}. The phonon mean free path $l_{\rm pi}$ obtained from this procedure is about two orders of magnitude smaller than the mean free path $l_{\rm mi}$ for magnon-impurity scattering. To allow for a comparison with the theory of Refs.\ \cite{Kikkawa-2016,Flebus-2017}, which infers a smaller magnon-impurity mean free path from the low-temperature spin Seebeck effect measurements, we also consider parameter values in which the orders of magnitude for phonon-impurity and magnon-impurity scattering are interchanged, which corresponds to a sample that is of higher acoustic than magnetic quality. These values are shown in the rightmost column of Table \ref{tab:impurity}. 

To make is easier to separate different contributions to the spin Seebeck effect, in some of our calculations we also use a phenomenological white-noise model for the impurity scattering rates, for which the mean free paths $l_{\rm pi}$ and $l_{\rm mi}$ have a weaker frequency dependence than for the microscopic model of Eqs.\ (\ref{eq:magimp}) and (\ref{eq:phoimp}). The phenomenological white-noise model is defined by setting
\begin{align}
  \langle |U^{\rm mi,0}_{\vq;\vq'}|^2 \rangle =u_{\rm mi}^2 / V , \ \
  \langle |U^{\rm pi,0}_{\vq,\lambda;\vq', \lambda'}|^2 \rangle = u_{\rm pi}^2 /  3V.
  \label{eq:whitenoise}
\end{align}
The variances $u_{\rm mi}^2$ and $u_{\rm pi}^2$, which determine the mean free paths
\begin{align}
l^{-1}_{\rm mi}(\omega) = \frac{u^2_{\rm mi}}{4\pi a^4 J^2 S^2} \ , \ \
l^{-1}_{\rm pi,\lambda}(\omega) = \frac{u^2_{\rm pi}}{3 \pi c_{\lambda}} \sum_{\lambda'} \frac{1}{c^3_{\lambda'}} \omega^2 ,
\label{eq:mfp_whitenoise}
\end{align}
are adjusted to low temperature measurements of the magnon and phonon thermal mean free paths in Ref.\ \cite{Heremans-2014}.
Numerical values for the case of a sample with higher magnetic than acoustic quality and alternative values for the case of a higher acoustic than magnetic quality are shown in Table \ref{tab:impurity}.

\begin{table}
\begin{ruledtabular}
\begin{tabular}{lcc}
\textrm{Quantity} & \textrm{Value} & \textrm{Alternative value with $l_{\rm mi} < l_{\rm pi}$} \\
\colrule\\
$\sqrt{\langle\delta m^2\rangle}/m $ & $0.09$ & $0.009$ \\
$\sqrt{\langle\delta S^2\rangle} / S $ & $0.05$ & $0.16$ \\
$\sqrt{\langle\delta B^2\rangle}$ & $0.07\,{\rm T}$ & $0.22\,{\rm T}$\\ \colrule
$u_{\rm mi} / \sqrt{a^3}$ & $65\,{\rm GHz}$ & 650\,{\rm GHz}\\
$u_{\rm pi} / \sqrt{a^3}$ & $332\,{\rm GHz}$ & 33.2\,{\rm GHz}\\
\end{tabular}
\caption{\label{tab:impurity} Numerical values for the model parameters for impurity scattering.}
\end{ruledtabular}
\end{table}

{\em Inelastic scattering---}  At low temperatures the elastic scattering with impurities dominates the relaxation of the magnon polarons. With increasing temperature or decreasing impurity concentration the relative importance of inelastic scattering increases. Here we supply the leading-order contributions to the corresponding magnon-polaron Hamiltonians and collision integrals. Except for the three-phonon interaction, the underlying magnon-magnon and magnon-phonon Hamiltonians can be derived from the microscopic model presented in Section \ref{sec:mp1}, by expanding the Heisenberg and pseudo-dipolar interactions to higher orders in the magnon amplitudes $\vn_j$ and the displacement vectors $\vu_j$. The leading inelastic interaction involving phonons only is the three-phonon interaction. It arises from anharmonicities of the lattice potential and can effectively be derived within a continuum elastic strain model \cite{Ziman-1960}, see appendix \ref{app:inel}.

In terms of magnon polarons the three-polaron interaction Hamiltonian may be written as
\begin{align}
H^{\rm in,3} =&\, \frac{1}{\sqrt{V}} \sum_{\vq,\nu} \sum_{\vq',\nu'} \sum_{\vq'',\nu''} ( U^{\rm in,3}_{\vq \nu;\vq'\nu',\vq'' \nu''} a^*_{\vq,\nu} a_{\vq',\nu'} a_{\vq'',\nu''} \nonumber\\
&\, + U^{\rm in,3}_{\vq \nu,\vq'\nu';\vq''\nu''} a^*_{\vq,\nu} a^*_{\vq',\nu'} a_{\vq'',\nu''}) \label{eq:Hamagnonpolaron}
\end{align}
where we left out contributions that create or annihilate three magnon polarons, because these do not contribute to the collision integral. The matrix element $U^{\rm in,3}_{\vq \nu;\vq'\nu',\vq'' \nu''} = U^{\rm in,3}_{\vq \nu;\vq''\nu'',\vq' \nu'}$ of this general three-polaron interaction is assumed to be symmetric. It has contributions from three-phonon scattering (with matrix element $U^{\rm 3p,0}_{\vq \lambda;\vq' \lambda' , \vq'' \lambda''}$ in the absence of magnon-phonon coupling), three-magnon scattering (with matrix element $U^{\rm 3m,0}_{\vq;\vq',\vq''}$ in the absence of magnon-phonon coupling), one-phonon-two-magnon scattering (with matrix element $U^{\rm mp,0}_{\vq;\vq' \lambda,\vq''}$ in the basis of phonon and magnon states), and relativistic one-phonon-two-magnon scattering (with matrix element $U^{\rm rel,0}_{\vq \lambda;\vq',\vq''}$ in the basis of phonon and magnon states),
\begin{widetext}
\begin{align}
	U^{\rm in,3}_{\vq \nu;\vq' \nu',\vq'' \nu''} =&\, \left\{\vphantom{frac12} \sum_{\lambda} \sum_{\lambda'} \sum_{\lambda''} U^{\rm 3p,0}_{\vq \lambda;\vq' \lambda' , \vq'' \lambda''} (V^*_{\vq, \lambda,\nu} - V^*_{\vq,\lambda+4,\nu}) (V_{\vq',\lambda',\nu'} - V_{\vq',\lambda'+4,\nu'}) (V_{\vq'',\lambda'',\nu''} - V_{\vq'' , \lambda''+4,\nu''}) \right. \nonumber\\
	&\, \mbox{} +  \sum_{\lambda} U^{\rm mp,0}_{\vq;\vq' \lambda,\vq''}
        \left[ \frac{1}{2} (V^*_{\vq,4,\nu} V_{\vq'',4,\nu''} + V^*_{\vq,8,\nu} V_{\vq'',8,\nu''}) (V_{\vq',\lambda,\nu'} - V_{\vq',\lambda+4,\nu'}) \right. \nonumber\\
	&\, \ \ \ \ \left. \mbox{} + \frac{1}{2} (V^*_{\vq,4,\nu} V_{\vq',4,\nu'} + V^*_{\vq,8,\nu} V_{\vq',8,\nu'}) (V_{\vq'',\lambda,\nu''} - V_{\vq'',\lambda+4,\nu''}) \right] \nonumber\\
	&\, \left. \mbox{} + \sum_{\lambda} U^{\rm rel,0}_{\vq\lambda;\vq',\vq''} (V^*_{\vq,\lambda,\nu} - V^*_{\vq,\lambda+4,\nu}) (V_{\vq',4,\nu'} V_{\vq'',4,\nu''} + V_{\vq',8,\nu'} V_{\vq'',8,\nu''})   \vphantom{\frac12}\right.\nonumber\\
	&\, \left. \mbox{} + U^{\rm 3m,0}_{\vq;\vq',\vq''} (V^*_{\vq,4,\nu} V_{\vq',4,\nu'} V_{\vq'',4,\nu''} - V^*_{\vq,8,\nu} V_{\vq',8,\nu'} V_{\vq'' , 8,\nu''}) \vphantom{\frac12}\right\} \delta_{\vq,\vq'+\vq''}.
\end{align}
Here the indices $\lambda$, $\lambda'$, and $\lambda''$ label the polarization state of phonons and, hence, take the values $1$, $2$, $3$. The corresponding linearized three-polaron transition rate is
\begin{align}
  \Gamma^{\rm in,3}_{\vq,\nu;\vq',\nu'} =&\, \frac{2\pi}{\hbar^2} \frac{\omega_{\vq',\nu'}}{\omega_{\vq,\nu} (1+n^0_{\vq,\nu})} \sum_{\vq'', \nu''} \left\lbrace\vphantom{\frac12} |U^{\rm in,3}_{\vq' \nu' ; \vq \nu , \vq'' \nu''}|^2 \delta(\omega_{\vq,\nu}-\omega_{\vq',\nu'}+\omega_{\vq'',\nu''}) (1+ n^0_{\vq',\nu'}) n^0_{\vq'',\nu''} \right. \nonumber\\
&\, \mbox{} - |U^{\rm in,3}_{\vq''\nu'';\vq \nu, \vq' \nu'}|^2 \delta(\omega_{\vq,\nu}+\omega_{\vq',\nu'}-\omega_{\vq'',\nu''}) n^0_{\vq',\nu'} (1+n^0_{\vq'',\nu''})  \nonumber\\
&\, \left. \mbox{} + |U^{\rm in,3}_{\vq\nu;\vq' \nu', \vq'' \nu''}|^2 \delta(\omega_{\vq,\nu}-\omega_{\vq',\nu'}-\omega_{\vq'',\nu''}) (1+ n^0_{\vq',\nu'})(1+ n^0_{\vq'',\nu''})  \vphantom{\frac12}\right\rbrace . \label{eq:collisions_3in}
\end{align}

The main contribution to the four-polaron Hamiltonian results from the Heisenberg exchange interaction (\ref{eq:hex}). As the underlying four-magnon Hamiltonian conserves the magnon number, see Eq.\ (\ref{eq:h4m}), the four-polaron scattering is also dominated by processes that conserve the polaron number. These processes are described by the Hamiltonian
\begin{equation}
H^{\rm in,4} = \frac{1}{V} \sum_{\vq,\nu} \sum_{\vq_2,\nu_2} \sum_{\vq',\nu'} \sum_{\vq'_2,\nu'_2} U^{\rm in,4}_{\vq \nu,\vq_2 \nu_2 , \vq' \nu', \vq_2' \nu_2'} a^*_{\vq,\nu} a^*_{\vq_2,\nu_2} a_{\vq',\nu'} a_{\vq_2',\nu_2'} + {\rm H.c.},
\end{equation}
where the matrix element $U^{\rm in,4}_{\vq \nu,\vq_2 \nu_2 , \vq' \nu', \vq_2' \nu_2'}$ is expressed in terms of the matrix element $U^{\rm 4m,0}_{\vq \nu,\vq_2 \nu_2 , \vq' \nu', \vq_2' \nu_2'}$ of the four-magnon Hamiltonian in the absence of magnon-phonon coupling as
\begin{equation}
  U^{\rm in,4}_{\vq \nu,\vq_2 \nu_2 , \vq' \nu', \vq_2' \nu_2'} = U^{\rm 4m,0}_{\vq \nu,\vq_2 \nu_2 , \vq' \nu', \vq_2' \nu_2'} (V^*_{\vq,4,\nu} V^*_{\vq_2,4,\nu_2} V_{\vq',4,\nu'} V_{\vq_2',4,\nu_2'} + V^*_{\vq,8,\nu} V^*_{\vq_2,8,\nu_2} V_{\vq',8,\nu'} V_{\vq_2' , 8,\nu_2'}) \delta_{\vq+\vq_2,\vq'+\vq_2'} .
\end{equation}
The contribution of four-polaron processes to the inelastic collision rate reads
\begin{align}
\Gamma^{\rm in,4}_{\vq,\nu;\vq', \nu'} =&\, \frac{2\pi}{\hbar^2} \frac{\omega_{\vq',\nu'}}{\omega_{\vq,\nu} (1+n^0_{\vq,\nu})} \sum_{\vq_2,\nu_2} \sum_{\vq_2',\nu_2'} n^0_{\vq_2',\nu_2'} \left\lbrace (1+n^0_{\vq',\nu'}) n^0_{\vq_2,\nu_2} |U^{\rm in,4}_{\vq\nu,\vq_2 \nu_2; \vq' \nu', \vq_2' \nu_2'}|^2 \delta(\omega_{\vq,\nu}+\omega_{\vq_2,\nu_2}-\omega_{\vq',\nu'}-\omega_{\vq'_2,\nu'_2}) \right. \nonumber\\
&\ \left. - \frac{1}{2} n^0_{\vq',\nu'} (1+n^0_{\vq_2,\nu_2}) |U^{\rm in,4}_{\vq\nu,\vq' \nu'; \vq_2 \nu_2, \vq_2' \nu_2'}|^2 \delta(\omega_{\vq,\nu}+\omega_{\vq',\nu'}-\omega_{\vq_2,\nu_2}-\omega_{\vq'_2,\nu'_2}) \right\rbrace . \label{eq:collisions_4in}
\end{align}
\end{widetext}
Explicit expressions for the matrix elements $U^{{\rm 3p},0}$, $U^{{\rm mp},0}$, $U^{{\rm 3m},0}$, and $U^{{\rm 4m},0}$ in the framework of the microscopic model of Sec.\ \ref{sec:model} are given in App.\ \ref{app:inel}. The inelastic rates $\Gamma_{\vq,\nu;\vq',\nu'}$ can then be calculated using the numerical values given in Table \ref{tab:values}. To obtain the inelastic magnon and phonon relaxation lengths $\lambda_{\rm m,p}$ of Fig.\ \ref{fig:relaxationlengths}, we first calculate the frequency-resolved scattering rate $\tau_{\rm m,p}^{-1}(\omega)$ of magnons and phonons using Fermi's Golden rule and then determine the corresponding relaxation lengths $\lambda_{\rm m,p}(\omega)$ as
\begin{equation}
  \lambda_{\rm m,p}(\omega) = l_{\rm mi,pi}(\omega) \sqrt{\frac{\tau_{\rm m,p}(\omega)}{\tau_{\rm mi,pi}(\omega)}},
\end{equation}
where $l_{\rm mi}$ and $l_{\rm pi}$ are the mean free paths for magnon-impurity and phonon-impurity scattering, see Eqs.\ (\ref{eq:mfpm}) and (\ref{eq:mfpp}), and $\tau_{\rm mi}$ and $\tau_{\rm pi}$ are the corresponding lifetimes.

{\em Simplified angular dependence.---} In Eq.\ (\ref{eq:psimoments}) a simplified ansatz for the linearized distribution function, with one isotropic moment $\psi_{0,\nu}(\omega)$ and one anisotropic moment $\psi_{1,\nu}(\omega)$ per mode $\nu$ and frequency $\omega$ was introduced. Because of the symmetry property (\ref{eq:Vsymmetry}) of the matrix elements $V_{\vq,\lambda,\nu}$, the linearized transition rates $\Gamma_{\vq,\nu;\vq',\nu'}$ for the magnon polarons satisfy the symmetry property
\begin{equation}
  \Gamma_{\vq,\nu;\vq',\nu'} = \Gamma_{-\vq,\nu;-\vq',\nu'}
\end{equation}
if the microscopic rates in the absence of magnon-phonon coupling also satisfy this symmetry. Combining this symmetry property with the antisymmetry of the magnon-polaron velocity $v_{\vq,\nu x}$, one finds that the general form (\ref{eq:Jgeneral}) of the linearized collision integral results in two coupled equations for the isotropic and anisotropic moments $\psi_{0,\nu}(\omega)$ and $\psi_{1,\nu}(\omega)$,
\begin{align}
\frac{\partial \psi_{1,\nu}(\omega)}{\partial x} =&\, - \int d\omega' \sum_{\nu'} \cG^{0}_{\nu,\nu'}(\omega,\omega') \psi_{0,\nu'}(\omega'), \nonumber\\
\frac{\partial \psi_{0,\nu}(\omega)}{\partial x} =&\, - \int d\omega' \sum_{\nu'} \cG^{1}_{\nu,\nu'}(\omega,\omega') \psi_{1,\nu'}(\omega'). \label{eq:boltzmannElastic}
\end{align}
Here the $4 \times 4$ matrices $\cG^{0}$ and $\cG^{1}$ are defined as
\begin{align}
  \cG^0_{\nu,\nu'}(\omega,\omega') =&\, \frac{1}{V^2 \cE_{2,\nu}(\omega)}
  \sum_{\vq,\vq''}
  \sum_{\nu''} \Gamma_{\vq,\nu;\vq',\nu''} \delta(\omega - \omega_{\vq,\nu})
  \nonumber \\ &\, \mbox{} \times
  [\delta_{\nu,\nu'} \delta(\omega'-\omega) -
    \delta_{\nu',\nu''} \delta(\omega' - \omega_{\vq'',\nu''})], \nonumber \\
  \cG^1_{\nu,\nu'}(\omega,\omega') =&\, \frac{1}{V^2 \cE_{2,\nu}(\omega)}
  \sum_{\vq,\vq''}
  \sum_{\nu''} \Gamma_{\vq,\nu;\vq',\nu''} \delta(\omega - \omega_{\vq,\nu})
  \nonumber \\ &\, \mbox{} \times
  [\delta_{\nu,\nu'} \delta(\omega'-\omega) v_{\vq,\nu x}^2
  \nonumber \\ &\, \ \ \mbox{} -
  \delta_{\nu',\nu''} \delta(\omega' - \omega_{\vq'',\nu''})
  v_{\vq,\nu x} v_{\vq'',\nu''x}].
    \label{eq:lambdaAverage}
\end{align}
The normalization coefficients $\cE_{2,\nu}(\omega)$ are defined in Eq.\ (\ref{eq:norm}).
For elastic scattering from impurities the transition rate $\Gamma_{\vq,\nu;\vq',\nu'}$ contains an additional delta function $\delta (\omega_{\vq,\nu} - \omega_{\vq',\nu'})$, which nullifies the frequency integration in Eq.\ (\ref{eq:boltzmannElastic}), so that one arrives at a set of eight coupled differential equations, which can be solved for each frequency separately.

The matrix $\cG^{0}_{\nu,\nu'}$ in the isotropic Boltzmann moment obeys the condition
\begin{equation}
  \int d\omega' \sum_{\nu'} \cG^{0}_{\nu,\nu'}(\omega,\omega') = 0.
  \label{eq:lambdaCondition}
\end{equation}
This ensures that the uniform isotropic solution $\psi_{0,\nu} = \psi$ and $\psi_{1,\nu}= 0$, with $\psi$ a constant, is a solution of the equations. In the case of elastic scattering only, $\psi_{0,\nu}(\omega) = \psi(\omega)$ is a solution of the equations, with $\psi(\omega)$ an arbitrary function of $\omega$.

\subsection{Reflection coefficients at interfaces}
\label{sec:reflection}

To describe the boundary conditions at the IF and FN interfaces at $x=0$ and $x=L$, we note that the frequency $\omega$ and the transverse wavevector $\vq_{\perp} = q_y \ve_y + q_z \ve_z$ are conserved at the interface. Hence, instead of using the wavevector $\vq$ to label the magnon-polaron modes, we use the triple $(\omega,\vq_{\perp},\sigma_{\vq})$, where the sign $\sigma_{\vq} = \mbox{sign}\, (v_{\vq,\nu x})$ is the sign of the propagation direction. The range of the transverse wavevector $\vq_{\perp}$ is restricted to those values of $\vq_{\perp}$ for which the magnon-polaron mode $\nu$ is propagating at frequency $\omega$, {\em i.e.}, for which $q_{x}$ is real.

\begin{figure*}
\includegraphics[width=\linewidth]{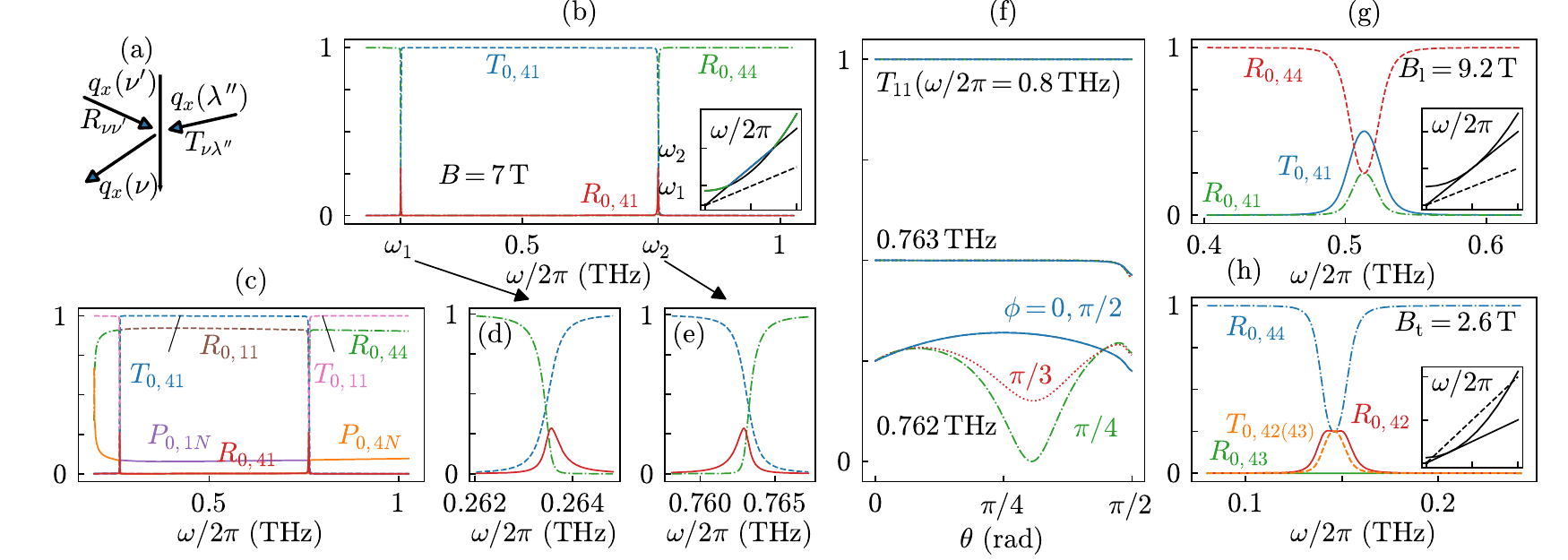}
\caption{\label{fig:interface}(a) Schematic picture of the interface reflection of a magnon polaron from branch $\nu'$, and transmission of a phonon from branch $\lambda''$, into a magnon polaron of branch $\nu$ with the respective reflection and transmission coefficients $R_{\nu \nu'}$ and $T_{\nu \lambda''}$. (b) Angle-averaged reflection and transmission coefficients $R_{0,\nu\nu'}$ and $T_{0,\nu\lambda''}$ at the IF interface for scattering into magnon-polaron mode $\nu = 4$ as a function of the frequency. (c) Coefficients $R_{0,\nu\nu'}$ and $T_{0,\nu\lambda''}$ at the FN interface for magnon-polaron modes $\nu=1,4$, as well as the probability $P_{0,\nu{\rm N}}$ that the magnon-polaron mode $\nu$ at the FN interface is excited by a spinful excitation of the conduction electrons in the normal metal. Panels (d) and (e) show magnifications of the resonant regions in (b). The coefficients shown in panels (b)--(e) are for incident longitudinal-phonon-like or magnon-like modes $\nu'=1$, $4$ (in F) and $\lambda''=1$ (in I or N). Coefficients for the transverse-phonon-like incident modes $\nu'=2$, $3$ or $\lambda'=2$, $3$ are approximately zero (not shown). Within the accuracy of the figure, the probability $P_{0,{\rm N}\nu}$ that the magnon-polaron mode $\nu'$ incident on the FN interface excites a spinful excitation of the conduction electrons in N is equal to $P_{0,\nu{\rm N}}$. In (f) the transmission amplitude $T_{11}$ is shown as a function of the polar angle $\theta$ and azimuthal angles $\phi$ for different frequencies $\omega$ in the vicinity of the resonance frequency. Panels (b)-(f) are evaluated for an applied magnetic field $B=7\,{\rm T}$. The non-zero reflection and transmission coefficients at the critical magnetic fields $B_{\rm l} \approx 9.2\,{\rm T}$ and $B_{\rm t}\approx\,2.6{\rm T}$ are shown as a function of the frequency in panels (g) and (h), respectively. In panels (b)--(h) material parameters for YIG are taken from Table \ref{tab:values}.}
\end{figure*}

The boundary condition at the IF interface at $x=0$ relates the distribution function of magnon-polaron modes moving away from the interface to the distribution of magnon-polaron modes and phonon modes moving towards the interface,
\begin{align}
n_{\nu}(\omega,\vq_{\perp},+) =&\ \sum_{\nu'} R_{\nu\nu'}(\omega,\vq_{\perp}) n_{\nu'}(\omega,\vq_{\perp},-) \nonumber\\
&\ + \sum_{\lambda''} T_{\nu \lambda''}(\omega,\vq_{\perp}) n_{\rm I}(\omega).
\label{eq:bboundaryL}
\end{align}
Here $R_{\nu\nu'}(\omega,\vq_{\perp})$ is the probability that a magnon polaron $\nu'$ incident on the IF interface reflects as a magnon-polaron mode $\nu$, see Fig.\ \ref{fig:interface}(a). Similarly, $T_{\nu\lambda''}(\omega,\vq_{\perp})$ is the probability that a phonon mode $\lambda''$ incident on the IF interface from I is transmitted as a magnon-polaron mode $\nu$. Only modes that are propagating at frequency $\omega$ and transverse wavevector $\vq_{\perp}$ enter the summations over $\nu'$ and $\lambda''$. The distribution function of phonons approaching the interface from I is $n_{\rm I}(\omega)$, see Eq.\ (\ref{eq:nIN}). Energy conservation at the IF interface at $x=0$ implies the unitarity condition \cite{Kamra-2014,Kamra-2015}
\begin{align}
  1 =&\, \sum_{\nu'} R_{\nu\nu'}(\omega,\vq_{\perp}) + \sum_{\lambda''} T_{\nu\lambda''}(\omega,\vq_{\perp}) \nonumber \\
  =&\, \sum_{\nu'} R_{\nu'\nu}(\omega,\vq_{\perp}) + \sum_{\lambda''} T_{\lambda''\nu}(\omega,\vq_{\perp})
  \label{eq:unitarityL}
\end{align}
With the help of the unitarity condition (\ref{eq:unitarityL}), the boundary condition (\ref{eq:bboundaryL}) is written in terms of the reflection coefficients $R_{\nu\nu'}(\omega,\vq_{\perp})$ for magnon-polaron modes only,
\begin{align}
  \label{eq:bboundary}
  n_{\nu}(\omega,\vq_{\perp},+) =&\,  \sum_{\nu'} R_{\nu\nu'}(\omega,\vq_{\perp}) n_{\nu'}(\omega,\vq_{\perp},-) \nonumber\\
  &\, \mbox{} +  \sum_{\nu'} [\delta_{\nu \nu'} - R_{\nu\nu'}(\omega,\vq_{\perp})] n_{\rm I}(\omega) .
\end{align}

Similarly, for the FN interface at $x=L$ the boundary condition relates the distribution function of magnon-polaron modes moving away from the interface to the distribution of magnon-polaron modes and phonon modes moving towards the interface. In addition to considering reflection coefficients $R_{\nu\nu'}(\omega,\vq_{\perp})$ and transmission coefficients $T_{\nu\lambda''}(\omega,\vq_{\perp})$ at the FN interface, one also has to account for the possibility that a magnon polaron incident on the FN interface excites a spinful excitation of the conduction electrons in the normal metal, and for the inverse process. Since the total energy current at the FN interface is conserved, we find the probability $P_{\nu{\rm N}}(\omega,\vq_{\perp})$ that a magnon polaron in mode $\nu$ emerging from the FN interface was excited there by an incident a spinful excitation of the conduction electrons as
\begin{equation}
  \sum_{\nu'} R_{\nu\nu'}(\omega,\vq_{\perp}) + \sum_{\lambda''} T_{\nu\lambda''}(\omega,\vq_{\perp}) = 1 - P_{\nu{\rm N}}(\omega,\vq_{\perp}). \label{eq:unitarityR1}
\end{equation}
Similarly, the probability $P_{{\rm N}\nu}(\omega,\vq_{\perp})$ that a magnon polaron in mode $\nu$ incidents on the FN interface excites a spinful excitation of the conduction electrons is
\begin{equation}
  \sum_{\nu'} R_{\nu'\nu}(\omega,\vq_{\perp}) + \sum_{\lambda''} T_{\lambda''\nu}(\omega,\vq_{\perp}) = 1 - P_{{\rm N}\nu}(\omega,\vq_{\perp}). \label{eq:unitarityR2}
\end{equation}
Using Eq.\ (\ref{eq:unitarityR1}), the boundary condition at the FN interface then reads
\begin{align}
  \label{eq:bboundary2}
  n_{\nu}(\omega,\vq_{\perp},-) =&\ \sum_{\nu'} R_{\nu\nu'}(\omega,\vq_{\perp}) n_{\nu'}(\omega,\vq_{\perp},+) \nonumber \\
  &\, \mbox{} + \sum_{\nu'} [\delta_{\nu \nu'} - R_{\nu\nu'}(\omega,\vq_{\perp}) ] n_{\rm N}(\omega) ,
\end{align}
where $n_{\rm N}(\omega)$ is the equilibrium distribution function for the non-magnetic normal metal, see Eq.\ (\ref{eq:nIN}).

The reflection and transmission coefficients $R_{\nu\nu'}$ and $T_{\nu\lambda''}$ can be computed from the equations of motion for the magnon-polaron modes and the boundary conditions at the interfaces at $x=0$ and $x=L$, see Eqs.\ (\ref{eq:boundaryInsulator1})--(\ref{eq:boundaryF}). Details of this calculation, which follows the ideas of the Landauer-B\"uttiker formalism \cite{Datta-2003}, can be found in App.\ \ref{app:c}.

{\em Simplified angular dependence.---}
Boundary conditions for the isotropic moment $\psi_{0,\nu}(\omega)$ and the anisotropic moment $\psi_{1,\nu}(\omega)$ of the linearized distribution function are obtained from Eqs.\ (\ref{eq:bboundary}) and (\ref{eq:bboundary2}) by enforcing consistency for the frequency-resolved energy current density $j_{\nu x}(\omega)$ at the interface carried by the magnon-polaron mode $\nu$, which is uniquely linked to the anisotropic moment via
\begin{align}
j_{\nu x}(\omega) =&\,
  \cE_{2,\nu}(\omega) \hbar \omega^2 \left(-\frac{\partial n^0}{\partial \omega}\right) \psi_{1,\nu}(\omega) .
\label{eq:energycurrent}
\end{align}
The consistency condition for the anisotropic moment at the IF interface reads
\begin{align}
& \frac{\cE_{2,\nu}(\omega)}{\cE_{1,\nu}(\omega)} \sum_{\nu'}[\delta_{\nu \nu'} + \cR_{1,\nu\nu'}(\omega)] \psi_{1,\nu'}(\omega) = \nonumber \\
& \ \ \ \ \sum_{\nu'} [ \delta_{\nu\nu'} - \cR_{0,\nu\nu'}(\omega)] [ \psi_{\rm I}(\omega) - \psi_{0,\nu'}(\omega) ] ,  
\label{eq:bcL}
\end{align}
where 
\begin{align}
 \label{eq:Rdef}
\cR_{0,\nu\nu'}(\omega) =&\, \frac{2}{(2\pi)^3} \int \frac{d\vq_{\perp}}{\cE_{1,\nu}(\omega)} R_{\nu\nu'}(\omega,\vq_{\perp}) , \\
\cR_{1,\nu\nu'}(\omega) =&\, \frac{2}{(2\pi)^3} \int \frac{d\vq_{\perp}}{\cE_{2,\nu}(\omega)} R_{\nu\nu'}(\omega,\vq_{\perp}) | v_{\nu' x}(\omega,\vq_{\perp},-) | . \nonumber
\end{align}
The boundary condition at the interface at $x = L$ is derived in the same way and reads
\begin{align}
& \frac{\cE_{2,\nu}(\omega)}{\cE_{1,\nu}(\omega)} \sum_{\nu'}[\delta_{\nu \nu'} + \cR_{1,\nu\nu'}(\omega)] \psi_{1,\nu'}(\omega) = \nonumber \\
& \ \ \ \ \sum_{\nu'} [ \delta_{\nu\nu'} - \cR_{0,\nu\nu'}(\omega)] [ \psi_{0,\nu'}(\omega) - \psi_{\rm N}(\omega) ] .
  \label{eq:bcR}
\end{align}
Although the boundary conditions (\ref{eq:bcL}) and (\ref{eq:bcR}) are not quantitatively exact implementations of the microscopic boundary conditions (\ref{eq:bboundary}) and (\ref{eq:bboundary2}) --- an exact implementation of these boundary conditions is not compatible with the ansatz (\ref{eq:psiansatz}) --- they are good quantitative approximations. For example, for an interface with perfect transparency and a single mode, they correctly take into account interface effects in the ballistic and diffusive limits, and deviate less than $2.5 \%$ from the exact result in the ballistic-to-diffusive crossover \cite{DeJong-1994}. In App.\ \ref{app:ballistic} we compare distribution functions for a ballistic system of length $L \ll l_{\rm mi}$, $l_{\rm pi}$ and find that the difference between an exact calculation and a calculation based on the angle-averaged boundary conditions (\ref{eq:bcL}) and (\ref{eq:bcR}) differs less than $5\%$.

{\em Numerical values.---}
The angle-averaged reflection and transmission coefficients (\ref{eq:Rdef}) at the IF and FN interfaces are shown in Figure \ref{fig:interface}(b)--(e) for $B = 7\,{\rm T}$. Material parameters are as listed in Table \ref{tab:values}. At this value of the magnetic field, the magnon dispersion crosses that of longitudinal phonons, but not the disperson of transverse phonons, as shown in the inset of Fig.\ \ref{fig:interface}(b). The resonant frequencies at which the dispersions cross are at $\omega/2\pi \approx 0.26\,{\rm THz}$ and $\omega/2\pi \approx 0.76\,{\rm THz}$. The magnon-polaron modes are labeled such that $\nu=1$ refers to a magnon-like mode for $0.26\,{\rm THz} < \omega/2\pi < 0.76\,{\rm THz}$ and to a longitudinal-phonon-like mode otherwise. The mode $\nu=4$ is longitudinal-phonon-like for $0.26\,{\rm THz} < \omega/2\pi < 0.76\,{\rm THz}$ and magnon-like otherwise. The modes $\nu=2$ and $\nu=3$ are transverse-phonon-like for all frequencies. The label $\lambda''=1$ refers to the longitudinal phonon mode in the non-magnetic insulator I or the normal metal N. Figure \ref{fig:interface}(b) shows the angle-averaged reflection and transmission probabilities $R_{0,\nu\nu'}$ and $T_{0,\nu\lambda''}$ into the magnon-polaron mode $\nu=4$ at the IF interface, with close-ups near the resonance frequencies in panels (d) and (e). Figure \ref{fig:interface}(c) shows $R_{0,\nu\nu'}$, $T_{0,\nu\lambda''}$ at the FN interface for outgoing modes $\nu=1$ and $\nu=4$, as well as the probability $P_{0,\nu{\rm N}}$ that the magnon polaron mode $\nu$ was excited by a spinful excitation of the conduction electrons in N. The incident modes in panels (b)--(e) are $\nu'=1$ and $4$ (corresponding to longitudinal-phonon-like and magnon-like magnon polaron modes in F) and $\lambda''=1$ (longitudinal phonon mode in I or N); Reflection and transmission coefficients $R_{0,\nu\nu'}$ and $T_{0,\nu\lambda''}$ with $\nu=1$ or $4$ and $\nu'$ and $\lambda''$ equal to $2$ or $3$ (which corresponds to transverse-phonon-like incident modes in F and I or N, respectively) are approximately zero (not shown). In contrast to the IF interface, where the reflection and transmission probabilities obey a unitary condition, see Fig.\ \ref{fig:interface}(b), the spin pumping into the normal metal at the FN interface at $x=L$ significantly influences the reflection of magnon-like magnon polarons, see Fig.\ \ref{fig:interface}(c). Figure \ref{fig:interface}(f) shows the angle-dependence of the transmission coefficient $T_{11}$, as a function of the polar angle $\theta$ and the azimuthal angle $\phi$ of the wavevector $\vq$ of the outgoing mode. A significant angle dependence exists only in the immediate vicinity of the crossing points of the magnon and phonon dispersions. 

Significant mixing between different magnon-polaron modes takes place only for frequencies near the crossings of the magnon and phonon dispersions at the resonant frequencies $\omega/2\pi \approx 0.26\,{\rm THz}$ and $\omega/2\pi \approx 0.76\,{\rm THz}$ \cite{Kamra-2014,Kamra-2015}. The frequency range for which the interfaces mix different modes is strongly enhanced near the critical magnetic fields at which the magnon and phonon dispersions touch. This is illustrated in Figs.\ \ref{fig:interface}(g) and (h), which show reflection and transmission coefficients into mode $\nu=4$ at the two critical magnetic fields $B \approx 9.2\,{\rm T}$ and $B \approx 2.6\,{\rm T}$. The magnon and phonon dispersions are shown schematically in the insets. For the critical field $B \approx 2.6\,{\rm T}$, at which the magnon dispersion touches that of the transverse phonons mode, the interface mixes magnon-like, longitudinal-phonon-like and transverse-phonon-like modes at the resonance frequency.

{\em Spin current and spin Seebeck voltage.---}
The spin current $j^{\rm s}_{x}$ into the normal metal is calculated as the difference between the current of magnon polarons that excite a spin-$1$ excitation of the conduction electrons in the normal metal 
and the backflow of magnon polarons into the insulator that were excited by an incident spinful excitation of the conduction electrons (which are assumed to be at the same temperature as the phonons in N),
\begin{equation}
  j^{\rm s}_{x} = \frac{\hbar}{V} \sum_{\vq_{+},\nu} v_{\vq,\nu x} (n_{\vq,\nu} - n_{\rm N}(\omega)) P_{\nu{\rm N}}(\omega,\vq_{\perp}).   \label{eq:jmag}
\end{equation}
Substituting the linear-response form (\ref{eq:linearresponse}) of the distribution function and the simplified angular dependence, the spin current can be expressed in terms of the isotropic and anisotropic moments $\psi_{0,\nu}(\omega)$ and $\psi_{1,\nu}(\omega)$ of the distribution function,
\begin{align}
  j^{\rm s}_x = \int d\omega j^{\rm s}_{x}(\omega)
\end{align}
with
\begin{align}
  j^{\rm s}_{x}(\omega) = \hbar \omega \left( - \frac{\partial n^0}{\partial \omega} \right) \sum_{\nu} \sum_{n=0}^{1} {\cal P}_{n,\nu}(\omega) \psi_{n,\nu}(\omega)
  \label{eq:jsomega}
\end{align}
where
\begin{align}
  {\cal P}_{n,\nu}(\omega) = \int \frac{d\vq_{\perp}}{(2\pi)^3} P_{{\rm N}\nu}(\omega,\vq_{\perp})
  |v_{\nu,x}(\omega,\vq_{\perp},+)|^n .
\end{align}
The spin current at the interface is related to the spin Seebeck voltage in the normal metal via the inverse spin Hall effect \cite{Xiao-2010} and equals, averaged over the width of the normal metal \cite{Schmidt-2018},
\begin{align}
  V_{\rm sse} = \theta_{\rm sh} \varrho \frac{2e}{\hbar} \lambda_{\rm sf} \frac{w}{l} \tanh\left(\frac{l}{2\lambda_{\rm sf}}\right) \,j^{\rm s}_x(L). \label{eq:vsse}
\end{align}
Here $w$, $l$, $\varrho$, $\theta_{\rm sh}$, and $\lambda_{\rm sf}$ are the width, length, electrical resistivity, the spin Hall angle, and the spin-flip diffusion length of the metal contact. Numerical values for these parameters of Pt can be found in Table \ref{tab:values}.

\section{Results}
\label{sec:results}

We use the theoretical formalism outlined in the previous two Sections to describe the steady-state longitudinal spin Seebeck effect of ferromagnetic insulator--normal metal structures. We consider the geometry of Fig.\ \ref{fig:setup} and solve the linearized Boltzmann equation with the boundary conditions
\begin{equation}
  \psi_{\rm I} = \Delta T / T ,\ \
  \psi_{\rm N} = 0,
  \label{eq:INboundaryconditions}
\end{equation}
corresponding to a temperature difference $\Delta T$ applied across the magnetic insulator F. Numerical values for the material parameters are taken from Table \ref{tab:values}.

The long-wavelength approximation is used throughout. We found that for the range of temperatures $T \lesssim 30$ K considered by us all frequency integrals converge for $\omega \sim 6\, c_{\rm l} q^0_{\rm l}$, where $c q^0_{\rm l}$ is the frequency at which the magnon and longitudinal phonon dispersions cross at zero magnetic field, see Eq.\ (\ref{eq:crossing}). Comparing with realistic parameters for YIG, this corresponds to a convergence for wavenumbers $q \lesssim 1/a$, which {\em a posteriori} justifies the use of the long-wavelength approximation.

To evaluate the reflection and transmission probabilities at the IF and FN interfaces and the angular averages entering in the transition rates for the isotropic and anisotropic moments of the distribution function $\psi_{0,\nu}(\omega)$ and $\psi_{1,\nu}(\omega)$ we use a grid of $30 \times 30$ reference points for the polar angles $(\phi,\theta)$ parameterizing the direction of $\vq$. If necessary, reflection and transmission coefficients are interpolated linearly between grid points.

We first describe our results for the case of elastic impurity scattering only. We then discuss how these results are changed upon inclusion of inelastic scattering. 

\subsection{Elastic scattering only}
\label{sec:results_elastic}

{\em Distribution functions.--- }
After restriction of the distribution function to the simplified angular dependence with one isotropic moment $\psi_{0,\nu}(\omega)$ and one anisotropic moment $\psi_{1,\nu}(\omega)$, the steady-state Boltzmann equation reduces to a set of $2 \times 4$ coupled linear equations for these moments, see Eq.\ (\ref{eq:boltzmannElastic}). For impurity scattering, different frequencies decouple, which makes an efficient solution of the equations possible. For convenience of notation, we here introduce a matrix notation in which $\psi_{0,\nu} \to \vpsi_0$ and $\psi_{1,\nu} \to \vpsi_1$ are four-component column vectors. Then the linearized Boltzmann equation (\ref{eq:boltzmannElastic}) takes the simple form
\begin{equation}
\frac{\partial \vpsi_1}{\partial x} = - \cG_{0} \vpsi_0 , \ \ \ \frac{\partial \vpsi_0}{\partial x} = - \cG_{1} \vpsi_1 \label{eq:boltzmannVectorized},
\end{equation}
where the $4 \times 4$ matrices $\cG_{0}$ and $\cG_{1}$ were defined in Eq.\ (\ref{eq:lambdaAverage}) (a delta function for frequencies is factored out and we switch notation from superscripts $0$ and $1$ to subscripts). The condition (\ref{eq:lambdaCondition}) ensures that the uniform isotropic solution $\psi_{0,\nu} = \psi$ and $\psi_{1,\nu}= 0$, with $\psi$ an arbitrary function of $\omega$, is a solution of the equations.

\begin{figure}
  \includegraphics[width=\linewidth]{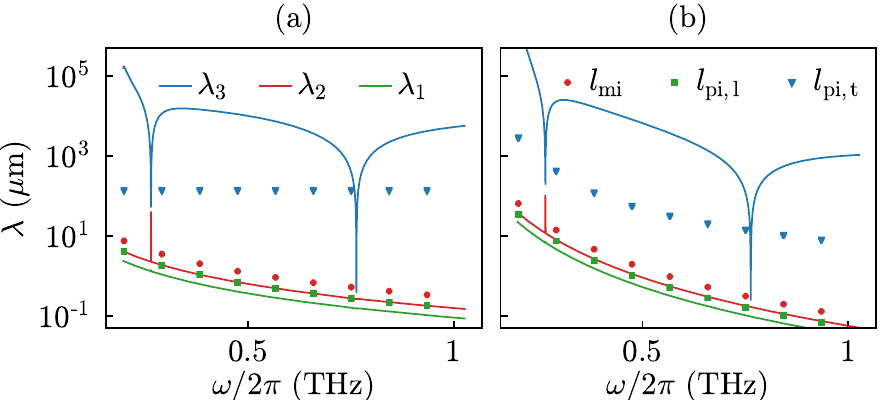}
  \caption{Relaxation lengths $\lambda_i(\omega)$, $i=1,2,3$, of Eq.\ (\ref{eq:lambdaSolution}) for the material parameters given in Table \ref{tab:values} and impurity scattering parameters from Table \ref{tab:impurity} (center column) for the white-noise impurity model (\ref{eq:whitenoise}) with the mean free paths (\ref{eq:mfp_whitenoise}) (a) and for the microscopic model (\ref{eq:magimp}), (\ref{eq:phoimp}) with mean free paths from Eqs.\ (\ref{eq:mfpm}) and (\ref{eq:mfpp}) (b). The dots indicate the magnon-impurity (blue) and phonon-impurity scattering lengths (green and red).
    \label{fig:lengthscales}}
\end{figure}

The matrix product $\cG_{1} \cG_{0}$, which describes the relaxation towards local equilibrium, is non-negative.
We diagonalize the matrix $\cG = \sqrt{\cG_{1} \cG_{0}}$ as
\begin{equation}
  \cG = \cU\, \mbox{diag}\, (0,\lambda_1^{-1},\lambda_2^{-1},\lambda_3^{-1})\, \cU^{-1}, \label{eq:lambdaSolution}
\end{equation}
with $\cU$ a $4 \times 4$ matrix, $\lambda_0 = 0$ and $0 < \lambda_1 < \lambda_2 < \lambda_3$. The three non-zero relaxation lengths $\lambda_i$, $i=1,2,3$, are shown in Fig.\ \ref{fig:lengthscales} for the parameter values of Table \ref{tab:values} and the impurity scattering parameters of Table \ref{tab:impurity}, center column. We also show the relaxation lengths $l_{\rm mi}$ and $l_{\rm pi,\lambda}$ for the white-noise impurity model of Eqs.\ (\ref{eq:whitenoise}), for which the relaxation lengths have a weaker systematic dependence on the frequency $\omega$. 

The longest relaxation length $\lambda_3$, which is the length scale for full equilibration of all four magnon-polaron modes, is the impurity-mediated inter-mode scattering length for magnon polarons,
\begin{equation}
  \limp(\omega) = \lambda_3(\omega).
\end{equation}
For generic frequencies, magnon-phonon hybridization is weak, so that impurity scattering has a very small inter-mode component. This explains why $\limp$ is much larger than the mean free paths $l_{\rm mi}$ and $l_{\rm pi}$ for magnon-impurity and phonon-impurity scattering, see the dotted lines in Fig.\ \ref{fig:lengthscales}.
For the parameter values of Table \ref{tab:impurity}, for which phonon-impurity scattering is stronger than magnon-impurity scattering, the two shorter relaxation lengths $\lambda_{1,2} \approx l_{\rm pi}$ describe the equilibration of the three phonon modes among each other. The frequency dependence of the relaxation lengths $\lambda_{1,2}$ reflects the frequency dependence of the phonon-impurity scattering.

To obtain a formal solution of the coupled equations (\ref{eq:boltzmannVectorized}), we use $\vu_{i}$ to denote the $i$-th column of $\cU$. Condition (\ref{eq:lambdaCondition}) implies that we may set the first column $\vu_{0} = (1,1,1,1)^{\rm T}$. The general solution of the isotropic distribution moment then reads
\begin{equation}
  \vpsi_{0}(x) = \sum_{i=0}^{3} \sum_{j=0}^{1} \beta_{ij}(x) c_{ij} \vu_{i} ,
\label{eq:gensol}
\end{equation}
where $c_{ij}$ are constants that are determined by the boundary conditions, and
\begin{align}
  \beta_{i0}(x) =&\, \left\{ \begin{array}{ll}
  x/L, & i=0, \\ e^{(x-L)/\lambda_i}, & i=1,2,3, \end{array} \right. \nonumber \\
  \beta_{i1}(x) =&\, \left\{ \begin{array}{ll}
  1-x/L, & i=0, \\ e^{- x/\lambda_i}, & i=1,2,3. \end{array} \right. 
  \label{eq:beta}
\end{align}
The anisotropic moment is obtained via Eq.\ (\ref{eq:boltzmannVectorized}) by taking the matrix inverse,
\begin{equation}
	\vpsi_1 = - \cG_1^{-1} \frac{\partial \vpsi_0}{\partial x} . \label{eq:gensol1}
\end{equation}
In matrix notation, the boundary conditions (\ref{eq:bcL}) and (\ref{eq:bcR}) at the IF interface at $x=0$ and the FN interface at $x=L$ read
\begin{align}
	 \vpsi_1(0) =&\, \cS_{\rm I} [\vpsi_{\rm I} - \vpsi_0(0)] , \nonumber\\
	 \vpsi_1(L) =&\, \cS_{\rm N} [\vpsi_0(L) - \vpsi_{\rm N}] ,
  \label{eq:bc1}
\end{align}
where $\vpsi_{\rm I,N} = \psi_{\rm I,N} \vu_{0}$ are four-component vectors describing the equilibrium distributions in the reservoirs I and N, see Eq.\ (\ref{eq:INboundaryconditions}). The matrices $\cS_{\rm I,N}$ are defined as
\begin{align}
  \cS_{\rm I,N} =&\ \frac{\cE_{1}}{\cE_{2}} (\openone_4+\cR_{1 \rm I,N})^{-1} (\openone_4-\cR_{0 \rm I,N}), 
\end{align}
where the angle-averaged reflection coefficient matrices $\cR_{0,\nu \nu'}$ and $\cR_{1,\nu \nu'}$ are defined in Eq.\ (\ref{eq:Rdef}).

These equations have a formal solution near the FN interface at $x=L$ if $L \gg \limp$. The vicinity of the FN interface is what is relevant for the calculation of the spin Seebeck effect, see Eq.\ (\ref{eq:jsomega}). Using the vector notation, one has
\begin{align}
  \label{eq:iso_thick}
  \vpsi_0(x) =&\, \frac{(L-x) \Delta T}{L T} \vu_{0} +
  e^{(x-L) \cG} \vpsi_0(L), \\ \nonumber 
  \vpsi_1(x) =&\, \frac{\Delta T}{L T} \cG_1^{-1} \vu_{0}
  - \cG_1^{-1} \cG e^{(x-L) \cG} \vpsi_0(L),
\end{align}
where
\begin{equation}
  \vpsi_0(L) = \frac{\Delta T}{L T}
  (\cG + \cG_1 R_{\rm N})^{-1} \vu_{0}.
\end{equation}
A solution near the IF interface at $x=0$ can be obtained in a similar way. If the condition $L \gg \limp$ is not satisfied, the boundary conditions at the two interfaces at $x=0$ and $x=L$ have to be implemented simultaneously, which is easily carried out numerically.

\begin{figure*}
\includegraphics[width=\linewidth]{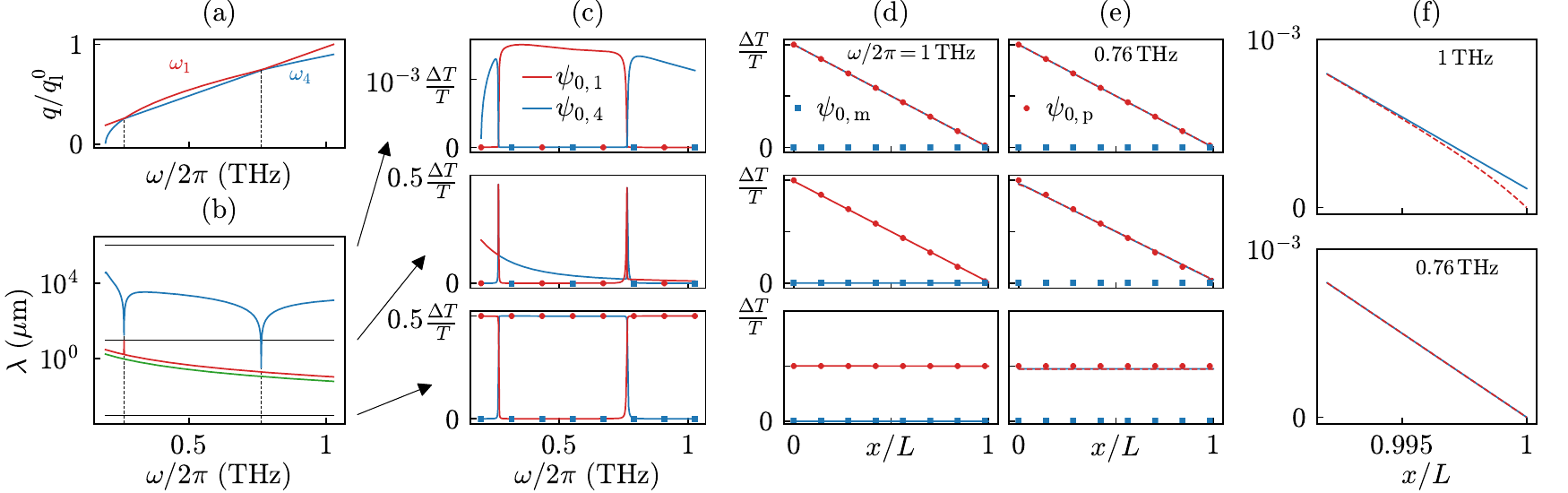}
\caption{\label{fig:isomoment} Isotropic moment for impurity scattering only. We take material parameters from Table \ref{tab:values} and set the magnetic field equal to $B=7\,{\rm T}$, so that the magnon dispersion crosses that of the longitudinal phonons, but not the transverse phonons, see panel (a). To keep the presentation of the frequency dependence of the isotropic moment (\ref{eq:psiansatz}) as simple as possible we use white-noise impurity scattering rates (\ref{eq:whitenoise}) with potentials according to Table \ref{tab:impurity}, center column. Panels (c)--(e) show the isotropic moment $\psi_0(\omega)$ of the distribution function for three choices of the system length $L$ as compared to the relaxation lengths $\lambda_i$, as schematically indicated in (b). Panel (c) shows the frequency dependence of the isotropic moment $\psi_0(\omega)$ at the FN interface at $x=L$. Panels (d) and (e) show the dependence on position $x$ for a generic frequency ($\omega/2\pi = 1\,{\rm THz}$, panel (d)) and a frequency at the crossing point of the magnon and phonon dispersions ($\omega/2\pi = 0.76\,{\rm THz}$, panel (e)). Panel (f) shows a close-up of the spatial dependences near the FN interface for the longest system length considered.}
\end{figure*}

In Fig.\ \ref{fig:isomoment} we show the frequency dependence of the isotropic moment $\psi_{0,\nu}(\omega)$ at the FN interface, {\em i.e.}, for $x=L$, as well as the spatial dependence of $\psi_{0,\nu}(\omega)$ for a generic frequency away from the crossing points of longitudinal phonon and magnon dispersions and at a crossing point. To keep the discussion simple, we have chosen the magnetic field $B = 7$ T. At this value of the magnetic field the dispersions of magnons and longitudinal phonons cross, but not of magnons and transverse phonons, see Fig.\ \ref{fig:dispersion}, so that there are only two ``resonance frequencies'' $\omega/2\pi \approx 0.26\,{\rm THz}$ and $\omega/2\pi \approx 0.76\,{\rm THz}$. To avoid artifacts from the strong frequency dependence of the impurity scattering lengths in the microscopic model, we use the white-noise model (\ref{eq:whitenoise}) for the impurity potential. We consider three system sizes $L$: a length $L$ much larger than the impurity-mediated inter-mode scattering length $\limp$ (top panels), an intermediate length $L$ smaller than $\limp$, but still much larger than the individual elastic magnon and phonon mean free paths $l_{\rm mi}$ and $l_{\rm  pi}$ (center panels), and a very short system size $L \ll \lambda_1 \approx l_{\rm pi}$, in which the only source of equilibration between magnon-like and phonon-like magnon-polaron modes is at the interfaces (bottom panels). In the intermediate regime the three phonon modes equilibrate among each other, but magnon-like and phonon-like modes remain out of equilibrium, except for the immediate vicinity of the resonance frequencies. The three system sizes considered are indicated by horizontal lines in Fig.\ \ref{fig:isomoment}(b) (reproduced from Fig.\ \ref{fig:lengthscales}). Taking material parameters from Tables \ref{tab:values} and the center column of \ref{tab:impurity}, only the intermediate range of length scales $l_{\rm mi}$, $l_{\rm pi} \lesssim L \lesssim \limp$ is of experimental relevance. 

We first discuss the isotropic moment $\psi_{0,\nu}(\omega)$ for the case without magnon-phonon coupling, shown by the dots in Fig.\ \ref{fig:isomoment}(c)--(e). Without magnon-phonon coupling, the isotropic moment $\psi_{0,{\rm m}}(\omega)$ of the magnon distribution is at the equilibrium value corresponding to the temperature of the conduction electrons in N. This is the case, because the magnons are coupled to the conduction electrons in N via the spin mixing conductance at the FN interface, see Sec.\ \ref{sec:reflection}. In the top and center panels of Fig.\ \ref{fig:isomoment} (c)--(e), which describe system lengths $L$ large in comparison to the phonon-impurity mean free path $l_{\rm pi}$, the distribution function $\psi_{0,{\rm p}}(\omega)$ of the phonon linearly interpolates between its values at the warm and cold reservoirs I and N and $x=0$ and $x=L$. In the bottom panels, which are for a length $L \ll l_{\rm pi}$, the isotropic moment $\psi_{0,{\rm p}}(\omega)$ of phonon distribution is at a characteristic temperature $T + \Delta T/2$ precisely between the temperatures of the hot and cold reservoirs. A full discussion of the case of zero magnon-phonon coupling using our formalism can be found in App.\ \ref{app:magpho}.

To interpret the curves in Fig.\ \ref{fig:isomoment}(c)--(e) that show the isotropic moment $\psi_{0,\nu}(\omega)$ in the presence of magnon-phonon coupling (solid curves), one should take into account that the label $\nu$ of the ``magnon-like'' magnon-polaron mode is ``4'' (blue curves) for frequencies $\omega/2\pi$ between the crossing points at $0.26$ THz and $0.76$ THz, and ``1'' otherwise, see Fig.\ \ref{fig:isomoment}(a). The character of the magnon-polaron mode changes at the crossing points of the dispersion.

For the largest system size $L \gg \limp$ (top panels in Fig.\ \ref{fig:isomoment}(c)--(e)), all magnon-polaron modes are in local equilibrium in the bulk of the sample, as shown in Fig.\ \ref{fig:isomoment}(d) and (e). At the FN interface, the magnon-like mode has a weakly elevated population, when compared to the temperature of the N reservoir, whereas the temperature of the phonon-like modes is very close to that of the N reservoir, as shown in detail in Fig.\ \ref{fig:isomoment}(f). This difference occurs, because magnon-like modes are mostly reflected at the FN interface, whereas the phonon-like modes are almost perfectly transmitted. At the resonance frequencies there is a sharp ``dip'' in the population of magnon-like magnon-polaron modes at the FN interface, reflecting the increased equilibration between different magnon-polaron modes at those frequencies. As a result, close to the resonance frequencies the thickness of the layer near the FN interface in which different magnon-polaron modes have different distribution functions is much thinner than for generic frequencies, see Fig.\ \ref{fig:isomoment}(f).

For the intermediate system size (center panels in Fig.\ \ref{fig:isomoment}(c)--(e)), the magnon-like magnon-polaron modes are no longer in equilibrium with the phonon modes for generic frequencies $\omega$ and, hence, have an occupation consistent with that of the cold reservoir N, to which they are coupled via the spin mixing conductance. For frequencies in the close vicinity of the crossing points, they still equilibrate with the phonons, and their occupation is essentially the same as in the case $L \gg \limp$ discussed above.

At the shortest system length $L \ll \lambda_1 \approx l_{\rm pi}$, the isotropic moments are constants as a function of position. From the boundary conditions (\ref{eq:bc1}) at the IF and FN interfaces one finds
\begin{equation}
\vpsi_0 = (\cS_{\rm I} + \cS_{\rm N})^{-1}  (\cS_{\rm I} \vpsi_{\rm I} + \cS_{\rm N}  \vpsi_{\rm N} ). \label{eq:iso_thin}
\end{equation}
At generic frequencies, magnon-like modes have an occupation corresponding to the temperature of the cold reservoir, whereas the isotropic moment of the distribution of phonon-like modes corresponds to a temperature $T + \Delta T/2$ precisely intermediate between the temperatures of cold and warm reservoirs. For frequencies close to the crossing points, reflection at the IF and FN interfaces still leads to an equilibration of magnons and phonons. Since the magnon-like modes are coupled to the cold reservoir, but not to the warm reservoir, the occupation corresponds to a temperature slightly below $\Delta T/2$, see Fig.\ \ref{fig:isomoment}(e). 

\begin{figure}
\includegraphics[width=\linewidth]{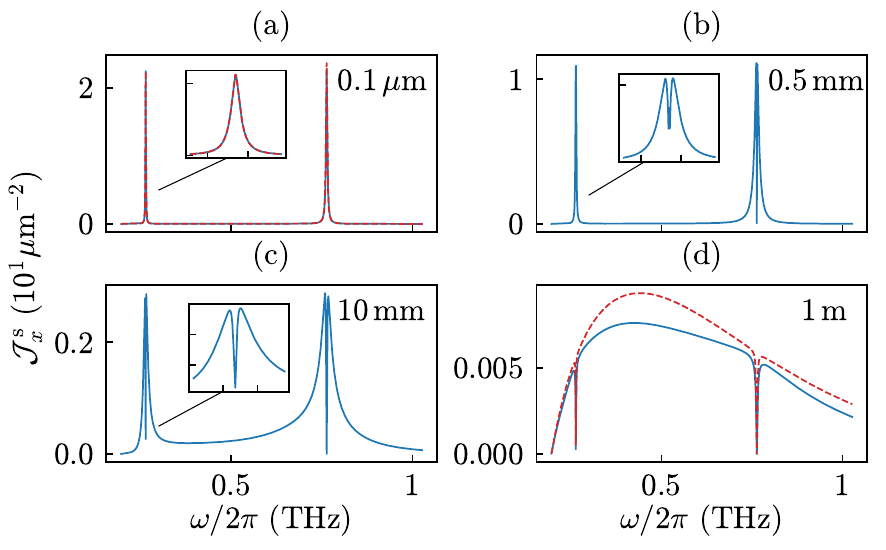}
\caption{\label{fig:spincurrent} Frequency-resolved spin current $j^{\rm s}_x(\omega) =\hbar {\cal J}^{\rm s}_x(\omega) \Delta T/T$ at the FN interface (blue, solid) for $B=7\,{\rm T}$ and $T = 10\,{\rm K}$ at sample length $L = 0.1\,\mu{\rm m}$ (a), $L = 0.5\,{\rm mm}$ (b), $L = 10\,{\rm mm}$ (c), and $L = 1\,{\rm m}$ (d). System parameters are taken from Table \ref{tab:values}. Impurity scattering of magnon and phonons is described by the white-noise model (\ref{eq:whitenoise}), with parameter values taken from the center column of Table \ref{tab:impurity}. For comparison, panels (a) and (d) also show the frequency-resolved spin current calculated from the short-length approximation of Eq.\ (\ref{eq:iso_thin}) and the long-length approximation (\ref{eq:iso_thick}), respectively (red, dashed).
}
\end{figure}

{\em Spin current.---} 
Results for the spin current are shown in Fig.\ \ref{fig:spincurrent}, using the white-noise model (\ref{eq:whitenoise}) with parameter values from the center column of Table \ref{tab:impurity} for the impurity scattering rates. The relaxation lengths $\lambda_i(\omega)$ for this set of parameters are shown in Fig.\ \ref{fig:lengthscales}(a). Figure \ref{fig:spincurrent} shows the frequency-resolved spin current $j^{\rm s}_x(\omega)$ for system lengths $L = 0.1\,\mu{\rm m}$, $L = 0.5\,{\rm mm}$, $L = 10\,{\rm mm}$, and $L = 1\,{\rm m}$. At the shortest length $L = 0.1\,\mu{\rm m}$, magnon and phonon dynamics are ballistic and no equilibration between magnon-like and phonon-like magnon-polaron modes takes place, except in the immediate vicinity of the resonance frequencies. Consequentially, the frequency-resolved spin current differs appreciably from zero only near these frequencies. At the intermediate sample length $L= 0.5\,{\rm mm}$ magnon and phonon transport is diffusive, whereas for generic frequencies the sample size is still larger than $\limp$. Here, the spin current is still carried at frequencies close to the crossing points, although the frequency-resolved spin current shows a narrow dip precisely at the crossing point because the strong mixing of magnons and phonons, combined with the large transparency of the FN interface for phonons, suppresses an accumulation of magnon polarons at the FN interface at that frequency. To illustrate what happens upon further increasing the system length $L$, Fig.\ \ref{fig:spincurrent} also shows the case $L = 10\,{\rm mm}$ and the unrealistically large length $L = 1\,{\rm m}$, at which all frequencies contribute to the spin current, except for narrow dips at the frequencies at which magnon and phonon dispersions cross. Figure \ref{fig:spincurrent} also contains a comparison with the asymptotic large-$L$ result (\ref{eq:iso_thick}), showing that the comparison becomes quantitatively accurate only for lengths $L$ far outside the experimentally relevant range.

Magnon and phonon dynamics are ballistic for the shortest system length considered in Fig.\ \ref{fig:spincurrent}, $L \ll \lambda_1 \approx l_{\rm pi}$. With ballistic dynamics, the validity of the parameterization (\ref{eq:psimoments}) breaks down. To estimate the error, in App.\ \ref{app:ballistic} we compare the spin current obtained using the parameterization (\ref{eq:psimoments}) with an exact calculation for the ballistic short-$L$ limit.

\begin{figure}
\includegraphics[width=\linewidth]{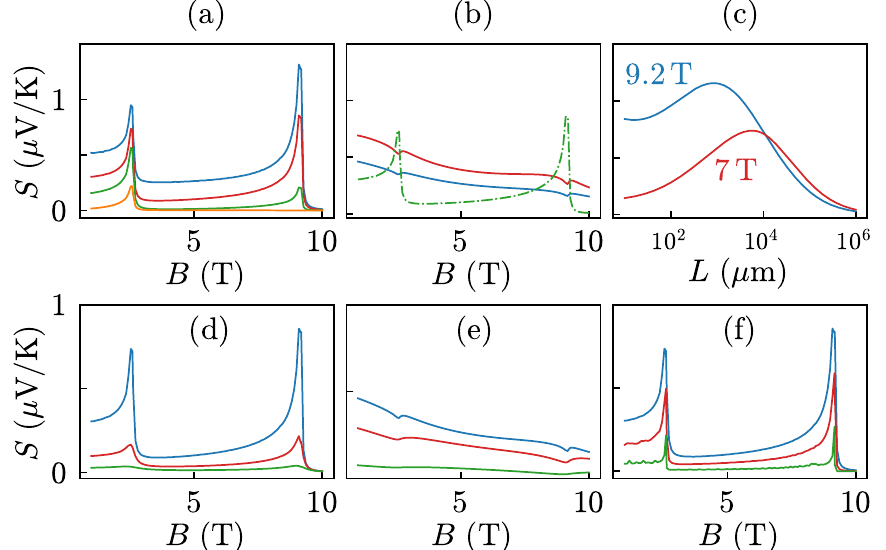}
\caption{\label{fig:elastic} Spin Seebeck coefficient for elastic scattering only. Material parameters are taken from Table \ref{tab:values} and impurity scattering of magnon and phonons is described by the microscopic model of Eqs.\ (\ref{eq:magimp}) and (\ref{eq:phoimp}) with parameter values taken from the center column of Table \ref{tab:impurity}, unless noted otherwise. Panel (a) shows the spin Seebeck coefficient at $L=10\,\mu{\rm m}$ for temperatures $T=2\,{\rm K}$, $5\,{\rm K}$, $10\,{\rm K}$, and $20\,{\rm K}$ (bottom to top). Panel (b) shows the spin Seebeck coefficient for $T=10\,{\rm K}$ at different system lengths $L=10\,{\mu \rm m}$ (green), $5\,{\rm cm}$ (red), and $10\,{\rm cm}$ (blue). Panel (c) shows $S$ for $B=7\,{\rm T}$ and $B=9.2\,{\rm T}$ as a function of the system length $L$. Panels (d) and (e), which are evaluated at $T = 10\, {\rm K}$ and  at $L = 10\,\mu{\rm m}$ and $L = 1\,{\rm cm}$, respectively, show the spin Seebeck coefficient $S$ vs.\ magnetic field $B$ for spin mixing conductivity $\sigma_{\uparrow\downarrow}$ equal to $1$, $10^{-1}$, and $10^{-2}$ times the value listed in Table \ref{tab:values}, from top to bottom, respectively. Panel (f) shows the spin Seebeck coefficient for $L = 10\,\mu{\rm m}$, $T = 10\,{\rm K}$, and magnon-phonon coupling parameters $D$ and $D'$ a factor $1$, $0.5$, and $0.1$ times the values listed in Table \ref{tab:values}, from top to bottom.}
\end{figure}

{\em Spin Seebeck coefficient.---}
The spin Seebeck voltage $V_{\rm SSE}$ and the spin Seebeck coefficient
\begin{equation}
  S = \frac{V_{\rm sse}}{\Delta T}
\end{equation}
can be obtained from Eq.\ (\ref{eq:vsse}). Figure \ref{fig:elastic} shows the spin Seebeck coefficient $S$ as a function of the applied magnetic field $B$. We use the microscopic model of Eqs.\ (\ref{eq:magimp}) and (\ref{eq:phoimp}), with parameter values taken from Table \ref{tab:impurity} (center column), to describe magnon-impurity and phonon-impurity scattering. The spin Seebeck coefficient shows sharp features near the critical magnetic fields at which the magnon and phonon dispersions touch, reflecting the drastic enhancement of the frequency window of strong magnon-phonon coupling at those magnetic fields. For short and intermediate lengths $L \lesssim \limp$, the magnitude of the spin Seebeck effect, including the singular features at the critical magnetic field, increases with the temperature, see Fig.\ \ref{fig:elastic}(a). The spin Seebeck coefficient and the singular features at the critical magnetic fields depend non-monotonically on the length $L$ or the impurity concentration, as shown in Fig.\ \ref{fig:elastic}(b). Whereas $S$ shows peaks at the critical magnetic fields if the system size is $\lesssim \limp$, the spin Seebeck coefficient exhibits a dip at the critical magnetic fields if $L \gtrsim \limp$. This dip originates from the narrow dip in the frequency-resolved spin current for frequencies close to the resonance frequencies, see Fig.\ \ref{fig:spincurrent}, and reflects the increase of the resonant frequency range at the critical magnetic fields. The non-monotonous length dependence of the spin Seebeck coefficient is also shown in panel (c). For short lengths, the spin Seebeck coefficient is small, because magnons are excited only in a very narrow frequency window around the resonance frequencies. Upon increasing $L$, the width of this frequency window is increased, which is what leads to an increase of $S$ with $L$. The spin Seebeck coefficient reaches a maximum as a function of $L$, when all frequencies contribute to the spin current. A further increasing of $L$ leads to a suppression of $S$ because of the increase of the thermal resistance with $L$. The maximum occurs at smaller lengths (but higher spin Seebeck coefficient) for magnetic fields close to the critical field, because there the widths of the peaks of the frequency-resolved spin current at the resonant frequency is largest.

Figure \ref{fig:elastic} also addresses the dependence of the spin Seebeck coefficient on the spin mixing conductivity $\sigma_{\uparrow\downarrow}$ and on the strength of the magnon-phonon coupling. A decrease of the spin mixing conductivity --- a lower transparency of the FN interface --- leads to a smaller spin Seebeck voltage and less pronounced features as a function of magnetic field, see Fig.\ \ref{fig:elastic}(d) and (e) for a system short and long in comparison to $\limp$, respectively. A decrease of the magnon-phonon coupling below the value listed in Table \ref{tab:values} leads to a smaller spin Seebeck coefficient only if the system is short enough, see Fig.\ \ref{fig:elastic}(f). If $L \gg \limp$, a decrease of the magnon-phonon coupling has no significant effect on the spin Seebeck coefficient (data not shown).

\begin{figure*}
\includegraphics[width=\linewidth]{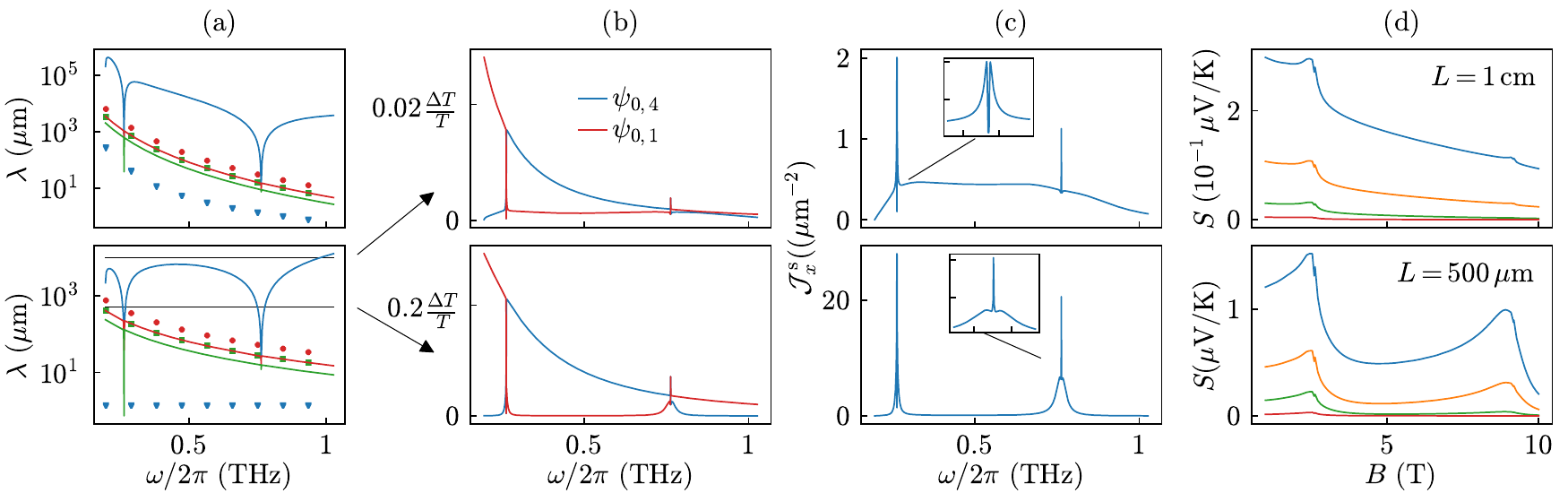}
\caption{\label{fig:elastic2} Relaxation lengths (a), isotropic distribution function $\psi_0(\omega)$ at the FN interface (b), frequency-resolved spin current $j^{\rm s}_x(\omega) = \hbar {\cal J}^{\rm s}_x(\omega) \Delta T/T$ at $T = 10\,{\rm K}$ (c), and spin Seebeck coefficient $S$ vs.\ magnetic field $B$ (d) for a system in which magnon-impurity scattering is stronger than phonon-impurity scattering. Material parameters are taken from Table \ref{tab:values}. Impurity scattering is described by the microscopic model of Eqs.\ (\ref{eq:magimp}) and (\ref{eq:phoimp}) (a, top panel, and d) and the white-noise model of Eq.\ (\ref{eq:whitenoise}) (a, bottom panel, and b, c) with parameters taken from the right column in Table \ref{tab:impurity}. Dots in panel (a) indicate the magnon-impurity mean free path (blue) and the phonon-impurity mean free paths (green and red) in the absence of magnon-phonon coupling. The system length $L$ for panels (b)--(d) is indicated by the horizontal lines in panel (a). The spin Seebeck coefficient in panel (d) is evaluated for temperatures $T=2\,{\rm K}$, $5\,{\rm K}$, $10\,{\rm K}$, and $20\,{\rm K}$ (bottom to top).}
\end{figure*}

{\em Relative strength of magnon-impurity and phonon-impurity scattering.---} The numerical values used for Figs.\ \ref{fig:isomoment}--\ref{fig:elastic} have a magnon-impurity scattering length that is a factor $\sim 10$--$100$ larger than the phonon-impurity length. For comparison, Fig.\ \ref{fig:elastic2} shows the relaxation lengths $\lambda_i$, the distribution function at the FN interface, the frequency-resolved spin current, and the spin Seebeck coefficient for a system that is of higher acoustic quality than of magnetic quality. The parameters for the impurity model are such that the orders of magnitude of the corresponding scattering rates are interchanged (see Table \ref{tab:impurity}, right column). This interchange leads to a significant increase in the two smallest relaxation lengths $\lambda_{1,2} \approx l_{\rm pi}$, which describe the relaxation between the three phonon-like polaron modes, but hardly affects the longest relaxation length $\lambda_3 = \limp$, which describes the impurity-mediated inter-mode scattering of magnon polarons. For short system lengths $L \lesssim \limp$ (but $L$ larger than the phonon and magnon mean free paths $l_{\rm pi}$ and $l_{\rm mi}$), the distribution function at the FN interface has the same singular frequency dependence near the resonant frequencies as in the phonon-impurity-dominated case discussed above. As a result, the frequency-resolved spin current $j^{\rm s}_x(\omega)$ is strongly peaked near the resonance frequencies, and the spin Seebeck coefficient $S$ has a peak at the ``critical'' values of the magnetic field for which the magnon and phonon dispersions are tangential to each other. For large lengths $L \gtrsim \limp$, the frequency-resolved spin current shows sharp peaks at the resonant frequencies (as opposed to sharp dips in the case of stronger phonon-impurity scattering). Correspondingly, the spin Seebeck coefficient continues to exhibit peaks as a function of $B$ for the critical magnetic fields in the limit of large sample length $L$. 

\subsection{With inelastic scattering}
\label{sec:results_inelastic}

For temperatures $T \lesssim 30\,{\rm K}$, inelastic relaxation lengths remain well above the mean free paths $l_{\rm mi}$ and $l_{\rm pi}$ for magnon-impurity and phonon-impurity scattering for all frequencies of interest. In the vicinity of the resonance frequencies, the impurity-mediated inter-mode scattering length $\limp$ is comparable to the mean free path for impurity scattering. Hence, near the resonance frequencies impurity-mediated inter-mode scattering remains the dominant scattering mechanism coupling phonon and magnon degrees of freedom. For generic frequencies, however, depending on the temperature and the magnetic field, $\limp$ may well be larger than the inelastic relaxation length. If this is the case, the inclusion of inelastic scattering may lead to a modification of the elastic-scattering-only results that were derived in the previous Subsection, as we now discuss.

To describe inelastic processes in the Boltzmann equation (\ref{eq:boltzmannElastic}), the collision integrals (\ref{eq:collisions_3in}) and (\ref{eq:collisions_4in}) are included in the transition matrices $\cG_0(\omega,\omega')$ and $\cG_1(\omega,\omega')$. Whereas these transition matrices were proportional to $\delta(\omega-\omega')$ for the elastic-scattering-only case, they acquire contributions off-diagonal in frequency if inelastic processes are included. Apart from this, the general structure of the Boltzmann equations for the isotropic and anisotropic moments (\ref{eq:boltzmannElastic}) stays untouched.

The transition matrix $\cG_0(\omega,\omega')$ governs the relaxation of the isotropic moment $\psi_0(\omega)$. Without inelastic scattering, $\cG_0(\omega,\omega') = \cG_0(\omega) \delta(\omega-\omega')$. As discussed in the previous Subsection, the $4 \times 4$ matrix $\cG_0$ has three nonzero eigenvalues and one zero eigenvalue. The smallest nonzero eigenvalue is typically several orders of magnitude smaller than the two larger eigenvalues, because it contains the magnon-phonon coupling. (Without magnon-phonon coupling, $\cG_0$ has two zero eigenvalues, see App.\ \ref{app:magpho}.) Inclusion of inelastic processes is a significant perturbation to $\cG_0$, because their strength has to be compared to the two smallest eigenvalues of $\cG_0$. In contrast, for the elastic-scattering-only case the transition matrix $\cG_1(\omega,\omega')=\cG_1(\omega)\delta(\omega-\omega')$ has four nonzero eigenvalues at each frequency, of the order of the rates $\tau_{\rm pi}$ and $\tau_{\rm mi}$ for phonon-impurity and magnon-impurity scattering length, see App.\ \ref{app:magpho}. For $T \lesssim 30\,{\rm K}$, these are much larger than the inverse inelastic scattering lengths, so that inelastic processes do not significantly affect $\cG_1$.
Motivated by these observations, we calculate $\cG_1(\omega,\omega')$ without inclusion of inelastic processes, whereas to simplify the calculation of $\cG_0(\omega,\omega')$ we replace the microscopic models for the various sources of inelastic scattering, see Sec.\ \ref{sec:scattering} and App.\ \ref{app:inel}, by an effective model in which the matrix elements depend on the frequencies of the magnon and phonon modes involved, but not on the polarization and (directions of the) wavevectors.
The magnitude of the matrix elements in the effective models is chosen such that the angle average is the same as in the original microscopic model, so that the inelastic contribution to $\cG_0(\omega,\omega')$ is still exact without the elastic magnon-phonon coupling. We keep the full wavevector dependence of the transformation matrix $V_{\vq}$ between magnon and phonon modes and magnon-polaron modes.

Once inelastic processes are included, the Boltzmann equation for the isotropic and anisotropic moments (\ref{eq:boltzmannElastic}) couples distribution functions at different frequencies. To solve these resulting integro-differential equations, we choose a frequency grid with $(N+1)$ frequencies $\omega_n$ with $0 \equiv \omega_0 < \omega_1 < \dots < \omega_N$ and approximate the distribution functions $\psi_{0,\nu}(\omega_n)$ and $\psi_{1,\nu}(\omega_n)$ by linear interpolation between the grid frequencies. The density of reference frequencies is chosen to be high in regions in which $\psi_{0,\nu}(\omega_n)$ and $\psi_{1,\nu}(\omega_n)$ are strongly frequency dependent, such as in the vicinity of the crossing points of magnon and phonon branches. Frequency integrals over a function $f(\omega)$ can be discretized by a trapezoidal rule,
\begin{equation}
\int d\omega f(\omega) \to \sum_{n = 1}^{N} [f(\omega_n) + f(\omega_{n-1})] \frac{\omega_n - \omega_{n-1}}{2} . \label{eq:trapez}
\end{equation}
The resulting discretized equations then have the same structure as the equations for elastic scattering (\ref{eq:boltzmannVectorized}) --- but with matrices $\cG_0$ and $\cG_1$ of dimension $4(N+1)$ instead of $4$ --- and they can be efficiently solved with the same methods as discussed in the previous Subsection. The boundary conditions at the IF and FN interfaces do not mix frequencies and can be implemented in the same way as for the elastic-scattering-only case, see Eq.\ (\ref{eq:bc1}).

\begin{figure}
\includegraphics[width=\linewidth]{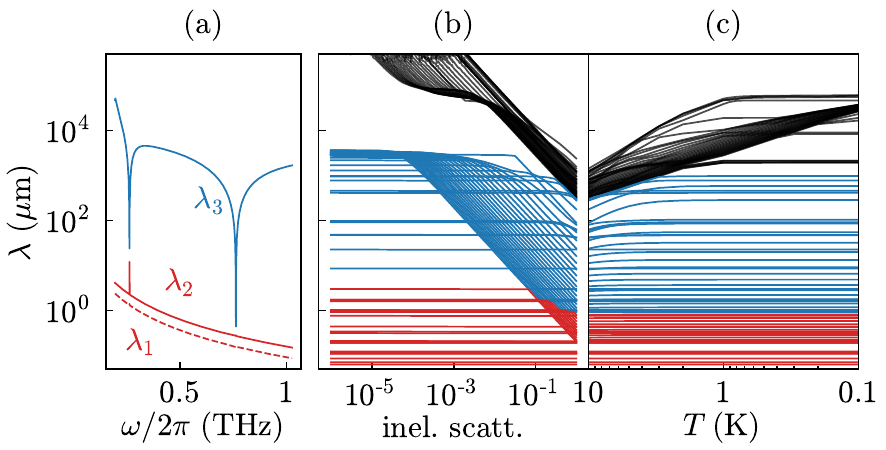}
\caption{\label{fig:lambda_inelastic} (a) Frequency-dependent relaxation lengths $\lambda_i(\omega)$, $i=1,2,3$, for the case of elastic scattering only. (b) and (c) Relaxation lengths $\lambda_j$, $j=1,\ldots,4(N+1)-1$ in the presence of inelastic scattering, calculated for a frequency grid with $N = 200$ frequencies. For clarity, only every $20$th relaxation length is shown. In (b) the relaxation lengths are shown vs.\ the strength of the inelastic scattering processes, normalized to the inelastic processes at $10\,{\rm K}$. In (c) the relaxation lengths are shown vs.\ temperature. The black curves represent those relaxation lengths that are divergent in absence of inelastic scattering.}
\end{figure}

The eigenvalues of the matrix $\cG^2=\cG_1 \cG_0$, which are the inverse square relaxation lengths $\lambda_j^{-2}$, compare with Eq.\ (\ref{eq:lambdaSolution}), again play an important role in the construction of the solution of the Boltzmann equation, which is given by Eqs.\ (\ref{eq:gensol})--(\ref{eq:gensol1}), with the summations over $i$ and $j$ running from $0$ to $4(N+1)-1$ instead of $0$ to $3$. In contrast to elastic impurity scattering, which relaxes only differences between polaron branches $\nu$ at the same frequency, the matrix condition (\ref{eq:lambdaCondition}) for inelastic three and four-polaron interaction ensures that only the globally uniform solution $\psi_{0,\nu}(\omega) = \psi$, $\psi_{1,\nu}(\omega) = 0$ is an equilibrium solution. As a consequence the matrix $\cG$ has only one zero eigenvalue. The nonzero eigenvalues give the inverse relaxation lengths. As an illustration, Fig.\ \ref{fig:lambda_inelastic} shows the relaxation lengths $\lambda_j$, $j=1,\dots,4(N+1)-1$ for the case of material parameters listed in Table \ref{tab:values} as a function of the inelastic scattering strength. Panel (b) shows how the relaxation lengths at temperature $10\,{\rm K}$ are connected with the frequency-dependent equilibration lengths for the case without inelastic processes if one smoothly ``switches on'' the inelastic processes; panel (c) shows the relaxation length as a function of temperature $T$. Without inelastic scattering, one of the four inverse square relaxation lengths $\lambda_j^{-2}$ for each frequency $\omega_n$ is zero, $n=0,\ldots,N$, so that there are $N+1$ divergent relaxation lengths in that limit. Upon switching on inelastic scattering, all but one of these relaxation lengths become finite (black curves in Fig.\ \ref{fig:lambda_inelastic}b). 

\begin{figure*}
\includegraphics[width=\linewidth]{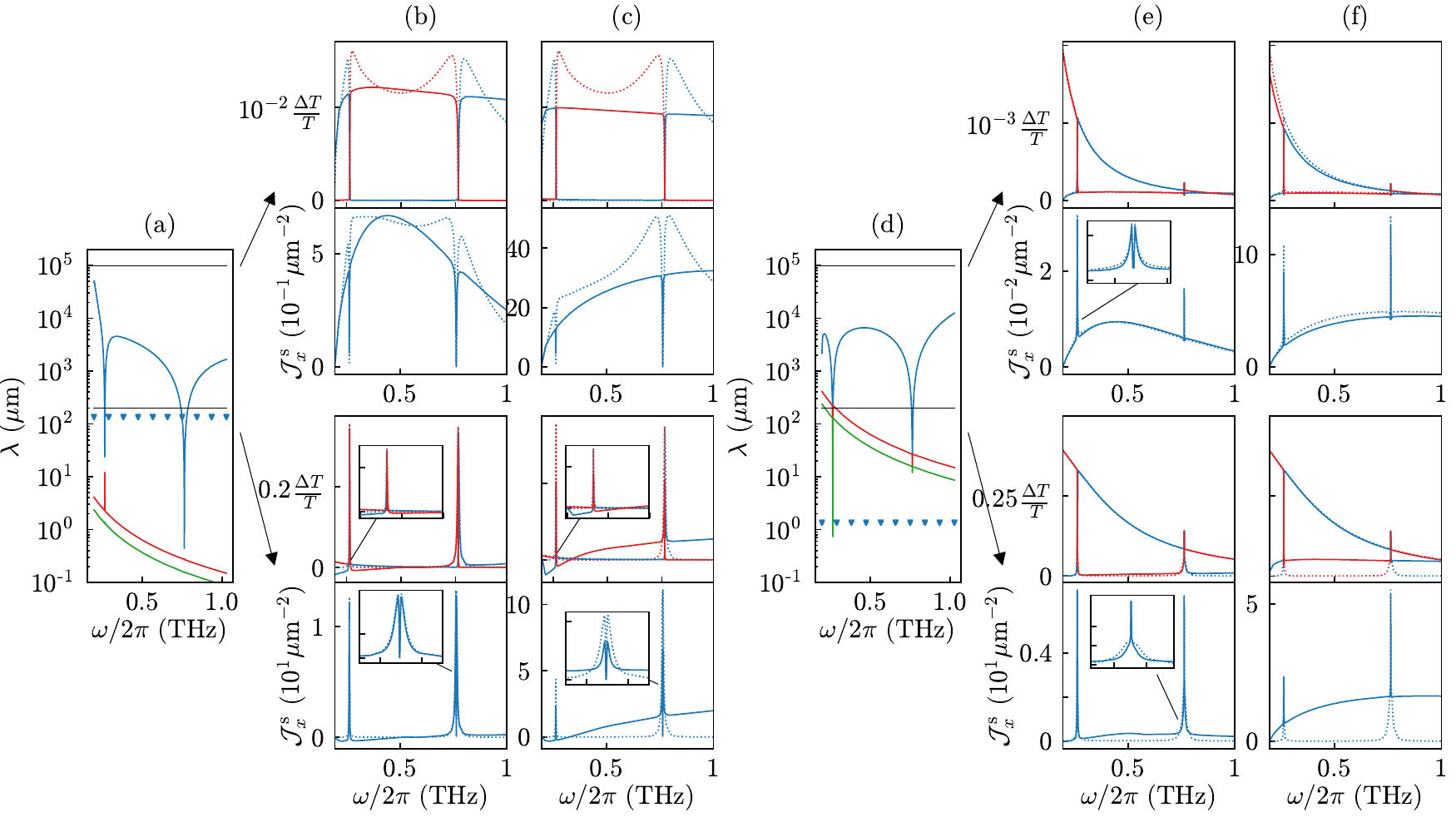}
\caption{\label{fig:inelastic1} Relaxation lengths (a), (d) as in Fig. \ref{fig:lengthscales} with material parameters taken from Table \ref{tab:values}. Frequency-resolved isotropic moment $\psi_0(\omega)$ and frequency-resolved spin current $j^{\rm s}_x(\omega) = \hbar {\cal J}^{\rm s}_x(\omega) \Delta T/T$ at the FN interface for $T = 10\,{\rm K}$ (b), (e) and $T=30\,{\rm K}$ (c), (f) together with the values obtained from a theory with elastic scattering only (dotted lines). Impurity scattering is modeled by the white-noise model (\ref{eq:whitenoise}) with parameters taken from the center column (panels (a)--(c)) and rightmost column (panels (d)--(e)) in Table \ref{tab:impurity}, corresponding to the cases of higher magnetic and acoustic quality, respectively.}
\end{figure*}

The dominant inelastic processes are inelastic magnon-magnon and phonon-phonon scattering and exchange-based magnon-phonon scattering. Their main effect is to equilibrate the distributions of magnon-like and phonon-like magnon polarons between different frequencies. Except in the vicinity of the resonance frequencies, the exchange-based magnon-phonon interaction allows for the exchange of energy between the lattice and spin subsystems, but it cannot change the number of magnons. In the same way, at generic frequencies the phonon-phonon and magnon-magnon interactions cannot exchange energy between the subsystems, although the three-magnon interaction can change the number of magnons. The relativistic and dipole-dipole contributions to the inelastic magnon-phonon interaction, which can exchange energy between subsystems and change the magnon number, are weaker than the impurity-mediated inter-mode scattering processes for $T \lesssim 30\,{\rm K}$, see Fig.\ \ref{fig:relaxationlengths}, so that these do not play a role in the spin Seebeck effect. We verified that all results shown in this Section are the same with or without inclusion of the relativistic and dipole-dipole contributions to the inelastic magnon-phonon interaction (data not shown).

{\em Distribution functions and frequency-resolved spin current.---}
For system lengths $L$ below the shortest inelastic length, an effect of inelastic processes on the distribution functions is trivially absent and the elastic-scattering-only theory of the previous Subsection applies. Taking the parameters listed in Tables \ref{tab:values} and \ref{tab:impurity}, for $T \lesssim 30\,{\rm K}$ this includes the experimentally relevant range $L \sim 10\,\mu{\rm m}$. To illustrate the effect of inelastic processes on the distribution functions, in Fig.\ \ref{fig:inelastic1} we show the distribution function $\psi_0(\omega)$ at the FN interface and the frequency-resolved spin current for two larger system sizes $L = 200\,\mu{\rm m}$ and $L=100\,{\rm mm}$.

The smaller system length $L = 200\,\mu{\rm m}$ is below the impurity-mediated inter-mode scattering length $\limp$, so that $\psi_0(\omega)$ is sharply peaked at the resonance frequencies in the absence of inelastic processes. In this case, the main effect of inelastic processes is to change the occupation of magnon-like modes at generic frequencies, which, for $L = 200\,\mu{\rm m}$, outweighs the effect of impurity-mediated processes for temperatures $T \gtrsim 10\,{\rm K}$. At the same time the weight of the peaks at the resonance frequencies is decreased upon inclusion of inelastic processes. The exchange-based magnon-phonon interaction transfers energy from the phonon system into the magnon system without changing the magnon number. This leads to a population shift of the magnon-like modes from low to high frequencies, which can be seen in Fig.\ \ref{fig:inelastic1}(b). Although via this mechanism inelastic processes can cause a small decrease of the frequency-resolved spin current $j^{\rm s}_x(\omega)$ for low frequencies, for system lengths $L \lesssim \limp$ their over-all effect after integration over all frequencies is to {\em increase} the spin current above the value previously evaluated for elastic scattering. This increase may be quite substantial at the highest temperatures we consider (panel (c) in Fig.\ \ref{fig:inelastic1}).

The larger system length $L=100\,{\rm mm}$ is well above $\limp$. In this case impurity-mediated inter-mode scattering processes already fully equilibrate phonon-like and magnon-like modes at all frequencies. Inclusion of inelastic processes leads to a quantitative, but not to a qualitative change of the distribution functions compared to the elastic-scattering-only case. For this large system length, inclusion of inelastic processes leads to a small {\em decrease} of the spin current. These conclusions apply to the impurity scattering parameters taken from Table \ref{tab:impurity} (Fig.\ \ref{fig:inelastic1}(a)--(c)), for which the phonon-impurity length $l_{\rm pi}$ is much shorter than the magnon-impurity length $l_{\rm mi}$, as well as to the case in which the strengths of phonon-impurity and magnon-impurity scattering are interchanged such that $l_{\rm pi} \gg l_{\rm mi}$ (Fig.\ \ref{fig:inelastic1}d--f).

\begin{figure*}
\includegraphics[width=\linewidth]{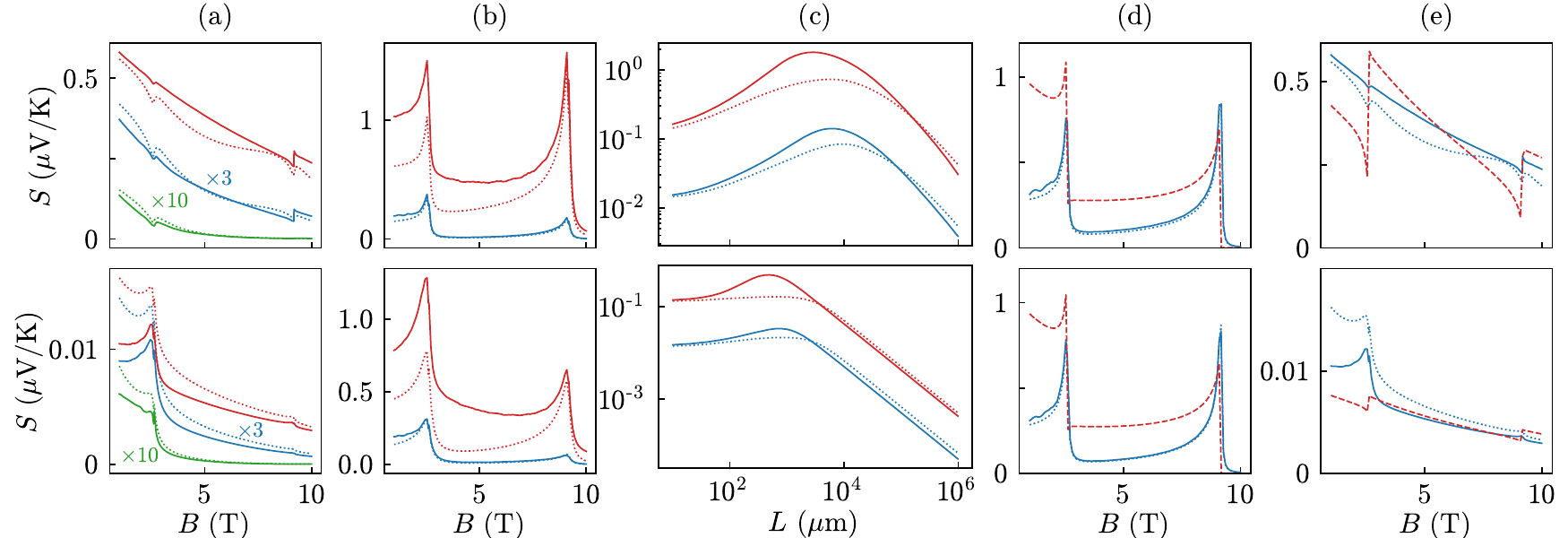}
\caption{\label{fig:inelastic} Spin Seebeck coefficient $S$ for systems of higher magnetic quality (upper panels) and higher acoustic quality (lower panels), as a function of magnetic field for $L=10\,{\rm cm}$ and temperatures $T=2\,{\rm K}$, $5\,{\rm K}$, and $10\,{\rm K}$ from bottom to top (a) and for $L=200\,\mu {\rm m}$ and $T= 5\,{\rm K}$ and $10\,{\rm K}$ from bottom to top (b). Panel (c) shows $S$ as a function of length $T=10\,{\rm K}$ (red) and $T=5\,{\rm K}$ (blue). Panels (d) and (e) show a comparison to the incoherent theory of Ref.\ \cite{Schmidt-2018} (red, dashed) for $L=10\,\mu{\rm m}$ (d) and $L=10\,{\rm cm}$ (e) and $T=10\,{\rm K}$. In all panels solid curves show results of the full theory, including inelastic scattering, whereas dotted curves include elastic scattering only. Material parameters are taken from Table \ref{tab:values}. Impurity scattering of magnons and phonons is described by the microscopic model of Eqs.\ (\ref{eq:magimp}) and (\ref{eq:phoimp}), with parameter values taken from Table \ref{tab:impurity}.}
\end{figure*}

{\em Spin Seebeck coefficient.---}
The spin Seebeck coefficient in the presence of inelastic scattering is shown in Fig.\ \ref{fig:inelastic}. As can be anticipated from the above results, inclusion of inelastic scattering leads to a small decrease of the spin Seebeck coefficient for ultralong system lengths $L \gg \limp$ (panel (a), $L = 100\,{\rm mm}$) and to an increase of the spin Seebeck coefficient for $L = 200\,\mu{\rm m}$, which is below $\limp$, provided the temperature is sufficiently high that inelastic processes become relevant at this system length. In both cases, inelastic processes reduce the visibility of the singular features at the critical values of the magnetic field. Because of the strong enhancement of the spin Seebeck coefficient for lengths $L \lesssim \limp$, the non-monotonicity of the length dependence of $S$ is more pronounced with inelastic processes than without, see Fig.\ \ref{fig:inelastic}(c).

In our previous work \cite{Schmidt-2018}, as well as in Ref.\ \cite{Flebus-2017}, inelastic magnon-magnon and phonon-phonon collisions were only accounted for indirectly as a ``background interaction'' that ensures the fast relaxation of the magnon and phonon distributions (in Ref.\ \cite{Schmidt-2018}) or magnon-polaron distributions (in Ref.\ \cite{Flebus-2017}) towards Planck and Bose-Einstein distributions, respectively, characterized by a temperature and chemical potential. In Fig.\ \ref{fig:inelastic}(d) and (e) we compare the predictions of the present theory with that of the Boltzmann theory of Ref.\ \cite{Schmidt-2018}. For an experimentally relevant system length $L = 10\,\mu{\rm m}$, which is well below the inelastic scattering lengths at $T \lesssim 30\,{\rm K}$, the two theories nevertheless give very similar predictions for the magnetic-field dependence of the spin Seebeck coefficient, see Fig.\ \ref{fig:inelastic}(d). This is remarkable, because the two theories have vastly different predictions for the frequency-resolved spin current. That the agreement is not always as good can be seen from Fig.\ \ref{fig:inelastic}(f), which compares the two theories for a much larger system length, showing that while both theories predict the same order of magnitude for the spin Seebeck coefficient, the singular features at the critical magnetic field are qualitatively different.

\begin{figure}
\includegraphics[width=\linewidth]{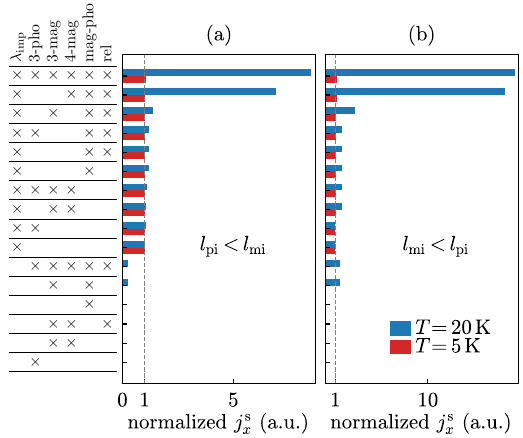}
\caption{\label{fig:contributions} Spin current $j^{\rm s}_x$, normalized to the
spin current for elastic scattering only at $T = 5\,{\rm K}$ and $T=20\,{\rm K}$, $L = 200\,\mu{\rm m}$, and $B = 7\,{\rm T}$,
with or without physical processes as indicated on the left. The physical processes
are: Impurity-mediated inter-mode scattering ($\limp$), inelastic
three-phonon scattering (3-pho), inelastic three-magnon scattering (3-mag),
inelastic four-magnon scattering (4-mag), inelastic exchange-based magnon-phonon
scattering (mag-pho), and relativistic/dipole-based inelastic magnon-phonon
scattering (rel). Phonon-impurity and magnon-impurity scattering are included in all
cases, using the microscopic model of Eqs.\ (\ref{eq:magimp}) and (\ref{eq:phoimp}) with
values taken from the center and right column of Table \ref{tab:impurity} (panels (a) and (b), respectively).}
\end{figure}

{\em Anatomy of the spin Seebeck effect.---}
Given the multitude of physical processes contributing to the spin Seebeck effect, it is instructive to determine, which process or which combination of processes is responsible for the observed spin current. This is addressed in Fig.\ \ref{fig:contributions}, in which we compare the spin currents at $L = 200\, \mu{\rm m}$ ({\em i.e.}, for a length below $\limp$, but much larger than the length $L \approx 10\,\mu{\rm m}$ relevant for experiments) at $T=5\,{\rm K}$ and $T=20\,{\rm K}$ with each of the contributing physical processes switched on or off. While we see that at the lowest temperatures impurity-mediated inter-mode scattering is the dominant cause of the spin Seebeck effect, at $T = 20\,{\rm K}$ it is the combination of impurity-mediated inter-mode scattering, exchange-based magnon-phonon scattering, and inelastic four-magnon scattering. Removing each one of these processes leads to a decrease of the spin Seebeck coefficient by an order of magnitude.

At the experimentally relevant length $L = 10\,\mu{\rm m}$, impurity-mediated inter-mode scattering is the sole cause of the spin Seebeck effect at all temperatures we considered ($T$ up to $30\,{\rm K}$) (data not shown).

\section{Conclusion}
\label{sec:conclusions}

In this article, we constructed a Boltzmann theory of the spin Seebeck effect at low temperatures. Our theory treats quadratic terms in the magnon-phonon Hamiltonian exactly, including terms that couple spin and lattice degrees of freedom. Such terms lead to the formation of magnon polarons, coherent superpositions of collective excitations of the spin and lattice subsystems \cite{Kittel-1958,Akhiezer-1959,Schloemann-1960}. Elastic scattering from impurities and inelastic relaxation processes are included via a collision integral. To accommodate the strong frequency dependence of the degree of magnon-phonon mixing, which predominantly takes place near the resonant frequencies at which the magnon and phonon dispersions cross, we keep the full frequency dependence of the distribution function at all stages of the calculation. In this respect, our calculation goes beyond previous theories of the spin Seebeck effect, which approximate the magnon and phonon distribution functions using a Planck-type or Bose-Einstein-type ansatz \cite{Kikkawa-2016,Flebus-2017,Cornelissen-2017,An-2016,Schmidt-2018,Agrawal-2013,Troncoso-2020,Rueckriegel-2020}. Our theory treats the boundary between the magnetic insulator and the normal metal (into which the spin current is emitted) non-perturbatively, which allows us to treat bulk effects and the accumulation of magnon polarons at the interface to the normal metal on equal footing.

The magnon-polaron dispersion and the collision integrals are obtained from a simplified microscopic model of a ferromagnetic insulator. This model consists of spins placed on a simple cubic lattice, with isotropic Heisenberg exchange interaction and Zeeman coupling to an external magnetic field, a harmonic lattice potential, and a pseudo-dipolar anisotropic exchange interaction that couples the spin and lattice sub-systems. Although the model is highly simplified in comparison to the complexity of the synthetic ferrimagnetic insulator Yttrium Iron Garnet Y$_3$Fe$_5$O$_{12}$ (YIG) used in most experiments, having a true model at hand offers the possibility to have a faithful description of dependences on external parameters, such as the magnetic field or the temperature. This is an essential requirement for a description of the singular magnetic-field dependent features of the spin Seebeck coefficient observed at the critical magnetic fields at which magnon and phonon dispersions touch \cite{Kikkawa-2016}. The characteristic momentum dependence of matrix elements of the magnon-phonon coupling from the anisotropic pseudodipolar interaction also sheds light on the relative importance of relativistic processes as a function of temperature. 

The use of collision integrals obtained from a microscopic model (with parameters adjusted to reproduce known properties of YIG) has the advantage that no phenomenological relaxation terms are needed, such as a stochastic magnetic field \cite{Xiao-2010,Hoffman-2013,Adachi-2013,Rameshti-2019} or a relaxation-approximation-type exponential relaxation to the equilibrium form \cite{Rezende-2014,Flebus-2017}. This is important, because the use of such phenomenological relaxation terms may violate conservation laws that apply to the underlying microscopic processes. To see that such conservation laws can be a true barrier for an efficient coupling between the lattice and spin systems, we note that only the weakest of all elastic and inelastic interactions of the spin and lattice system, the relativistic and dipole-dipole-based conversion of a phonon into a pair of magnons or vice versa, is by itself capable of sustaining a spin Seebeck effect with contributions from all frequencies. Taken separately, none of the other (stronger) interactions, such as the exchange-based magnon-phonon coupling (which conserves spin), elastic impurity scattering (which conserves energy), or inelastic three-magnon scattering (which does not couple to the lattice subsystem) can fully equilibrate lattice and spin subsystems. It is the interplay of these elastic and inelastic processes that eventually allows lattice excitations to create an excitation of the spin system that carries a steady-state spin current. A simple microscopic modeling that obeys the relevant conservation laws is better suited to describe the magnetic-field and temperature dependences characteristic of this combined effect than a phenomenological approach.

The predictions of our theory differ in an essential way from previous studies of the magnetic-field dependence of the spin Seebeck effect at low temperatures. We find that at experimentally relevant length scales, which are larger than the mean free paths for magnon-impurity and phonon-impurity scattering, but smaller than the inelastic relaxation lengths at low temperatures, the dominant mechanism coupling lattice and spin degrees of freedom is impurity-mediated scattering between different magnon-polaron modes. This impurity-mediated inter-mode scattering is strongest near the resonance frequencies at which magnon and phonon dispersions cross. Consequentially, the spin current is carried almost entirely by magnon polarons at those frequencies. Previous theories of the spin Seebeck effect explicitly or implicitly assumed a strong equilibration by inelastic magnon-magnon and phonon-phonon scattering processes \cite{Kikkawa-2016,Flebus-2017,Cornelissen-2017,An-2016,Schmidt-2018,Agrawal-2013,Troncoso-2020,Rueckriegel-2020}, which implies that the spin current carried by excitations at all frequencies, insofar as they are accessible for thermal excitation. Such assumption of strong relaxation is valid for high temperatures and large system sizes, but the equilibration length exceeds the system size at the low temperatures at which the singular features in the magnetic-field dependence of the spin Seebeck effect are observed \cite{Kikkawa-2016}. In this parameter regime, our theory predicts a robust peak at the critical magnetic fields at which magnon and phonon dispersions cross, irrespective of the type of the underlying elastic relaxation mechanism. The peak arises, because at these magnetic fields the frequency window that contributes to the spin Seebeck effect is maximized. Our prediction differs from that of Refs.\ \cite{Kikkawa-2016,Flebus-2017}, which predict a peak or a dip, depending on whether magnon-impurity scattering or phonon-impurity scattering dominates at low temperatures. Our predictions also differ from our own previous theory for the fully equilibrated regime \cite{Schmidt-2018}, although the differences are more subtle.

Whereas the theory presented here treats the frequency dependence of the distribution function exactly, it does make approximations in other respects. The use of the Boltzmann equation misses coherences between different magnon-polaron modes, which can lead to an overestimate of impurity-mediated inter-mode scattering near the resonance frequencies. This could be a relevant issue at very low temperatures, at which magnons and phonons are coherent over long distances. If elastic scattering dominates, such coherence effects can be treated theoretically within a calculation of the spin Seebeck effect using, {\em e.g.}, diagrammatic perturbation theory \cite{Akkermans-2010} or the Lindblad approach \cite{Breuer-2002}. At system lengths shorter than the elastic mean free path or at temperatures high enough that the inelastic relaxation length become shorter than the elastic mean free paths, the approximation that we only take one isotropic moment and one anisotropic moment of the distribution function may no longer be strictly valid. In this case, a systematic quantitative improvement of the theory can be obtained by keeping higher moments in an expansion in spherical harmonics. However, as our comparison for the ballistic regime has shown, see App.\ \ref{app:ballistic}, the quantitative error associated with the use of a simplified angular dependence with one isotropic and one anisotropic moment is numerically small.

In the implementation of our formalism, we assumed that the magnon and phonon dispersions are isotropic. This is consistent with known properties of YIG in the long-wavelength limit \cite{Cherepanow-1993,Clark-1961}. Inclusion of magnon-phonon coupling gives a small anisotropy, which our calculations do account for. When going to higher frequencies and temperatures, in principle, there can also be anisotropy of the magnon and phonon dispersions in the absence of magnon phonon coupling. If that is the case, the frequency region in which magnon and phonon dispersions cross will be enhanced. As a result, the impurity-mediated inter-mode scattering will be strong over a larger range of frequencies, leading to an enhancement of the spin Seebeck effect at short system sizes. At high temperatures, excitation of zone-boundary phonons or optical phonons (and, eventually, excitation of zone-boundary magnons and optical magnons and the appearance of Umklapp processes) will become important. Our lattice-based formulation of the Boltzmann theory should in principle be able to deal with this regime, although the simplified microscopic lattice model of Sec.\ \ref{sec:model} will need to be refined \cite{Barker-2017,Shen-2019,Simensen-2019,Troncoso-2020}.

Our formalism can also be used to study the spin Seebeck effect in the ``nonlocal'' geometry \cite{Cornelissen-2015,Shan-2016,Cornelissen-2017,Liu_2018}, in which the driving source is the injection of magnons from a second normal metal, instead of the injection of phonons from an insulating non-magnetic heat bath. Low-temperature measurements of the nonlocal spin Seebeck effect, showing anomalous features at the critical magnetic fields, were recently reported \cite{Oyanagi-2020}. The frequency-resolved theory can further be applied to model other manifestations of magnon-polaron formation, such as the accumulation of magnon polarons in the spectral region near the resonant frequencies after parametric excitation of magnons \cite{Bozhko-2017}, anomalies in the spin pumping efficiency at resonance frequencies \cite{Hayashi-2018}, or the direct observation of wave-like excitation in the lattice subsystem after excitation of the spin subsystem \cite{Holanda-2018}, where a frequency-resolved description is natural and essential. 

\section*{Acknowledgements}

The authors acknowledge stimulating discussions with Tobias Kampfrath, Unai Atxitia, Benedetta Flebus, and Gerrit E. W. Bauer. This work was financially supported by the Deutsche Forschungsgemeinschaft (DFG) via TRR 227 ``Ultrafast Spin Dynamics'' (project B03).

\appendix

\section{Lattice model for magnon polarons}
\label{app:a}

In this appendix we give the expressions for the Fourier representation of the lattice model in Section \ref{sec:mp1}. The equations of motion for the Fourier-transformed lattice displacement $\vu_{\vq}$, momentum $\vp_{\vq}$, and magnetization amplitude $\vn_{\vq}$ are
\begin{align}
  \dot \vu_{\vq} = \frac{\partial H}{\partial \vp_{-\vq}}, \ \ \dot \vp_{\vq} = -\frac{\partial H}{\partial \vu_{-\vq}}, \ \  \dot \vn_{\vq} = - \ve \times \frac{\partial H}{\partial \vn_{-\vq}}.
\end{align}
Without the restriction of long wavelengths $q\to 0$ the general expressions for the interaction matrices ${\cal B}(\vq)$, ${\cal K}(\vq)$, and ${\cal D}(\vq)$ in the Fourier-transformed magnon-phonon Hamiltonian \ (\ref{eq:HFourier}) for the full lattice model read, for the magnon energy,
\begin{align}
  {\cal B}(\vq) = 2 J S \sum_{\alpha} (1 - \cos q_{\alpha} a) + \mu B, 
\end{align}
for the phonon energy,
\begin{align}
{\cal K}(\vq) =&\,
    2 \sum_{\alpha,\beta} \left[ {\cal K}_{\alpha\beta} (1 - \cos q_{\alpha} a) + {\cal K}_{\alpha\beta}'
        \sin(q_{\alpha} a) \sin(q_{\beta} a)
  \right. \nonumber \\ &\, \ \ \ \ \left. \mbox{} 
    + {\cal K}_{\alpha\beta}''
        (1 - \cos (q_{\alpha} a) \cos (q_{\beta} a))  \right],
\end{align}
with the indices $\alpha$ and $\beta$ summed over the three coordinate directions $x$, $y$, and $z$, and
\begin{align}
  {\cal K}_{\alpha} =&\, 
 ( K_1 \ve_{\alpha} \ve_{\alpha}^{\rm T} + K_2 \openone) \delta_{\alpha,\beta}, 
  \nonumber \\
  {\cal K}'_{\alpha\beta} =&\, \frac{1}{2} K_1 (\ve_{\alpha} \ve_{\beta}^{\rm T} + \ve_{\beta} \ve_{\alpha}^{\rm T}),
  \nonumber \\
  {\cal K}''_{\alpha\beta} =&\, 
  \frac{1}{2} K_1 
    (\ve_{\alpha} \ve_{\alpha}^{\rm T} + \ve_{\beta} \ve_{\beta}^{\rm T}),
\end{align}
and for the magnon-phonon coupling
\begin{align}
{\cal D}(\vq) =&\, \frac{2 i S^{3/2}}{a^{5/2}} \sum_{\alpha} \left\{ \vphantom{\frac{M}{M}} D [ (\ve \cdot \ve_{\alpha}) \openone
  + \ve_{\alpha} \ve^{\rm T}] \right. \nonumber\\
&\, \left. \mbox{} + \left(\frac{D'}{a} - 2 D \right)
(\ve \cdot \ve_{\alpha}) \ve_{\alpha} \ve_{\alpha}^{\rm T} \right\} \sin q_{\alpha} a.
\end{align}
Applying the long-wavelength approximation $q \to 0$ recovers the continuum theory expressions (\ref{eq:Kdef}), (\ref{eq:Bdef}), and (\ref{eq:Ddef}) in the main text.

\section{Magnon-polaron Hamiltonian}
\label{app:b}

{\em Symmetries of the magnon-polaron Hamiltonian.---}
The $8$-component column vector $\vb_{\vq}$ of Eq.\ (\ref{eq:bqdef}) and the $8 \times 8$ hermitian matrix $H_{\vq}$ of Eq.\ (\ref{eq:Hq}) satisfies the symmetry conditions
\begin{equation}
  \vb_{-\vq} = \Sigma_1 \vb_{\vq}^*,\ \
  H_{-\vq} = \Sigma_1 H_{\vq}^* \Sigma_1,
\end{equation}
where
\begin{equation}
  \Sigma_1 = \begin{pmatrix} 0 & 1 \\ 1 & 0 \end{pmatrix}.
\end{equation}
The symplectic matrix $V_{\vq}$ and the diagonal matrix $\Omega_{\vq}$ that diagonalize $H_{\vq}$, see Eq.\ (\ref{eq:Vq}), satisfy the conditions
\begin{align}
  \Sigma_1 \Omega_{-\vq} \Sigma_1 =&\, \Omega_{\vq},\nonumber \\
  \Sigma_3 V_{\vq}^{\dagger} \Sigma_3 =&\, V_{\vq}^{-1},\nonumber \\
  \Sigma_1 V_{-\vq} \Sigma_1 =&\, V_{\vq}^*,
  \label{eq:Vprop}
\end{align}
with
\begin{equation}
  \Sigma_3 = \begin{pmatrix} 1 & 0 \\ 0 & -1 \end{pmatrix}.
\end{equation}

In addition to the symplectic structure outlined above, the Hamiltonian $H_{\vq}$ satisfy the symmetry condition
\begin{equation}
  H_{-\vq} = I H_{\vq} I^{\dagger},
\end{equation}
where 
\begin{equation}
  I = \begin{pmatrix} \openone_3 \\  & -1 \\ & & \openone_3 \\ & & & -1 \end{pmatrix}.
\end{equation}
Since $I$ commutes with $\Sigma_1$ and $\Sigma_3$, it follows that 
\begin{equation}
  \label{eq:Vinversion}
  V_{-\vq} =  I V_{\vq},\ \ 
  \Omega_{-\vq} = \Omega_{\vq}.
\end{equation}
Together with the conditions (\ref{eq:Vprop}), this implies that 
\begin{equation}
  V_{\vq}^* = I \Sigma_1 V_{\vq} \Sigma_1,\ \
  \omega_{\vq,\lambda} = \omega_{-\vq,\lambda}.
  \label{eq:Irelation}
\end{equation}

{\em Symplectic diagonalization.---} To (numerically) diagonalize the positive definite hermitian matrix $H_{\vq}$ using the symplectic matrix $V_{\vq}$ that satisfies the condition (\ref{eq:Vprop}), we first consider the non-hermitian matrix $\Sigma_3 H_{\vq}$ and diagonalize it with the $8 \times 8$ matrix $V'$ as (we omit the subscript $\vq$ in the following equations)
\begin{equation}
  \Sigma_3 H = \frac{1}{2} V' \Sigma_3 \Omega V'^{-1} . \label{eq:Sigma3Hdecomp}
\end{equation}
Since $H$ is hermitian, $\Sigma_3 H = \Sigma_3 H^{\dagger}$ and we have
\begin{equation}
  \Sigma_3 H = \frac{1}{2} \Sigma_3 (V'^{\dagger})^{-1} \Sigma_3 \Omega V'^{\dagger} \Sigma_3.
\end{equation}
Uniqueness of the diagonalization of a matrix then implies that
\begin{equation}
  D \Sigma_3 V'^{-1} = V'^{\dagger} \Sigma_3,
\end{equation}
where $D$ is a diagonal matrix. Moreover, $D$ is positive definite, so that we may write $D = P^{-2}$, with $P$ a diagonal matrix. It follows that
\begin{equation}
  \Sigma_3 = (\Sigma_3 V' P)^{\dagger} \Sigma_3 (\Sigma_3 V' P).
\end{equation}
One then easily verifies that the matrix $V = \Sigma_3 V' P$ diagonalizes $H$ via
\begin{equation}
 H = \frac{1}{2} V \Omega V^{\dagger}
\end{equation}
and satisfies the condition (\ref{eq:Vprop}).

\section{Magnon-phonon coupling from magneto-elastic theory}
\label{app:pheno}

Here, we review the magneto-elastic theory of magnon-phonon interaction and show how microscopic magnon-phonon Hamiltonians from magnetic dipole-dipole interaction and magnetic anisotropy compare to it.

\bigskip
{\em Magneto-elastic coupling---} Based only on symmetry considerations in a continuous medium Kittel \cite{Kittel-1949} and Kaganov {\em et al.} \cite{Kaganov-1959,Kaganov-1961-2} proposed a phenomenological magneto-elastic coupling energy. The leading-order contribution to magnon-phonon coupling in the presence of a cubic symmetry takes the form
\begin{equation}
H^{\rm me} = \int\! \frac{dV}{a^3} \sum_{\alpha,\beta} B_{\alpha\beta} e_{\alpha \beta} s_{\alpha}(\vr) s_{\beta}(\vr), \label{eq:hme}
\end{equation}
where $\vs(\vr_j)$ is the continuous spin density in the long-wavelength limit, $e_{\alpha\beta} = (1/2) \left( \partial_{\beta} u_{\alpha} + \partial_{\alpha} u_{\beta} \right)$ the symmetrized strain tensor, the integral is over the volume $V$ of the magnetic insulator, and $B_{\alpha\beta} = B_1 \delta_{\alpha\beta} + B_2 (1-\delta_{\alpha\beta})$ is the magneto-elastic coupling tensor. The constants $B_1$ and $B_2$ are material-specific constants describing the strength of the magneto-elastic coupling. For YIG these constants were fitted to magnetostriction experiments at 300 K \cite{Strauss-1968} as $B_{1} = 0.08 J$ and $B_{2} = 0.16 J$.

The magneto-elastic Hamiltonian (\ref{eq:hme}) can be written in terms of magnon and phonon creation and annihilation operators by applying Eqs.\ (\ref{eq:expansion}) and (\ref{eq:normal}). Taking the magnetization direction $\ve$ in the $(111)$ direction, we can compare the magneto-elastic Hamiltonian with the magnon-phonon Hamiltonian obtained for a simple cubic lattice model with nearest-neighbor pseudo-dipolar exchange interactions, see Sec.\ \ref{sec:mp1}. Comparing Eq.\ (\ref{eq:hme}) with Eq.\ (\ref{eq:hpd_mp}) shows that the two Hamiltonians agree if we set $B_1 = a D'$ and $B_2 = 2 D$.

We now discuss two additional contributions to the microscopic simple cubic spin model of Sec.\ \ref{sec:mp1} that also lead to magnon-phonon coupling and that can easily be included into the phenomenological magneto-elastic Hamiltonian: an on-site magnetic anisotropy and the long-range dipole-dipole interaction. The numerical evaluations reported in this article are obtained without these two additional contributions.

\bigskip
{\em On-site magnetic anisotropy.---}
A model with cubic symmetry allows an on-site magnetic anisotropy term of the form \cite{Kittel-1949}
\begin{align}
  H^{\rm a} =&\ \sum_{\langle i,j \rangle} K_{ij} (\vS_{i} \cdot \ve_{ij})^2,
  \label{eq:ha}
\end{align}
where $\ve_{ij}$ is the unit vector connecting nearest-neighbor lattice sites $i$ and $j$ and the coupling constant $K_{ij}$ depends on the direction of $\ve_{ij}$ and the distance $|\vr_i-\vr_j|$ between the sites $i$ and $j$. Although such a magnetic anisotropy term does not affect the magnon dispersion in a system with cubic symmetry, there is a magnon-phonon interaction resulting from it. Passing to the continuum limit, we find that the on-site anisotropy Hamiltonian (\ref{eq:ha}) yields the same magnon-phonon coupling as the pseudo-dipolar exchange coupling (\ref{eq:hpd_mp}) with the on-site anisotropy constants chosen as $K = D$ and $K'=D'$, where $K' = dK_{ij}/d|\vr_i-\vr_j|$.

\bigskip
{\em Dipole-dipole interaction---}
The weak but long-range dipole-dipole coupling between magnetic dipoles corresponds to the Hamiltonian
\begin{align}
H^{\rm di} =&\ \frac{1}{2} \sum_{i,j} \frac{\mu^2}{r^3_{ij}} \left[ \vS_{i} \cdot \vS_{j} - \frac{3}{r_{ij}^2} (\vS_{i} \cdot \vr_{ij}) (\vS_{j} \cdot \vr_{ij}) \right], \label{eq:hdi}
\end{align}
where $r_{ij} = |\vr_i - \vr_j|$ and $\mu = g \mub$ is the magnetic moment of the spins. The summation is over all pairs of lattice sites ($i,j$), irrespective of their distance. In contrast to the Heisenberg Hamiltonian, the interaction strength does not only depend on the length of the bond between the spins, but also on the angle of the spins with the connecting bonds. Expanding the dipole-dipole coupling to first order in the lattice displacements $\vu_i$ we find a Hamiltonian of the same structure as (\ref{eq:hme}). Evaluating the summation over pairs of lattice sites and taking the continuum limit, the coupling strength converges towards $B^{\rm d}_1 = (9\pi/4) \mu^2 / a^3$ and $B^{\rm d}_2 = - (3\pi/2) \mu^2 / a^3$. Comparing the strength of the dipole-dipole interaction to the magneto-elastic constants $B_{1,2}$ we see that the dipole-dipole interaction makes up for a fraction of roughly 5\% of the measured magneto-elastic constants in YIG, in agreement with the estimate of Ref.\ \cite{Keffer-1962}.

\section{Interface and boundary conditions}
\label{app:c}

{\em Solutions at fixed frequencies.---} The frequency $\omega$ and the transverse component $\vq_{\perp} = q_y \ve_y + q_z \ve_z$ of the wavevector are conserved at the interfaces. To prepare for the calculation of the transmission and reflection coefficients of the IF and FN interfaces, we therefore need to construct wave-like solutions of the equations of motion (\ref{eq:equationsOfMotionReal}) at fixed $\omega$ and $\vq_{\perp}$. In general such solutions are of the form
\begin{align}
  \vu_j(t) =&\, \sum_{\nu} c_{\nu} \vu_{\omega,\vq_{\perp},\nu} e^{i \vq(\nu) \cdot \vr_{j} - i \omega t}, \nonumber \\
  \vn_j(t) =&\, \sum_{\nu} c_{\nu} \vn_{\omega,\vq_{\perp},\nu} e^{i \vq(\nu) \cdot \vr_{j} - i \omega t} , \label{eq:generalSolutions}
\end{align}
where we write $\vq(\nu) = q_x(\nu) \ve_x + \vq_{\perp}$ for the wavevector of the mode $\nu$. In all cases the momentum amplitude $\vp_j$ follows from the equalities $\vp = -i m \omega \vu$, so that the momentum amplitude needs not be considered explicitly.

For each combination $(\omega,\vq_{\perp})$ there are ten such solutions, five of which are propagating or exponentially decaying in the positive $x$ direction and five are propagating or exponentially decaying in the negative $x$ direction. At least one out of each set of modes that propagate in the positive and negative directions, respectively, is decaying \cite{Kamra-2015}. Exponentially decaying solutions have a complex wavenumber $q_x$. We label the ten solutions by the composite index $\nu=(n,{\rm L/R})$ where $n=1,2,3,4,5$. The label L is used for solutions that are propagating or exponentially decaying in the positive $x$ direction; the label R is for solutions that are propagating or exponentially decaying in the negative $x$ direction.

The equations of motion determine the prefactors $\vu_{\omega,\vq_{\perp},\nu}$ and $\vn_{\omega,\vq_{\perp},\nu}$ up to an over-all factor. For the propagating modes, we fix this factor by requiring that the energy current carried by that mode is $|c_{\nu}|^2$ for right-moving modes and $-|c_{\nu}|^2$ for left-moving modes. The expression for the energy current is, after averaging over one period and using the Fourier representation (\ref{eq:Fourier}),
\begin{align}
  \label{eq:Jenergy}
  J =&\, 
  \frac{1}{2 N_x a} \sum_{\vq} \mbox{Im}\, \left[
  \dot \vn_{\vq}^* \cdot 
    \frac{\partial {\cal B}(\vq)}{\partial q_x} \vn_{\vq}
  + \dot \vu_{\vq}^* \cdot 
    \frac{\partial {\cal K}(\vq)}{\partial q_x} \vu_{\vq} 
  \right. \nonumber \\ &\, \left. \mbox{}
  + 2
    \dot \vu_{\vq}^* \cdot 
    \frac{\partial {\cal D}(\vq)^{\dagger}}{\partial q_x}
    \vs_{\vq}
  \right]
\end{align}
This equation is derived for the lattice model at the end of this appendix. A discussion in the continuum limit can be found in Refs.\ \cite{Kamra-2014,Kamra-2015,Akhiezer-1968}.

{\em IF and FN interfaces.---} The magnetic medium F exists for $0 < x < L$ with $L = N_x a$, $N_x$ being the number of lattice sites in the $x$-direction. At $x=L$ there is a boundary to a non-magnetic metal N; at $x=0$ there is a boundary to a non-magnetic insulator I. 

In both the non-magnetic insulator I and the normal metal N we consider phonon degrees of freedom only. The phonon degrees of freedom have displacement $\vu_j'$. The equation of motion for the phonon modes at fixed frequency $\omega$ and transverse wavevector $\vq_{\perp}'$ in I and N is given by the eigensystem solution of a reduced version of the Hamiltonian (\ref{eq:HF}) with $H = H^{\rm pho}$ and without the magnon amplitudes,
\begin{equation}
\vu_j(t) = \sum_{\nu'} c'_{\nu'} \vu'_{\omega,\vq_{\perp}',\nu'} e^{i \vq'(\nu') \cdot \vr_{j} - i \omega t}
\end{equation}
for $\vr_j$ in I or N with the mode index $\nu'=(n',{\rm L}/{\rm R})$, with $n'=1,2,3$. The factors $\vu'_{\omega,\vq_{\perp}',\nu'}$ of the propagating modes are required that the energy current is $|c_{\nu'}|^2$ for right-moving modes and $-|c_{\nu'}|^2$ for left-moving modes. The expression for the energy current in I and N is given by Eq.\ (\ref{eq:Jenergy}) without the terms containing the magnetization amplitude $\vn_{\vq}$.

The  boundary conditions at the IF interface are Eqs.\ (\ref{eq:boundaryInsulator1}) and (\ref{eq:boundaryInsulator2}) of the main text. The boundary condition (\ref{eq:boundaryInsulator2}) ensures that the energy current (\ref{eq:Jenergy}) is continuous at the interface. The boundary conditions for the FN interface are given in Eqs.\ (\ref{eq:boundaryF1}) and (\ref{eq:boundaryF}) of the main text. In this case, one has to also take into account that magnons can excite conduction electrons in the normal metal \cite{Tserkovnyak-2002,Hoffman-2013}, which leads to the condition that the energy current carried by the magnon mode is equal to the energy current by the spin current emitted into the normal metal N by the precessing magnetization at $x=L$, which is Eq.\ (\ref{eq:boundaryF}) of the main text.

{\em Reflection and transmission coefficients.---} Following the ideas of the Landauer-B\"uttiker formalism \cite{Datta-2003}, the amplitudes of the normalized coefficients $|c_{\nu}|^2$ and $|c'_{\nu'}|^2$ which are solutions of the boundary conditions at the interfaces at $x=0$ and $x=L$ can be recast in the form of reflection and transmission coefficients for the propagating fixed-frequency solutions (\ref{eq:generalSolutions}). These coefficients are written $R_{\nu\nu'}(\omega,\vq_{\perp})$ and $T_{\nu\lambda''}(\omega,\vq_{\perp})$, where only indices $\nu$, $\nu'$, and $\lambda''$ that correspond to propagating modes are considered.

Energy conservation at the IF interface at $x=0$ implies the unitarity conditions (\ref{eq:unitarityL}) of the main text. There is no unitarity condition for the FN interface at $x=L$ because of the possibility that magnons excite the conduction electrons in the normal metal, which are not accounted for explicitly in our theory. Instead, we define the probability $P_{\nu{\rm N}}(\omega,\vq_{\perp})$ that a magnon polaron in mode $\nu$ emerging from the FN interface was excited there by an incident spinful excitation of the conduction electrons in N, and the probability $P_{{\rm N}\nu}(\omega,\vq_{\perp})$ that a magnon polaron in mode $\nu$ incident on the FN interface excites a spinful excitation in N by the amount that the sum of reflection and transmission coefficients differ from one, see Eqs.\ (\ref{eq:unitarityR1}) and (\ref{eq:unitarityR2}) of the main text.

{\em Energy current density.---} To find the energy current $J$ flowing through an interface between $x=(j-1)a$ and $x=ja$, we write the lattice Hamiltonian $H$ as
\begin{equation}
  H = H_{<} + H_{j-1,j} + H_{>},
\end{equation}
where $H_{<}$ and $H_{>}$ consist of all terms in $H$ that contain on-site terms and nearest-neighbor bond terms entirely within the regions $x \le (j-1)a$ and $x \ge ja$, respectively, whereas the Hamiltonian $H_{j-1,j}$ contains the bond terms that connect these two regions. We then have
\begin{equation}
	J = \dot H_> = \{ H_>,H_{j-1,j} \},
\end{equation}
where $\{ \cdot, \cdot \}$ is the Poisson bracket. Using the equations of motion for the amplitudes $\vu_j$, $\vp_j$, and $\vn_j$, this can be recast as
\begin{align}
  J =&\, - \sum_{x_i = j a}
  \left(
  \dot \vp_i \cdot \frac{\partial }{\partial \vp_i} +
  \dot \vu_i \cdot \frac{\partial }{\partial \vu_i}
  + \dot \vn_{i} \cdot \frac{\partial }{\partial \vn_{i}} \right)H_{j-1,j},
\end{align}
where the sum is over all lattice sites $i$ with $x = j a$.
If we substitute the Fourier representation (\ref{eq:Fourier}) and omit contributions that average to zero after one period, we find Eq.\ (\ref{eq:Jenergy}). Alternatively, expressing the energy current in terms of the phasor amplitudes $b_{\vq,\lambda}$ gives
\begin{align}
  J =&\, \frac{\hbar}{L} \sum_{\vq} \mbox{Im}\, \left[ \sum_{\lambda,\lambda'=1}^{3} v_{\vq,\lambda\lambda'} \dot b^*_{\vq,\lambda} b_{\vq,\lambda'} 
  + v_{\vq,44} \dot b^*_{\vq,4} b_{\vq,4}  \right. \\ \nonumber &\, \left. \mbox{} 
  - 2  \sum_{\lambda=1}^{3} v_{\vq,4\lambda} \dot b^*_{\vq,4} (b_{\vq,\lambda} + b^*_{-\vq,\lambda}) \right],
\end{align}
with the velocities
\begin{align}
  v_{\vq,\lambda\lambda'} =&\,
    \ve_{\vq,\lambda}^* \cdot 
    \frac{\partial {\cal K}(\vq)}{\partial q_x}
    \ve_{\vq,\lambda'}
  \sqrt{\frac{a^6}{2 m^2 \omega^0_{\vq,\lambda}
    \omega^0_{\vq,\lambda'}}}, \nonumber \\
  v_{\vq,4\lambda} =&\,
    \ve_+ \cdot \frac{\partial {\cal D}(\vq)}{\partial q_x}
    \ve_{\vq,\lambda} \sqrt{\frac{a^3}{2 m \omega^0_{\vq,\lambda}}} .
\end{align}
The diagonal velocities $v_{\vq,\lambda} \equiv v_{\vq,\lambda\lambda}$ are equal to the phonon group velocities $v_{\vq,\lambda} = \partial \omega^0_{\vq,\lambda} / \partial q_x$ whereas $v_{\vq,44} =  \partial \omega^0_{\vq,4} / \partial q_x$ is the magnon group velocity.

\section{Inelastic scattering}
\label{app:inel}

The main contributions to inelastic magnon and phonon scattering are three-magnon, four-magnon, three-phonon, and two-magnon-phonon collisions. Here we show how these interactions can be obtained as sub-leading corrections to the isotropic and anisotropic exchange coupling, anisotropic corrections to the periodic lattice potential, and long-range dipole-dipole coupling. 

{\em Three-phonon interaction.---} Instead of deriving a microscopic three-phonon interaction from an expansion of the lattice potential, we here use an effective description based on a continuum model \cite{Ziman-1960},
\begin{align}
H^{\rm 3p} =& \frac{1}{\sqrt{V}} \sum_{\vq,\lambda} \sum_{\vq_2 , \lambda_2} \sum_{\vq', \lambda'} U^{\rm 3p,0}_{\vq \lambda , \vq_2 \lambda_2 ; \vq' \lambda'} (b^{\ast}_{\vq,\lambda} + b_{-\vq,\lambda}) \nonumber\\ &\, \mbox{}
\times (b^{\ast}_{\vq_2,\lambda_2} + b_{-\vq_2,\lambda_2}) (b_{\vq',\lambda'} + b^{\ast}_{-\vq' ,\lambda'})
\end{align}
with the phonon-phonon scattering potential
\begin{align}
  U^{\rm 3p,0}_{\vq \lambda,\vq_2 \lambda_2 ; \vq' \lambda'} =&\, \frac{1}{6} \left(\frac{\hbar a^3}{2 m}\right)^{3/2} K' \delta_{\vq+\vq_2,\vq'} \nonumber \\ &\, \mbox{} \times \frac{q q_2 q'}{\sqrt{\omega^0_{\vq,\lambda} \omega^0_{\vq_2,\lambda_2} \omega^0_{\vq',\lambda'}}} .
  \label{eq:U3p0}
\end{align}
In general the anharmonicity constant $K' = K'_{\vq \lambda, \vq_2 \lambda_2 ; \vq' \lambda'}$ can itself be a complex function that depends on the crystal geometry. However it is argued in Ref.\ \cite{Ziman-1960} that approximating the anharmonicity by a single momentum- and polarization-independent number
\begin{equation}
  K' \simeq \frac{1}{3a}(K_{\rm 1} + 2K_{\rm 2})
\end{equation}
is well suited for a simplified study. For our model parameters this gives $K' \simeq 2 \times 10^{10}$ J/m$^3$.

{\em Three-magnon interaction.---} The magnon number non-conserving three-magnon collisions can be obtained from the sub-leading contributions of the Holstein-Primakoff expansion (\ref{eq:expansion}) of the pseudodipolar anisotropic exchange (\ref{eq:hpd}) and the dipole-dipole interaction (\ref{eq:hdi}). Omitting terms that create or annihilate three magnons, this gives
\begin{equation}
H^{\rm 3m} = - \frac{1}{\sqrt{V}} \sum_{\vq} \sum_{\vq_2} \sum_{\vq'} U^{\rm 3m,0}_{\vq,\vq_2;\vq'} a^{\ast}_{\vq} a^{\ast}_{\vq_2} a_{\vq'}
\end{equation}
with
\begin{equation}
  U^{\rm 3m,0}_{\vq,\vq_2;\vq'} = U^{\rm pd,0}_{\vq,\vq_2;\vq'} + U^{\rm di,0}_{\vq,\vq_2;\vq'}.
  \label{eq:U3m0}
\end{equation}
The matrix element from anisotropic exchange reads, with the magnetization direction $\ve$ along the $(111)$ direction,
\begin{align}
  U^{\rm pd,0}_{\vq,\vq_2;\vq'} =&\, \frac{i D S^{1/2} a^{3/2}}{3} \delta_{\vq+\vq_2,\vq'}
  \nonumber \\ &\, \mbox{} \times 
  \sum_{\alpha} [\cos(a q_{\alpha}) + \cos(a q_{2\alpha})] w_{\alpha}^2,
\end{align}
where we defined
\begin{equation}
  w_x = e^{-2 i \pi/3},\ \
  w_y = e^{2 i \pi/3},\ \
  w_z = 1.
\end{equation}
The contribution from dipole-dipole interaction is \cite{Sparks-1964,Gurevich-1996}
\begin{equation}
U^{\rm di,0}_{\vq,\vq_2;\vq'} = \sqrt{2 \pi^2 M_{\rm s} \mu^3} \left(\frac{q_z q_{+}}{q^2} + \frac{q_{2z} q_{2+}}{q_2^2}\right) \delta_{\vq+\vq_2,\vq'},
\end{equation}
with $M_{\rm s} = \mu S/a^3$. Although the magnitude of the anisotropic exchange coupling is larger than the dipole-dipole interaction, due to the equality $\sum_{\alpha} w^2_{\alpha}=0$ the leading-order contribution of anisotropic exchange vanishes in the long-wavelength limit, so that the dipole-dipole contribution is dominant.

{\em Four-magnon interaction.---} The magnon-number conserving four-magnon collisions derive from the sub-leading contribution of the Holstein-Primakoff expansion of the Heisenberg exchange coupling (\ref{eq:hex}). This gives the well-known expression \cite{Gurevich-1996},
\begin{align}
H^{\rm 4m} = \frac{1}{V} \sum_{\vq} \sum_{\vq_2} \sum_{\vq'} \sum_{\vq_2'} U^{\rm 4m,0}_{\vq,\vq_2;\vq',\vq'_2} a^{\ast}_{\vq} a^{\ast}_{\vq_2} a_{\vq'} a_{\vq'_2} \label{eq:h4m}
\end{align}
with the symmetrized matrix element in the long-wavelength limit
\begin{equation}
U^{\rm 4m,0}_{\vq, \vq_2 ; \vq' , \vq_2'} = \sqrt{2} J a^4 (\vq \cdot \vq_2) \delta_{\vq+\vq_2,\vq'+\vq_2'} .   \label{eq:u4m}
\end{equation}

{\em Two-magnon-phonon interaction.---} Upon inclusion of exchange-based magnon-phonon coupling, the magneto-elastic coupling (\ref{eq:hme}) is extended by spatial derivatives of the spin density \cite{Gurevich-1996},
\begin{equation}
H^{\rm me} = \int\!\frac{dV}{a^3} \sum_{\alpha,\beta} (A_{\alpha\beta} e_{\alpha\beta} \partial_{\alpha} \vs \cdot \partial_{\beta} \vs + A'_{\alpha\beta} e_{\alpha\alpha} |\partial_{\beta} \vs|^2 ), \label{eq:mel_ex}
\end{equation}
where again all processes allowed by symmetry in a cubic crystal are taken into account. The coupling tensors are $A_{\alpha\beta} = A_1 \delta_{\alpha\beta} + A_2 (1-\delta_{\alpha\beta})$ and $A'_{\alpha\beta} = A' (1-\delta_{\alpha\beta})$.

In a simple model these processes can be reproduced by nearest and next-nearest-neighbor Heisenberg exchange coupling (\ref{eq:hex}) as well as a next-nearest-neighbor transverse exchange coupling due to super-exchange via nearest-neighbors. As explained in detail in Sec. \ref{sec:mp1}, magnon-phonon coupling appears when one takes into account that the magnitudes of the exchange couplings $J_{ij}$, $D_{ij}$ depend on the precise positions of the lattice atoms. In a simple cubic lattice, symmetry requires that for nearest-neighbor sites $i$ and $j$ the isotropic exchange coupling $J_{ij}$ depends on the distance $r_{ij} = |\vr_i-\vr_j|$ only. For small displacements $\vu_j$ of the lattice atoms, the nearest-neighbor exchange coupling can then be approximated as
\begin{align}
  J \to J + J' \vu_{ij} \cdot \ve_{ij},  
  \label{eq:Japprox}
\end{align}
where $\ve_{ij} = (\vr_i - \vr_j)/|\vr_i-\vr_j|$ is the unit vector connecting lattice sites $i$ and $j$ and $J' = dJ/d r_{ij}$. We note that in general the strength of the exchange coupling $J_2$ between next-nearest-neighbor spins $1$ and $2$ not only depends on the displacements $\vu_1$ and $\vu_2$ of the positions of the spins $1$ and $2$, but (via super-exchange) also on the displacements $\vu_3$ and $\vu_4$ of the two lattice site positions intermediate between $1$ and $2$. Denoting the corresponding spatial derivatives by $J'_{2\parallel}$ and $J'_{2\perp}$, respectively, we find that upon inclusion of next-nearest-neighbor Heisenberg exchange the simple cubic model reproduces the full magneto-elastic magnon-phonon Hamiltonian (\ref{eq:mel_ex}), with 
\begin{align}
  A_1 =&\, \frac{a^3}{2} J' + \frac{a^3}{\sqrt{2}} (2 J'_{2\parallel} + J'_{2\perp}), \nonumber \\ 
  A_2 =&\, \frac{a^3}{\sqrt{2}} (2 J'_{2\parallel} - J'_{2\perp}), \nonumber \\
  A_3 =&\, \frac{a^3}{2 \sqrt{2}} (2 J'_{2\parallel} + J'_{2\perp}).
\end{align}
We specialize to an isotropic medium \cite{Kaganov-1961-2} with the choice $A_1 = A_2$ and $A_3=0$. This can be achieved by choosing the Heisenberg exchange constants as $J'_{2 \perp} = -2J'_{2\parallel}$ and $J'_{2\parallel} = J' / 4\sqrt{2}$. We find a magnon-phonon Hamiltonian of the form
\begin{align}
H^{\rm mp} =&\ \frac{1}{\sqrt{V}} \sum_{\vq,\lambda} \sum_{\vq_2} \sum_{\vq'} U^{\rm mp,0}_{\vq, \vq_2 \lambda ; \vq'} b^{\ast}_{\vq_2,4} (b^*_{\vq,\lambda}+b_{-\vq,\lambda}) b_{\vq',4} \label{eq:he}
\end{align}
with
\begin{equation}
  U^{\rm mp,0}_{\vq\lambda,\vq_2 ; \vq'} = U^{\rm ex,0}_{\vq\lambda,\vq_2 ; \vq'} + U^{\rm pd,0}_{\vq\lambda,\vq_2 ; \vq'}. \label{eq:Ump0}
\end{equation}
In the continuum limit the matrix element from Heisenberg exchange reads
\begin{align}
U^{\rm ex,0}_{\vq\lambda,\vq_2;\vq'} =&\, J' S a^3 \sqrt{\frac{\hbar a^3}{8 m \omega^0_{\vq,\lambda}}} i [(\ve_{\vq,\lambda}\cdot\vq')(\vq\cdot\vq_2) \nonumber\\
&\, + (\ve_{\vq,\lambda} \cdot \vq_2) (\vq\cdot\vq')] \delta_{\vq+\vq_2,\vq'}  , \label{eq:hex-ph}
\end{align}
whereas the contribution from anisotropic exchange is
\begin{equation}
U^{\rm pd,0}_{\vq\lambda,\vq_2;\vq'} = -i D \sqrt{\frac{2 \hbar S^2 a^3}{m \omega^0_{\vq,\lambda}}} \sum_{\alpha\neq \beta} q_{\alpha} e_{\vq\lambda,\beta} .
\end{equation}
To find a numerical value for the coupling constant $J'$, we approximate $J' \approx J/a$.

{\em Relativistic two-magnon-phonon interaction.---} 
As the anisotropic exchange Hamiltonian does not conserve spin, besides the spin-conserving two-magnon-phonon contribution it also yields processes where a phonon is converted into two magnons via
\begin{align}
H^{\rm rel} =&\ \frac{1}{\sqrt{V}} \sum_{\vq,\lambda} \sum_{\vq'} \sum_{\vq_2'} U^{\rm rel,0}_{\vq \lambda ;  \vq' , \vq_2'} b^{\ast}_{\vq,\lambda} b_{\vq',4} b_{\vq_2',4} .
\label{eq:hrel}
\end{align}
The matrix element in the long-wavelength limit is
\begin{align}
U^{\rm rel,0}_{\vq \lambda ;  \vq' , \vq_2'} =&\, -i \sqrt{\frac{\hbar S^2 a^3}{18 m \omega^0_{\vq,\lambda}}} \sum_{\alpha,\beta} q_{\alpha} w_{\alpha} w_{\beta} e_{\vq\lambda,\beta} \nonumber\\
&\, \ \ \ \ \ \ \times [2 D (1-\delta_{\alpha,\beta}) - a D' \delta_{\alpha,\beta}],
\label{eq:Urel0}
\end{align}
where we again chose the magnetization direction $\ve$ along the $(111)$ direction.

{\em Angle-independent models.---} For a numerical computation of the inelastic relaxation rates, we replace the matrix elements of these four inelastic interaction channels by a phenomenological model, for which the matrix elements become statistical quantities with zero mean and with a variance chosen in such a way that the angle-averaged transition rates are the same as for the microscopic model. Specifically, for the three-phonon interactions, we set
\begin{align}
  \langle |U^{{\rm 3p},0}_{\vq\lambda,\vq_2\lambda_2;\vq'\lambda'}|^2 \rangle = |u^{\rm 3p}(\omega^0_{\vq},\omega^0_{\vq_2};\omega^0_{\vq'})|^2,
\end{align}
where the variance $|u^{\rm 3p}(\omega^0_{\vq},\omega^0_{\vq_2};\omega^0_{\vq'})|^2$ is given by 
\begin{align}
  |u^{\rm 3p}(\omega,\omega_2;\omega')|^2 =&\,
  \frac{1}{V^3 \cE^{{\rm p},0}_0(\omega)
  \cE^{{\rm p},0}_0(\omega')
  \cE^{{\rm p},0}_0(\omega_2)}
   \\ &\, \mbox{} \times
  \sum_{\vq,\lambda} \sum_{\vq_2,\lambda_2} \sum_{\vq',\lambda'} 
  |U^{{\rm 3p},0}_{\vq\lambda,\vq_2\lambda_2;\vq'\lambda'}|^2
  \nonumber \\ &\, \mbox{} \times
  \delta(\omega - \omega^0_{\vq,\lambda})
  \delta(\omega_2 - \omega^0_{\vq_2,\lambda_2})
  \nonumber \\ &\, \nonumber \mbox{} \times
  \delta(\omega' - \omega^0_{\vq',\lambda'}) .
\end{align}
On the right-hand side of this equation, the matrix element $U^{{\rm 3p},0}_{\vq\lambda,\vq_2\lambda_2;\vq'\lambda'}$ is taken from the microscopic model, see Eq.\ (\ref{eq:U3p0}). The density of states $\cE_0^{{\rm p},0}$ is defined as
\begin{equation}
  \cE^{{\rm p},0}_0(\omega) = \frac{1}{V} \sum_{\vq,\lambda} \delta(\omega - \omega^0_{\vq,\lambda}).
\end{equation}
By construction, in the absence of magnon-phonon coupling this change of the model results in the same relaxation matrix $\cG^0$ as the microscopic model (\ref{eq:U3p0}).

Similarly, for the three-magnon interaction, the four-magnon interaction, and the two-magnon-phonon interaction we set
\begin{align}
  \langle |U^{{\rm 3m},0}_{\vq,\vq_2;\vq'}|^2 \rangle =&\, |u^{\rm 3m}(\omega^0_{\vq},\omega^0_{\vq_2};\omega^0_{\vq'})|^2, \nonumber \\
  \langle |U^{{\rm 4m},0}_{\vq,\vq_2;\vq',\vq_2'}|^2 \rangle =&\, |u^{\rm 4m}(\omega^0_{\vq},\omega^0_{\vq_2};\omega^0_{\vq'},\omega^0_{\vq_2'})|^2, \nonumber \\
  \langle |U^{{\rm mp},0}_{\vq\lambda,\vq_2;\vq'}|^2 \rangle =&\, |u^{\rm mp}(\omega^0_{\vq},\omega^0_{\vq_2};\omega^0_{\vq'})|^2, \nonumber \\
   \langle |U^{{\rm rel},0}_{\vq\lambda;\vq' , \vq_2'}|^2 \rangle =&\, |u^{\rm rel}(\omega^0_{\vq};\omega^0_{\vq'},\omega^0_{\vq_2'})|^2,
\end{align}
with
\begin{widetext}
\begin{align}
  |u^{\rm 3m}(\omega,\omega_2;\omega')|^2 =&\,
  \frac{1}{V^3
  \cE^{{\rm m},0}_0(\omega)
  \cE^{{\rm m},0}_0(\omega')
  \cE^{{\rm m},0}_0(\omega_2)}
  \sum_{\vq} \sum_{\vq_2} \sum_{\vq'} 
  |U^{{\rm 3m},0}_{\vq,\vq_2;\vq'}|^2
  \delta(\omega - \omega^0_{\vq,4})
  \delta(\omega_2 - \omega^0_{\vq_2,4})
  \delta(\omega' - \omega^0_{\vq',4}), \nonumber \\
  |u^{\rm 4m}(\omega,\omega_2;\omega' , \omega_2')|^2 =&\,
  \frac{1}{V^3
  \cE^{{\rm m},0}_0(\omega)
  \cE^{{\rm m},0}_0(\omega')
  \cE^{{\rm m},0}_0(\omega_2)
  \cE^{{\rm m},0}_0(\omega_2')}
  \sum_{\vq} \sum_{\vq_2} \sum_{\vq'} \sum_{\vq_2'}
  |U^{{\rm 4m},0}_{\vq,\vq_2;\vq',\vq_2'}|^2
  \nonumber \\ &\, \ \ \ \ \mbox{} \times
  \delta(\omega - \omega^0_{\vq,4})
  \delta(\omega_2 - \omega^0_{\vq_2,4})
  \delta(\omega' - \omega^0_{\vq',4})
  \delta(\omega_2' - \omega^0_{\vq_2',4})
  , \nonumber \\
  |u^{\rm mp}(\omega,\omega_2;\omega')|^2 =&\,
  \frac{1}{V^3
  \cE^{{\rm p},0}_0(\omega)
  \cE^{{\rm m},0}_0(\omega')
  \cE^{{\rm m},0}_0(\omega_2)}
  \sum_{\vq,\lambda} \sum_{\vq_2} \sum_{\vq'} 
  |U^{{\rm mp},0}_{\vq\lambda,\vq_2;\vq'}|^2
  \delta(\omega - \omega^0_{\vq,\lambda})
  \delta(\omega_2 - \omega^0_{\vq_2,4})
  \delta(\omega' - \omega^0_{\vq',4}) , \nonumber \\
   |u^{\rm rel}(\omega; \omega' , \omega_2')|^2 =&\,
  \frac{1}{V^3
  \cE^{{\rm p},0}_0(\omega)
  \cE^{{\rm m},0}_0(\omega')
  \cE^{{\rm m},0}_0(\omega_2')}
  \sum_{\vq,\lambda} \sum_{\vq'} \sum_{\vq_2'} 
  |U^{{\rm rel},0}_{\vq\lambda;\vq',\vq_2'}|^2
  \delta(\omega - \omega^0_{\vq,\lambda})
  \delta(\omega' - \omega^0_{\vq',4})
  \delta(\omega'_2 - \omega^0_{\vq_2',4}) .
\end{align}
\end{widetext}
As before, the matrix elements $U^{{\rm 3m},0}_{\vq,\vq_2;\vq'}$, $U^{{\rm 4m},0}_{\vq,\vq_2;\vq',\vq_2'}$, $U^{{\rm mp},0}_{\vq\lambda,\vq_2;\vq'}$, and $U^{{\rm rel},0}_{\vq\lambda ; \vq',\vq_2'}$ on the right-hand side of this equation are taken from the microscopic models, see Eqs.\ (\ref{eq:U3m0}), (\ref{eq:u4m}), (\ref{eq:Ump0}), and (\ref{eq:Urel0}), respectively.

This replacement is motivated by practical considerations --- since the absence of an angular dependence considerably simplifies the calculation of the inelastic rates --- but also by the physical consideration that as long as impurity scattering is the dominant scattering process, the propagation direction of magnon-polaron modes is subject to fast fluctuations, calling for an effective description in terms of frequencies only. For the implementation in the angular summations in Eq.\ (\ref{eq:lambdaAverage}) we replace the frequency arguments $\omega_{\vq,\lambda}^0$ of the phonon modes and $\omega^0_{\vq,4}$ of the magnon mode by the frequency $\omega_{\vq,\nu}$ of the corresponding magnon-polaron mode.

\section{No magnon-phonon coupling}
\label{app:magpho}

Without magnon-phonon coupling and without inelastic processes, phonon and magnon modes obey separate linearized Boltzmann equations. 

{\em Magnons.---} For the magnon mode, we denote the isotropic and anisotropic moments of the linearized distribution function by $\psi_{0,{\rm m}}(\omega)$ and $\psi_{1,{\rm m}}(\omega)$, respectively. The linearized Boltzmann equation for these moments reads
\begin{align}
  \frac{\partial \psi_{1,\rm m}(\omega)}{\partial x} =&\ 0 , \nonumber\\
  \frac{\partial \psi_{0,\rm m}(\omega)}{\partial x} =&\ - \cG^1_{\rm m}(\omega) \psi_{1,\rm m}(\omega),
  \label{eq:bemag}
\end{align}
with
\begin{equation}
  \cG^1_{\rm m}(\omega) = \frac{v_{\rm m}(\omega)}{3 l_{\rm mi}(\omega)},
\end{equation}
where the magnon-impurity length $l_{\rm mi}(\omega)$ is given in Eq.\ (\ref{eq:mfpm}) and the magnon group velocity $v_{\rm m}(\omega) = \partial \omega_{\rm m}(q)/\partial q$, with $\omega_{\rm m}(q) = \mu B + J S a^2 q^2$ the magnon frequency. 

The magnon mode is fully reflected at the IF interface at $x=0$,
\begin{equation}
  R_{{\rm m},{\rm m}}(\omega,\vq_{\perp}) = 1,
\end{equation}
whereas at the FN interface at $x=L$,
\begin{align}
  R_{{\rm m},{\rm m}}(\omega,\vq_{\perp}) =&\, 1 - P_{4{\rm N}}(\omega,\vq_{\perp}) \nonumber \\ =&\,
      \left| \frac{4 \pi M_{\rm s} J S a^2 q_x(\omega,\vq_{\perp}) - \mu \omega \sigma_{\uparrow\downarrow}}{4 \pi M_{\rm s} J S a^2 q_x(\omega,\vp_{\perp}) + \mu \omega \sigma_{\uparrow\downarrow}} \right|^2,
\end{align}
where
\begin{equation}
  q_x(\omega,\vq_{\perp}) = \sqrt{\frac{\omega - \mu B}{J S a^2} - q_{\perp}^2}
\end{equation}
is the $x$ component of the wavevector of a magnon propagating towards the FN interface.
The boundary conditions (\ref{eq:bcL}) and (\ref{eq:bcR}) at the IF and FN interfaces become
\begin{align}
\psi_{1,\rm m}(\omega,0) =&\ 0 , \nonumber\\
\psi_{1,\rm m}(\omega,L) =&\ \frac{3}{2} \frac{1-\cR_{0,{\rm m}}}{1+\cR_{1,{\rm m}}} [\psi_{0,\rm m}(\omega,L) - \psi_{\rm N}] ,
\label{eq:bcmag}
\end{align}
where $\cR_{0,{\rm m}}$ and $\cR_{1,{\rm m}}$ are angular averages of the reflection coefficient $R_{{\rm m},{\rm m}}$, see Eq.\ (\ref{eq:Rdef}). The unique solution to these equations is
\begin{equation}
  \psi_{0,{\rm m}}(\omega,x) = \psi_{{\rm N}}(\omega),\ \
  \psi_{1,{\rm m}}(\omega,x) = 0.
\end{equation}

{\em Phonons.---}
For the phonon modes, we use the label $\lambda=1$ for the longitudinal mode and the labels $\lambda=2,3$ for the transverse phonon modes. The linearized Boltzmann equation for the isotropic and anisotropic moments $\psi_{0,\lambda}(\omega)$ and $\psi_{1,\lambda}(\omega)$ of the distribution function then reads
\begin{align}
\frac{\partial \psi_{1,\lambda}(\omega)}{\partial x} =&\ \sum_{\lambda'} \cG^0_{{\rm p},\lambda,\lambda'}(\omega) [ \psi_{0,\lambda'}(\omega) - \psi_{0,\lambda}(\omega) ] , \nonumber\\
\frac{\partial \psi_{0,\lambda}(\omega)}{\partial x} =&\ \sum_{\lambda'} \cG^1_{{\rm p},\lambda,\lambda'}(\omega) [ \psi_{1,\lambda'}(\omega) - \psi_{1,\lambda}(\omega) ] ,
\label{eq:bepho}
\end{align}
where the matrices $\cG^0_{\rm p}$ and $\cG^1_{\rm p}$ are
\begin{align}
  \cG_{\rm p}^0 =&\,
  \frac{3 \tau_{\rm pi}(\omega)^{-1}}{c_{\rm t}^2(2 c_{\rm l}^3+c_{\rm t}^3)}
  \begin{pmatrix} 2 c_{\rm l} c_{\rm t}^2 & - c_{\rm l} c_{\rm t}^2 & - c_{\rm l} c_{\rm t}^2 \\ -c_{\rm t}^3 & c_{\rm l}^3 + c_{\rm t}^3 & - c_{\rm l}^3 \\
-c_{\rm t}^3 & -c_{\rm l}^3 & c_{\rm l}^3 + c_{\rm t}^3 \end{pmatrix}, \\
  \cG_{\rm p}^1 =&\, \tau_{\rm pi}(\omega)^{-1} \begin{pmatrix} 1 & 0 & 0 \\ 0 & 1 & 0 \\ 0 & 0 & 1 \end{pmatrix}.
  \label{eq:Lambda1p}
\end{align}
Here $c_{\rm l}$ and $c_{\rm t}$ are the longitudinal and transverse phonon velocities and $\tau_{\rm pi}$ is the mean scattering time for phonon-impurity scattering, see Eq.\ (\ref{eq:tau_pi}). The equilibration lengths $\lambda_1$ and $\lambda_2$ corresponding to Eq.\ (\ref{eq:Lambda1p}) are
\begin{equation}
  \lambda_1 = \tau_{\rm pi} \sqrt{\frac{c_{\rm t}^{3} + 2 c_{\rm l}^{3}}{3(2 c_{\rm l} + c_{\rm t})}}, \ \
  \lambda_2 = \tau_{\rm pi} c_{\rm t} \sqrt{\frac{1}{3}}.
\end{equation}
Without magnon-phonon coupling, phonon modes are fully transmitted at the interfaces, {\em i.e.},
\begin{equation}
  T_{\lambda\lambda'}(\omega,\vq_{\perp}) = \delta_{\lambda\lambda'},\ \
  R_{\lambda\lambda'}(\omega,\vq_{\perp}) = 0,\ \
  \lambda=1,2,3.
\end{equation}
The boundary conditions at the IF and FN interfaces at $x=0$ and $x=L$ are then found to be
\begin{align}
\psi_{1,\lambda}(\omega,0) =&\ \frac{3}{2} [\psi_{\rm I} - \psi_{0,\lambda}(\omega,0)] , \nonumber \\
\psi_{1,\lambda}(\omega,L) =&\  \frac{3}{2} [\psi_{0,\lambda}(\omega,L) - \psi_{\rm N}] .
\label{eq:bcpho}
\end{align}
For short systems with length $L \ll \lambda_1$, the solution for the phonon distribution function is $\psi_{0,\lambda} = (\psi_{\rm I} + \psi_{\rm N})/2$. For a long system with length $L \gg \lambda_2$, the solution is $\psi_{0,\lambda} = \psi_{\rm I} + (\psi_{\rm N} - \psi_{\rm I}) x/L$. A small correction near the FN interface can be obtained from Eq.\ (\ref{eq:iso_thick}) upon taking Eq.\ (\ref{eq:Lambda1p}) for the matrices $\cG_1$ and $\cG = \sqrt{\cG_1 \cG_0}$ and replacing $\vu_{0,0}$ by the three-component vector $(1,1,1)^{\rm T}$.

\section{Ballistic systems}
\label{app:ballistic}

\begin{figure}
\includegraphics[width=\linewidth]{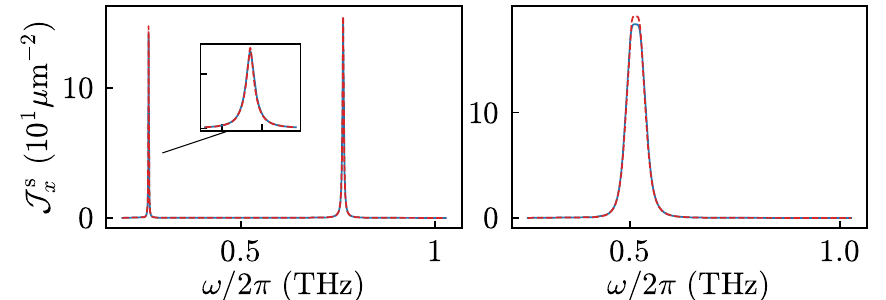}
\caption{\label{fig:spincurrent2} Frequency-resolved spin current $j^{\rm s}_x(\omega) =\hbar {\cal J}^{\rm s}_x(\omega) \Delta T/T$ at the FN interface for $T = 10\,{\rm K}$ and $B=7\,{\rm T}$ (left) and for $T = 10\,{\rm K}$ and $B = 9.2\,{\rm T}$ (right), calculated from the short-length prediction (\ref{eq:iso_thin}) based on the ansatz (\ref{eq:psimoments}) (blue, solid) and the exact result based on Eq.\ (\ref{eq:psi_ballistic}) (red, dashed). System parameters are taken from Table \ref{tab:values}. The magnetic field $B = 9.2\,{\rm T}$ is the highest magnetic field for which the dispersions of magnons and phonons touch.}
\end{figure}

Strictly speaking the linearization of the polaron distribution with one isotropic and one anisotropic moment as in Eq.\ (\ref{eq:psimoments}) is only applicable in the diffusive regime. For very short systems the transport is ballistic and the polarons only scatter off the interfaces. In this limit the distributions for left and right-moving polarons can be solved directly by using the boundary conditions (\ref{eq:bboundary}) and (\ref{eq:bboundary2}), without making the ansatz (\ref{eq:psimoments}). Parameterizing wavevectors $\vq$ via the frequency $\omega$, the transverse momentum $\vq_{\perp}$, and the propagation direction, the linear-response ansatz for the distribution function reads
\begin{align}
  n_{\nu}(\omega,\vq_{\perp},\pm) = n^0(\omega) + \omega \left(-\frac{\partial n^0}{\partial \omega}\right) \psi_{\nu}(\omega,\vq_{\perp},\pm).
\end{align}
To solve for the distribution function, we use a four-component vector notation,
\begin{align}
\label{eq:psi_ballistic}
  \vpsi(\omega,\vq_{\perp},+) =&\ (\openone_4 - R_{\rm I} R_{\rm N})^{-1}
  \\ &\, \nonumber \mbox{} \times
  [ (\openone_4-R_{\rm I}) \vpsi_{\rm I}(\omega) + R_{\rm I}(\openone_4-R_{\rm N}) \vpsi_{\rm N}(\omega) ] , \nonumber\\
  \vpsi(\omega,\vq_{\perp},-) =&\ (\openone_4 - R_{\rm I} R_{\rm N})^{-1}
  \\ &\, \nonumber \mbox{} \times
[ (\openone_4-R_{\rm N}) \vpsi_{\rm N}(\omega) + R_{\rm N}(\openone_4-R_{\rm I}) \vpsi_{\rm I}(\omega) ] .
\end{align}
Here we suppressed the arguments $(\omega,\vq_{\perp})$ of the matrices $R_{\rm I}$ and $R_{\rm N}$. The corresponding frequency-resolved spin current density $j^{\rm s}_x(\omega)$ is, compare with Eq.\ (\ref{eq:jsomega}),
\begin{align}
j^{\rm s}_{x}(\omega) =&\, \hbar\omega \left(-\frac{\partial n^0}{\partial \omega}\right) \sum_{\nu} \int \frac{d\vq_{\perp}}{(2\pi)^3} P_{\nu{\rm N}}(\omega,\vq_{\perp}) \nonumber\\
&\, \times \psi_{\nu}(\omega,\vq_{\perp},+)  . 
\label{eq:jballistic}
\end{align}
In Fig.\ \ref{fig:spincurrent2} we compare the frequency-resolved spin current density calculated using the exact solution (\ref{eq:psi_ballistic}) and the spin current density based on the ansatz (\ref{eq:psimoments}). While the approximate calculation differs quantitatively up to a factor $\sim 1.05$, all qualitative features are correctly reproduced.

%

\end{document}